\documentclass[12pt]{article}
\pdfoutput=1
\usepackage[nosort]{cite}

\usepackage{draft} 
\usepackage{hyperref}
 \usepackage{graphicx,color,subfig,upgreek,mathtools}
\usepackage{array,amsfonts,amssymb,amsmath,multirow}
\usepackage{mathabx,epsfig}
\usepackage{pifont}
\usepackage{cite}
\usepackage{mciteplus}
\usepackage{skak}
\DeclareFontFamily{OT1}{pzc}{}
\DeclareFontShape{OT1}{pzc}{m}{it}{<-> s * [1.10] pzcmi7t}{}
\DeclareMathAlphabet{\mathpzc}{OT1}{pzc}{m}{it}

\def\be#1\ee{\begin{align}#1\end{align}}

\def\SL{{\mathscr L}}

\def\CB{{\mathcal B}}

\def\CH{{\mathcal H}}

\def\CN{{\mathcal N}}

\def\CP{{\mathcal P}}

\def\CS{{\mathcal S}}
\def\CT{{\mathcal T}}

\def\CZ{{\mathcal Z}}

\def\Pol{\mathrm{Pol}}
\def\MCG{\mathrm{MCG}}
\def\C{\mathbb{C}}
\def\L{\Lambda}
\def\Z{\mathbb{Z}}
\def\R{\mathbb{R}}
\def\SL{{SL}}
\def\br{\mathrm{br}}

\def\acts{\mathrel{\reflectbox{$\righttoleftarrow$}}}

\def\tilde{\widetilde}
\renewcommand{\bar}{\overline}
\renewcommand{\hat}{\widehat}
\renewcommand{\d}{\mathrm{d}}
\def\^{{\wedge}}
\def\*{{\star}}

\definecolor{ao}{rgb}{0.13, 0.55, 0.13}

\usepackage{youngtab}
\usepackage{tikz-cd}
\usepackage{datetime}
\begin{document}

\begin{titlepage}

\preprint{CALT-TH-2020-045}

\begin{center}

\hfill \\
\hfill \\
\vskip 1cm

\title{Generalized Global Symmetries of $T[M]$ Theories. I}

\author{Sergei Gukov$^1$, Po-Shen Hsin$^1$, Du Pei$^{2}$
}

\address{$^{1}$Walter Burke Institute for Theoretical Physics, California Institute of Technology,
Pasadena, CA 91125, USA}

\address{$^2$Center of Mathematical Sciences and Applications, Harvard University, Cambridge, MA 02138, USA}

\end{center}

\abstract{
We study reductions of 6d theories on a $d$-dimensional manifold $M_d$, focusing on the interplay between symmetries, anomalies, and dynamics of the resulting $(6-d)$-dimensional theory $T[M_d]$. We refine and generalize the notion of ``polarization'' to \emph{polarization on $M_d$}, which serves to fix the spectrum of local and extended operators in $T[M_d]$. Another important feature of theories $T[M_d]$ is that they often possess higher-group symmetries, such as 2-group and 3-group symmetries. We study the origin of such symmetries as well as physical implications including symmetry breaking and symmetry enhancement in the renormalization group flow. To better probe the IR physics, we also investigate the ’t Hooft anomaly of 5d Chern--Simons matter theories.
The present paper focuses on developing the general framework as well as the special case of $d=0$ and 1, while an upcoming paper will discuss the case of $d=2$, 3 and 4.}

\vfill

\today

\vfill

\end{titlepage}

\eject

\tableofcontents

\unitlength = .8mm

\setcounter{tocdepth}{3}

\section{Introduction}

\hfill{\vbox{\hbox{\it Nothing in physics seems so hopeful to as the idea}
\hbox{\it that it is possible for a theory to have a high degree}	
\hbox{\it of symmetry was hidden from us in everyday life.}
\hbox{\it The physicist's task is to find this deeper symmetry.} }}

\hfill{\vbox{\hbox{Steven Weinberg}}}

\medskip

Throughout the history of physics, symmetries played a central role. They help to identify physical laws and principles that describe observable phenomena. In quantum physics, they guide us toward non-perturbative formulations of Quantum Field Theory and exact mathematical descriptions of strongly coupled systems.

In this paper, we study quantum field theories in 6d and their dimensional reductions. Considering such problems produced many interesting connections between low-dimensional topology and dynamics of quantum field theories. For example, on the physics side, it led to a very large collections of novel dualities and, on the math side, to novel invariants of manifolds.

A large class of 6d theories whose compactifications have been studied are the so-called ``relative theories'' \cite{Freed:2012bs}, which means they have an anomalous 2-form global symmetry with the anomaly captured by a 7d Topological Quantum Field Theory (TQFT) with possibly degenerate ground states. As a consequence, such 6d theories alone are not QFTs in the usual sense, {\it e.g.}~they do not have well-defined partition functions. In the literature, a notion of polarization was introduced in order to define a partition function on a given 6-manifold $M_6$ \cite{Witten:1996hc,Witten:1998wy,Witten:2009at,Freed:2012bs}.\footnote{
This is similar to the 2d chiral Wess--Zumino--Witten conformal field theories living on the boundary of 3d Chern--Simons theories, where the $2\d$ theory has chiral conformal blocks that correspond to vectors in the Hilbert space of the $3\d$ theory \cite{Witten:1988hf,Elitzur:1989nr}. See also \cite{Freed:2018cec} for a discussion on the interplay between polarization in 3d TQFTs and the boundary lattice models.
} 
 A closer inspection reveals, however, that a refinement is needed in order for the partition function to be fully specified and unambiguous.

One of the main goals of the present paper is to refine and generalize the notion of polarization, to the notion of  \emph{refined polarization on $M_d$}, so that not only it specifies the partition function but also defines a full-fledged quantum field theory upon partial compactification on a $d$-manifold $M_d$,
\begin{equation}
    \text{(refined) polarization $\CP$ on $M_d$ } \leadsto \text{  $(6-d)$ dimensional QFT $T[M_d,\CP]$} 
\end{equation}

\noindent
Questions we address in this paper (and in the sequel \cite{Gukov:2020inprepare}) include
\begin{itemize}
    \item How to classify polarizations on a given $M_d$?
    \item How does a polarization $\CP$ control the symmetries and anomalies of $T[M_d,\CP]$? 
    \item What is the spectrum of charged operators in the theory $T[M_d,\CP]$?
    \item How does the mapping class group of $M_d$ act on polarizations? And, what does this tell us about dualities of the theory $T[M_d,\CP]$? 
    \item What happens when $M_d$ itself has symmetries ({\it i.e.} isometries)? 
\end{itemize}

One useful way to tackle these questions about the reduction of 6d theory is to reduce the entire 6d-7d coupled system to lower dimensions. Then, it turns out that all of the above questions can be rephrased in terms of the $(7-d)$-dimensional TQFT $\CT^{\text{bulk}}[M_d]$. For example, one can shed much light onto the first question by relating polarizations with \emph{topological boundary conditions} of the TQFT $\CT^{\text{bulk}}[M_d]$. One advantage of this approach is that is it \emph{universal}, meaning that it does not depend on specific details of the boundary 6d theory. This is a typical feature in the study of quantum field theory via symmetries and anomalies. 

Moreover, this approach naturally leads to a systematic classification of polarizations on any $M_d$ by relating them to symmetry protected topological phases (SPT phases) in $6-d$ dimensions. It turns out that polarizations can be naturally sorted into two classes that we call \emph{pure polarizations} and \emph{mixed polarizations}, and they lead to theories with different types of extended operators. While the former have been used, sometimes implicitly, in the reduction of 6d theories on various $M_d$ in the literature (see {\it e.g.}~\cite{Vafa:1994tf,kapustin2007electric,Tachikawa:2013hya,Eckhard:2019jgg}), 
mixed polarizations give rise to novel dimensional reductions, enlarging the zoo of $T[M_d]$ theories even for a particular $M_d$ such as $S^1$ or a Riemann surface (thus, producing new theories of class $\CS$). For each type of polarization, we discuss how symmetries, anomalies, and spectrum of charged operators are constrained and determined by the polarization. 

Another interesting feature of $T[M_d]$ theories is the presence of \emph{higher-group symmetries} \cite{Kapustin:2013uxa,Sharpe:2015mja,Cordova:2018cvg,Benini:2018reh} when $M_d$ itself has isometries. In other words, the (0-form) symmetry coming from the isometry of $M_d$ often combines in a non-trivial way with higher-form symmetries of $T[M_d]$, whose gauge transformations are modified in the presence of a background gauge field for the isometry symmetry.
Higher-group symmetry has interesting implications for physics of the effective low-dimensional theory, including constraints on the symmetry breaking patterns in the renormalization group flows or enhancement of global symmetry. We will discuss these applications to $5\d$ theories in this paper,\footnote{
See also \cite{Morrison:2020ool,Albertini:2020mdx,Apruzzi:2020zot,Bhardwaj:2020phs,Cordova:2020tij} for some recent discussions of the higher-form symmetry (without non-trivial higher-group structures) in 5d and 6d.} relegating applications to lower dimensional theories to the upcoming work \cite{Gukov:2020inprepare}.

We also compute the 't Hooft anomaly for the one-form symmetry in $5\d$ $SU(N)$ gauge theory with  Chern--Simons term at level $k$, where the Chern--Simons term breaks the $\mathbb{Z}_N$ one-form center symmetry to a subgroup $\mathbb{Z}_{\gcd(N,k)}$ and contributes to the 't Hooft anomaly of the remaining one-form symmetry. Similarly, we compute the 't Hooft anomaly for the flavor symmetry when matter fields are included. 
The anomaly has interesting applications to the strongly-coupled dynamics and can be used to test dualities between different 5d gauge theories, which will be discussed in another work. 

The present paper is organized as follows. In Section~\ref{sec:compactification0d}, we discuss partition functions of 6d relative theories (which can be alternatively viewed as the special case of $T[M_d]$ with $d=6$, and $T[M_6]$ being a zero-dimensional theory), reviewing and refining the notion of polarization. In Section~\ref{sec:GeneralTheory}, we generalize the notion of polarization to arbitrary $M_d$, and build on it a theoretical framework for studying reductions of 6d theories.\footnote{In the past, certain decoupling procedures were used to construct different global forms of the theory $T[M_d]$ (see {\it e.g.}~\cite{Eckhard:2019jgg} where higher-form symmetries coming from 6d ${\cal N}=(2,0)$ theories also play an important role). Our approach does not make use of any decoupling procedure. See also \cite{Garcia-Etxebarria:2019cnb} for an alternative approach using boundary conditions of RR fields in type IIB string theory.} Section~\ref{sec:compactification6d} is devoted to a seemingly degenerate case of $d=0$ with $M_d$ being a point, which in fact turns out to be remarkably rich. Section~\ref{sec:compactification5d} focuses on the case of $d=1$ and $M_d=S^1$, where we will find many examples of mixed polarizations and higher-group symmetries. Section~\ref{sec:SO(8)} delves deeper into a special class of 6d and 5d theories associated with the Lie algebra $\frak{so}(8)$.

There are several appendices. In Appendix \ref{sec:7dCSlevel}, we discuss the level quantization of the $7\d$ three-form Chern--Simons theory. In Appendix \ref{sec:ArfK}, we discuss a class of $\mathbb{Z}_2$ higher-form gauge theories in various space-time dimensions. In Appendix \ref{sec:5ddiscretetheta}, we discuss a $5\d$ discrete two-form gauge theory with its symmetry and anomaly, which describes the discrete theta-angles in the $5\d$ gauge theories relevant to this paper. In Appendix \ref{sec:mixeddiscretetheta}, we discuss a discrete gauge theory that couples gauge fields of different degrees, which describes the mixed discrete theta-angles. In Appendix \ref{sec:linkingform}, we review some properties of linking forms in various dimensions. In Appendix \ref{sec:7ddualities}, we discuss dualities between $7\d$ three-form Chern--Simons theories. In Appendix \ref{sec:factorize}, we show that fermionic Abelian Chern--Simons theory factorizes into a bosonic Abelian TQFT and an invertible fermionic TQFT.

\subsection*{Summary of the upcoming work \cite{Gukov:2020inprepare}}

In a companion paper \cite{Gukov:2020inprepare}, we will apply the method developed here to systematically explore compactifications of 6d theories on manifolds $M_d$ with $d=2$, $3$, and $4$.
For instance, we will demonstrate how discrete theta angles in 4d gauge theories given by the Pontryagin square appear from choosing suitable polarizations, verifying a proposal in \cite{Tachikawa:2013hya}. We also 
find higher-group symmetries to be almost ubiquitous as long as one keeps 
enough degrees of freedom from the Kaluza--Klein tower on $M_d$ of finite size.
Just as in other dimensions, the global symmetry and 't Hooft anomaly of the compactified theory $T[M_d,\CP]$ depend on the polarization $\CP$ and can be obtained by studying the compactification of the $7\d$ TQFT $\CT^{\text{bulk}}$ coupled to the $6\d$ theory on the boundary.
In the special case of compactifying a free 6d $U(1)$ tensor multiplet to 3d, this gives a decoupled 3d Abelian Chern--Simons theory~\cite{Gadde:2013sca}, and we show that the mapping class group symmetry reproduces some of the anyon permutation symmetry in the TQFT that is studied systematically in \cite{Delmastro:2019vnj}. We will also show that symmetry consideration can help us study Gluck twist on $S^2\times S^1$, relevant for the construction of exotic smooth 4-manifolds.

\subsection{Frequently used notations}

We list some notations used throughout the paper for quick reference. 

\begin{itemize}
\item[$\CT^{\text{bulk}}$:] A seven-dimensional 3-form Abelian Chern--Simons theory.
    \item[$D$:] Defect group of the 7d TQFT that classifies  3-dimensional operators in the theory.
    \item[$M_d$:] A  connected $d$-dimensional (smooth) manifold.
    \item[{$\CT^{\text{bulk}}[M_d]$}:] A $(7-d)$-dimensional theory obtained by reducing the 7d TQFT on $M_d$.
    \item[$T$:] A six-dimensional quantum field theory living on the boundary of $\CT^{\text{bulk}}$. It has 2-form $D$ symmetry whose anomaly is described by $\CT^{\text{bulk}}$.
    \item[{$T[M_d]$}:] A $(6-d)$-dimensional theory obtained by reducing the 6d theory on $M_d$, which might be a relative theory living on the boundary of $\CT^{\text{bulk}}[M_d]$.
    \item[$H^i(M_d,D)$:] The $i$-th cohomology of $M_d$ with $D$ coefficients. It classifies $(3-i)$-dimensional topological operators in $\CT^{\text{bulk}}[M_d]$.
    
    \item[$\CH(M_6)$:] The Hilbert space of $\CT^{\text{bulk}}$ on $M_6$ or, alternatively, the Hilbert space of the 1d TQFT $\CT^{\text{bulk}}[M_6]$ on a single point.

    \item[$\langle\cdot,\cdot\rangle$:] An anti-symmetric bilinear form on $H^3(M_6,D)$ (with $M_6$ implicit from the context). It measures non-commutativity of operators (labeled by elements in $H^3(M_6,D)$) in the 1d TQFT $\CT^{\text{bulk}}[M_6]$ acting on $\CH(M_6)$.
    \item[$\L$:] A maximal isotropic subgroup of $H^3(M_6,D)$ with respect to $\langle\cdot,\cdot\rangle$, often referred to as a ``polarization.'' It is a set of maximal commuting operators in $\CT^{\text{bulk}}[M_6]$. 
    \item[$q$:] A quadratic function on $\L$ that refines certain (possibly degenerate) symmetric bilinear form on $\L$. Together with $\L$, it leads to a well-defined partition function of the 6d theory $T$ on $M_6$.
     
    \item[$\Pol(M_6)$] The set of polarizations on $M_6$. 
    \item[$\tilde \Pol(M_6)$] The set of refined polarizations $(\L,q)$ on $M_6$. It also classifies topological boundary conditions of $\CT^{\text{bulk}}[M_6]$.
    
    \item[{$T[M_6,(\L,q)]$}:] An absolute 0-dimensional theory constructed from $T[M_6]$ with the refined polarization $(\L,q)$.
    
    \item[$\L^\vee$:] The Pontryagin dual of $\L$. It is the group of $(-1)$-form symmetries of $T[M_6,(\L,q)]$. It is isomorphic to $H^3(M_6,D)/\L$.
    \item[$\bar\L$:] A lift of $\L^\vee$ to $H^3(M_6,D)$, which then can be decomposed into $\L\oplus\bar\L$. A choice of $\bar\L$ leads to an explicit set of basis for the partition vector of $T$  on $M_6$. 
    
    \item[$\tilde \Pol(M_d)$:] The set of refined polarizations on $M_d$. It also classifies topological boundary conditions of $\CT^{\text{bulk}}[M_d]$.
    
    \item[$\CP$:] A refined polarization on $M_d$ (with the manifold understood from the context).
    
    \item[{$T[M_d,\CP]$}:] An absolute $(6-d)$-dimensional theory constructed from $T[M_6]$ with refined polarization $\CP$.
    
    \item[$\CS(\CP)$:] A subgroup of $H^*(M_d,D)$ classifying charged objects in $T[M_d,\CP]$.
    
    \item[$\CS(\CP)_{\text{ind}}$:] A subgroup of $\CS(\CP)$ classifying charged objects that are independent, {\it e.g.}~those which
    exist without the need to be attached to higher-dimensional objects. (See Section~\ref{sec:RemainingSymmetry} for details.)
    
    \item[$L$:] A maximal isotropic subgroup of $H^{d-3\le*\le 3}(M_d,D)$. It is a sum of graded pieces $L^{(i)}$.
    
    \item[$\CP_L$:] A ``pure polarization'' labeled by $L$. It satisfies $\CS(\CP_L)=\CS(\CP)_{\text{ind}}=L$. The theory $T[M_d,\CP_L]$ has $(2-i)$-dimensional charged objects classified by $L^{(d-i)}$.
    
    \item[$L^\vee$:] The Pontryagin dual of $L$, which is isomorphic to $H^{d-3\le*\le 3}(M_d,D)/L$. It describes the symmetry of the theory $T[M_d,\CP_L]$. More precisely, the theory has a $(L^\vee)^{(i)}$ $(2-i)$-form symmetry.
    
    \item[$\bar L$:] A lift of $L^\vee$ to $H^{d-3\le*\le 3}(M_d,D)/L$. Existence of such a lift is equivalent to the $L^\vee$ symmetry of $T[M_d,\CP_L]$ that is anomaly-free.
    
\end{itemize}

\section{Partition functions of 6d relative theories}\label{sec:compactification0d}

The main objects of study in this paper are 7d/6d coupled  systems with the 7d theory being a 3-form abelian Chern--Simons theory and the 6d theory carrying 2-form symmetry whose anomaly is captured by the 7d TQFT. 
More precisely, 
the 7d theory has three-dimensional operators given by the holonomy of the three-form gauge field, which generate the two-form symmetry on the 6d boundary  \cite{Gaiotto:2014kfa}. In this paper, we will not use supersymmetry in any essential way, and the boundary theory can have either $(2,0)$ supersymmetry, $(1,0)$ supersymmetry, or no supersymmetry at all.

When the 7d theory is non-invertible, the 6d theory is said to be a relative theory \cite{Freed:2012bs}. For such theories, the notion of a partition function on a 6-manifold $M_6$ is not well-defined and requires choices of additional data, which serve to specify a state in the TQFT Hilbert space $\CH(M_6)$. 

Such additional data were best understood previously when the boundary theory has $(2,0)$ supersymmetry, coupling to a Wu Chern--Simons theory of three-form gauge field \cite{Witten:1998wy,Monnier:2017klz,Monnier:2016jlo}. It was found that part of the additional data is a choice of a ``polarization,'' which we will briefly recall below. 

The 6d $\mathcal{N}=(2,0)$ superconformal theory is labeled by a Lie algebra ${\frak g}$. And we denote the corresponding simply-connected Lie group by $\tilde G_{\text{6d}}$ and its center by $\mathcal{Z}$. On the tensor branch of the 6d theory, these three-form gauge fields are trivialized by the self-dual fields \cite{Monnier:2017klz} and they have non-trivial holonomy only in the presence of surface operators.
In 6d there are strings where the worldsheets have non-trivial linking with three-spheres that have holonomy of the three-form gauge field {\it i.e.}~they are operators charged under the two-form symmetry \cite{Gaiotto:2014kfa,Monnier:2017klz,Heckman:2017uxe}. The charges for the 6d strings are valued in the weight lattice of $\frak{g}$. In general, the Dirac quantization condition is not satisfied, and its violation is encoded in the  anti-symmetric bilinear forms on \begin{equation}
    H^3(M_6,\mathcal{Z})\times H^3(M_6,\mathcal{Z}) \rightarrow U(1).
\end{equation}

Then, the choice of a polarization is given by a maximal isotropic subgroup $\Lambda$ of $H^3(M_6,\mathcal{Z})$. Here, ``isotropic'' is with respect to the anti-symmetric bilinear form on $H^3(M_6,\mathcal{Z})$. Often it is assumed that there is a decomposition
\begin{equation}
H^3(M_6,\mathcal{Z})=\Lambda \oplus \bar\Lambda
\end{equation}
where $\bar\L$ is another maximal isotropic subgroup and is sometimes regarded part of the data for the choice of polarization.

In the following, we will re-examine and generalize the notion of polarization, both in the context of 6d $(2,0)$ theories and in a more general setting. Questions that we will answer inlcude: 
\begin{itemize}
    \item Is a choice of $\L$ enough to specify the partition function, or is more data need?
    \item For the same choice of $\L$, often there can be multiple compatible choices of $\bar\L$. What is the role played by different choices of $\bar\L$? 
    \item For some $\L$, there may be no choice of $\bar \L$. What is special about such $\L$?
    \item When polarization is changed, how do the partition functions change? 
\end{itemize}

We start by discussing in more detail the anomaly of the two-form symmetry and properties of the 7d TQFT.

\subsection{The 7d TQFT}

The action of the abelian 3-form Chern--Simons theory takes the following form  \cite{Witten:1998wy,Monnier:2017klz,Heckman:2017uxe,Gukov:2018iiq,Hsieh:2020jpj}
\begin{equation}\label{eqn:7dTQFT}
\frac{1}{4\pi}
\sum_{I,J}
K_{IJ} \int C^I d C^J,\quad I,J=1,\cdots r~,
\end{equation}
where $C^I$ are $U(1)$-valued three-form gauge fields, and $K_{IJ}=K_{JI}\in\Z$ can be assembled into an integral symmetric coupling matrix. This theory is a three-form analogue of the usual Abelian Chern--Simons theory. 

For a 6d ${\cal N}=(2,0)$ theory with Lie algebra ${\frak g}$ of ADE type, $K$ can be identified with the Cartan matrix of ${\frak g}$. In this case, the 7d bulk theory can be referred to as ``the 3-form Chern--Simons theory based on the root lattice of ${\frak g}$'' in the notation of \cite{Monnier:2017klz}. Indeed, the scalar product on the root lattice induced by the Killing form for simply-laced Lie algebra is given by the Cartan matrix. 

In the case of 6d $(2,0)$ theory on the world volume of a single M5 brane, the TQFT can also be obtained from 11 dimensions by reducing the Chern--Simons like theory for the 5-form gauge field. For more general 6d SCFTs that can be realized by F-theory on singular elliptic Calabi--Yau 3-folds, these matrices describe the intersection pairing between collapsing 2-cycles on the base of the elliptic fibration  \cite{Heckman:2017uxe}.

One important subclass of such 7d theories consists of those with even $K$, whose diagonal entries $K_{II}$ are all even. Such a theory is well-defined on any closed orientable 7-manifold. On the other hand, if any diagonal entry $K_{II}$ is odd, the theory requires the manifold to have additional structure such as a spin$^c$ structure. (This generalizes \cite{Witten:1996md,Witten:1996hc,Witten:1998wy} where a similar choice of spin structure is required, and \cite{Monnier:2016jlo,Monnier:2017klz} where a choice of Wu structure is required.) See Appendix~\ref{sec:7dCSlevel} for more details about the quantization of the coefficient matrix~$K$. We assume for a moment that $K$ is even.

The fusion rule of the three-dimensional operators $e^{i\oint C_3^I}$ can be derived from the equation of motion and is given by the quotient
\begin{equation}
D=\mathbb{Z}^r/K\mathbb{Z}^r~,
\end{equation}
which we will refer to as the ``defect group,'' following the terminology for the 6d boundary theory \cite{DelZotto:2015isa}.  It is isomorphic to $\CZ$, the center of the simply-connected Lie group $\tilde G_{\text{6d}}$ when $K$ is the Cartan matrix of an ADE Lie algebra $\frak{g}$.

The correlation function of the operators $e^{i\oint C_3}$ can be computed as in the usual Abelian Chern--Simons theory
\begin{equation}
\langle \; e^{i\oint_{V}\lambda_IC_3^I} \;
e^{i\oint_{V'}\lambda'_JC_3^J} \; \rangle \; = \; \exp\left(-2\pi i \lambda'^T(K^{-1})\lambda\cdot \text{Link}(V,V')\right)~.
\end{equation}
It gives a bilinear pairing on the defect group $D$
\begin{equation}
\langle\cdot,\cdot\rangle :\quad D\times D\rightarrow U(1)~.
\end{equation}

The bilinear form can be derived from the quadratic function
\begin{equation}\label{eqn:quadfun}
\mathbf{q}:\quad \lambda=\{\lambda_I\}\in D \mapsto \frac{1}{2}\lambda^TK^{-1}\lambda~.
\end{equation}
We will call the quadratic function the ``spin'' of the three-dimensional operator, by analogy with the spin of line operators in 3d TQFT. It is associated with the framing described by the free part of $\pi_7(S^4)=\mathbb{Z}\times \mathbb{Z}_{12}$ (see \cite{Gukov:2020inprepare} for further explanations of this perspective).
Just as in the 3d case, the spin of the operator is defined modulo 1. The braiding is related to the quadratic function by
\begin{equation}
\langle \lambda,\lambda'\rangle \; = \; \exp \, \left[-2\pi i\Big(\mathbf{q}(\lambda+\lambda')-\mathbf{q}(\lambda)-\mathbf{q}(\lambda')\Big)\right]~.
\end{equation}

Explicitly, for $K$ given by the Cartan matrix of ADE type Lie algebra, the operators in the 7d TQFT are as follows:\footnote{We note the discussion between (2.58)-(2.59) of \cite{Gukov:2018iiq} only consider the case without pairing between different cyclic factors in $D$, while here we consider the most general cases.
For instance, consider ${\frak g}=D_{4}$. The Cartan matrix is
\begin{equation}
K=\left(\begin{array}{cccc}
2 & -1 & 0 & 0\\
-1 & 2 & -1 & -1\\
0 & -1 & 2 & 0\\
0 & -1 & 0 & 2
\end{array}
\right)~.
\end{equation}
The operators $U_I=e^{i\oint C_I}$ satisfy
\begin{equation}
U_2=1,\quad
U_1U_3=U_4,\quad U_3U_4=U_1,\quad U_1U_4=U_3
\end{equation} 
and thus the fusion rule is
\begin{equation}
D=\mathbb{Z}_2\times\mathbb{Z}_2
\end{equation}
generated by $U_3,U_4$.
The operators $U_3$ and $U_4$ have trivial self-braiding but non-trivial mutual braiding.
Thus, one cannot relabel the operators to remove the mutual braiding between the two generators and in any basis the bilinear pairing on $D$ is non-trivial only between the two $\mathbb{Z}_2$ factors.
This theory is the analogue for the three fermion theory in the condensed matter literature, where the usual anyons are replaced by anyonic three-dimensional operators.}

\begin{itemize}
\item $A_{n-1}$. The three-dimensional operator has $\mathbb{Z}_n$ fusion algebra, and the generator has spin $\frac{n-1}{2n}$. 

\item $D_{2n}$. The fusion algebra is $\mathbb{Z}_2\times\mathbb{Z}_2$, generated by ${\cal U}$ and ${\cal V}$ of spin $\frac{1}{2}$ and $\frac{2n}{8}$.\footnote{Our convention is such that the spin is defined modulo 1. Namely, spin $\frac{2n}{8}$ is the same as spin $\frac{2n-8}{8}$, $\frac{2n-16}{8}$, {\it etc.}}

\item $D_{2n+1}$.  The operators have $\mathbb{Z}_4$ fusion algebra and the generator has spin $\frac{2n+1}{8}$.

\item $E_8$. The fusion algebra is trivial and there is only one trivial three-dimensional operator.

\item $E_7$. The fusion algebra is $\mathbb{Z}_2$ and the generator has spin $\frac{3}{4}$.

\item $E_6$. The fusion algebra is $\mathbb{Z}_3$ and the generator has spin $\frac{2}{3}$.

\end{itemize}
Indeed, fusion algebra of the three-dimensional operator in this theory is given by
\begin{equation}
D=\CZ({\tilde G_{\text{6d}}})~,
\end{equation}
where ${\tilde G_{\text{6d}}}$ is the simply-connected form of the group with the Lie algebra ${\frak g}$.

In particular, we have the following dualities between 7d TQFTs:
\begin{equation}\label{eqn:7ddualityADE}
E_N\leftrightarrow \overline{A}_{8-N},\quad
D_{2n+1}\leftrightarrow \left\{\begin{array}{cl}
A_3     & \text{odd }n \\
\overline{A}_3     & \text{even }n
\end{array}
\right., \quad
D_N\leftrightarrow D_{N+8}
~,
\end{equation}
where bar denotes the theory obtained by reversing the orientation, and the duality is up to multiplying with copies of the trivial TQFT associated with $E_8$.
Their counterparts in 3d are discussed in \cite{Aharony:2016jvv,Cordova:2018qvg}.

Since the three-dimensional operators obey non-trivial braiding relation, this leads to degenerate ground states on 6-manifold $M_6$ with non-trivial $H_3(M_6,D)$, as in the 3d Chern--Simons theory. More precisely, this leads to ground state degeneracy $|H_3(M_6,D)|^{1/2}$.\footnote{Since $H^3(M,D)$ has a non-degenerate skew-symmetric pairing, $|H^3(M,D)|^{1/2}$ is always an integer.}

\subsubsection*{``Fermionic'' TQFT and extended defect group}

When the coupling matrix $K$ is odd, {\it i.e.}~it has at least one odd diagonal entry, the theory has fermionic three-dimensional operator of spin $\frac{1}{2}$.
Denote a row that contains the odd diagonal entry $K_{II}$ by $K_I=(e_I^TK)_I$ with $(e_I)_J=\delta_{IJ}$, 
then the fermion operator can be written as $\exp\left(i \oint K_I C^I\right)$.
The operator has trivial braiding with other operators $\exp\left(i\oint \lambda_I C^I\right)$ and it is non-trivial only because of the self-statistics,
\begin{equation}
    \exp\left(-2\pi i e_I^T K K^{-1}\lambda\right)=1,\quad
    \frac{1}{2} e_I^T K K^{-1}Ke_I=\frac{1}{2}K_{II}=\frac{1}{2}\text{ mod }1~.
\end{equation}
The fusion algebra of the complete set of operators is
\begin{equation}
{\tilde D}=D\times \mathbb{Z}_2~,
\end{equation}
where $D=\mathbb{Z}^r/ K\mathbb{Z}^r$. See Appendix \ref{sec:factorize} for a proof that it always factorizes in this way.
Notice, however, that the non-degenerate pairing $\langle\cdot,\cdot\rangle$ on $D$ becomes degenerate on $\tilde D$. 

To understand better what ``fermionic'' means, consider the case $K=1$. We have 
\begin{equation}
\int_{\text{7d}}\frac{1}{4\pi}CdC+\frac{1}{2\pi}CdX
=\int_{\text{8d}}\pi\frac{dC}{2\pi}\frac{dC}{2\pi}+2\pi\frac{dC}{2\pi}\frac{dX}{2\pi}~,
\end{equation}
where we rewrite the action using a bounding 8-manifold, and we include a background 3-form $X$. To make the theory independent of the bounding 8-manifold, we require
\begin{equation}
    \frac{dX}{2\pi}\equiv\frac{1}{2}v_4 \pmod{\mathbb{Z}}~,
\end{equation}
where $v_4$ is the 4th Wu class. It is related to the Stiefel--Whitney classes $w_i$ of the tangent bundle by $v_4=w_4+w_2^2+w_1w_3+w_1^4$ and it reduces to $w_4+w_2^2$ on orientable manifolds (as we will consider here).
If we absorb $X$ by shifting $C\rightarrow C-X$, then the three-dimensional operator $\oint C$ depends on the bounding 4-surface:
\begin{equation}
\exp \left( i\int_{\text{4d}} dC \right)
\; =\; \exp \left( \pi i\int_{\text{4d}} v_4 \right)~.
\end{equation}
Namely, the operator $\oint C$ depends on the framing described by a trivialization of $v_4$. This is analogous to the neutral fermion particle in a spin TQFT, which depends on the framing specified by a trivialization of $w_2$ ({\it i.e.}~the spin structure). Similar higher fermion structures are discussed in \cite{Kapustin:2017jrc,Hsin:2021qiy}.

Consider the Hilbert space of the 7d TQFT on a Wu$_4$ manifold $M_6$. It decomposes as
\begin{equation}
    \CH(M_6)_{\text{full}}=\bigoplus_{B_f\in H^3(M_6,\Z_2)}\CH(M_6)_{B_f }.
\end{equation}
Different pieces labeled by $B_f$ are all isomorphic, with the isomorphisms given by the action of the central fermionic operator in 7d wrapped on a cycle dual to $B_f\in H^3(M_6,\Z_2)$. Alternatively, the theory quantized on a Wu$_4$ manifold $M_6$ depends on the trivialization of the 4th Wu class $v_4=\delta \sigma_3$, which is called the 4th Wu structure.\footnote{
The 2nd Wu structure that trivializes the second Wu class $v_2=w_2+w_1^2$ is the pin$^-$ structure. For an orientable pin$^-$ manifold {\it i.e. }a spin manifold it is the spin structure.
} Different trivializations $\sigma_3,\sigma_3'$ satisfy $\delta(\sigma_3-\sigma_3')=0$ and thus the trivializations are classified by $H^3(M_6,\mathbb{Z}_2)$.

As a consequence, one can focus on the subspace given by $B_f=0$. In the following, we will actually omit the subscript in $\CH(M_6)_{B_f=0}$, with the understanding that when $K$ is odd, there are other sectors, one for each elements in $H^3(M_6,\Z_2)$, obtained by the action of the central fermionic operator.

\begin{figure}[t]
  \centering
    \includegraphics[width=0.5\textwidth]{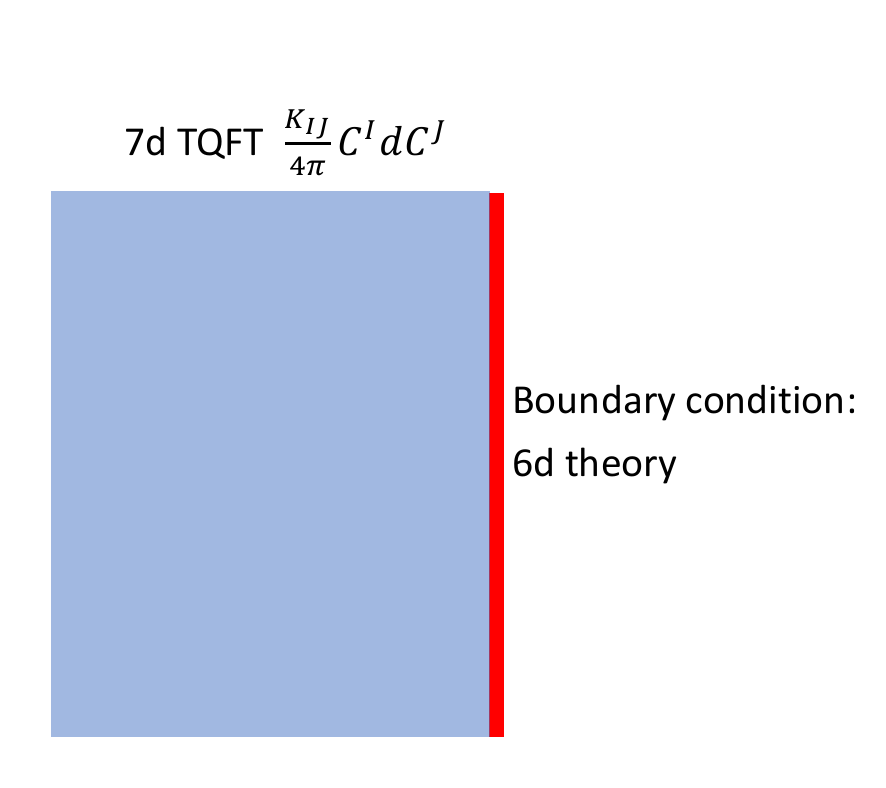}
      \caption{6d theories can be viewed as boundary conditions of a 7d TQFT.}\label{fig:6d7d}
\end{figure}

The 6d theory can be viewed as a boundary condition of the 7d theory, see Figure \ref{fig:6d7d}. Later, gapped boundary conditions will also play important role in our discussion. To specify such a 6d boundary we need to pick a boundary condition for the three-form gauge field, labeled by $H^3(M_6,D)$. As in the case of ordinary Chern--Simons theory \cite{Elitzur:1989nr}, different choices can be related by adding boundary terms, and in some cases it takes the form of a Legendre transformation on the boundary.\footnote{
In 3d TQFT the gapped boundary conditions are discussed in \cite{Kapustin:2010hk,Fuchs:2012dt,Levin:2013gaa,Barkeshli:2013yta,Kong:2013aya}, where they are classified by the maximal isotropic subalgebras in the set of operators.
} 
The three-form gauge field gives rise to three-dimensional operators in 6d that generate the two-form symmetry. There can also be open operators ending on the boundary as two-dimensional operator, and they are charged under the two-form symmetry.

\subsection{Polarizations from reduction to 1d TQFT}

As the boundary theory is coupled to the bulk TQFT, one cannot define its partition function entirely by itself. Instead, one has to specify a state in the Hilbert space $\CH(M_6)$ of the 7d TQFT on $M_6$. And such choices were conjectured to be labeled by maximal isotropic subgroups in $H^3(M_6,D)$,\footnote{By definition, $\Pol(M_6)$ also depends on $D$, the pairing on $D$, and possibly its quadratic refinement if $M_6$ is not spin. As such data is often clear from the context, we suppress it in the notation to avoid clutter.}
\begin{equation}
    \Pol(M_6):=\{\Lambda\subset H^3(M_6,D)| \Lambda\text{ maximal isotropic}\}.
\end{equation}
Such a choice of $\L$ is often referred to as a ``polarization.'' As we will later see, this is only part of the data, and one needs an additional piece of information to fully specify a state projectively in the Hilbert space $\CH(M_6)$. We will return to this problem at a later point. 

To prepare for the discussion in the next section that generalizes the notion of ``polarization on $M_d$'' with $d<6$, we will look at the maximal isotropic condition from a slightly different perspective. We will now argue that each polarization (together with one piece of additional data that we will specify later) corresponds to an \emph{absolute} 0-dimensional theory obtained from reduction of the 6d theory on $M_6$,
\begin{equation}\label{Pol6Ab}
    \Pol(M_6) \; \simeq \; \{\text{Absolute $0$-dim theory obtained by reduction on $M_6$}\}.
\end{equation}
Here, ``absolute'' means that a theory has a well-defined partition function (up to a phase), or, equivalently, the 7d TQFT reduces to an invertible 1d TQFT.\footnote{``Absolute theories" are opposite of ``relative theories.'' This terminology goes back to the work \cite{Freed:2012bs},
where it was used in a slightly stronger sense, requiring the bulk TQFT to be trivial, not only invertible. According to that terminology, some of the absolute theories studied in this paper are actually ``projective theories.'' They are consistent quantum field theories with an 't Hooft anomaly for global symmetry as described by the bulk invertible TQFT.} Also, the right-hand side of \eqref{Pol6Ab} is restricted to ones that are ``universal'' from the point of view of the 7d theory. For example, the same 7d TQFT can have different boundary 6d theories, which can have additional global symmetries either in 6d or when reduced to lower dimensions. And, when we reduce the theory on $M_6$, there are going to be additional choices to make, such as choices of holonomies/fluxes along various cycles on $M_6$. When they are included, strictly speaking, the map \eqref{Pol6Ab} is injective, but not in general an isomorphism.

The construction for \eqref{Pol6Ab} is as follows. First, reduce the coupled 7d/6d system on $M_6$. This gives a coupled 1d/0d system.
In the continuous notation, the reduction of the 7d TQFT (\ref{eqn:7dTQFT}) gives
\begin{equation}\label{1dAction}
    \frac{1}{4\pi}\sum K^{IJ} \int \phi_\gamma^Id\phi_{\gamma'}^J\eta^{\gamma,\gamma'}~,
\end{equation}
where $\phi_\gamma^I\sim \phi_{\gamma}^I+2\pi$ denotes the holonomy of the three-form fields $C^I$ over a three-cycle $\gamma$ in $M_6$, and $\eta$ is the intersection form on $M_6$.\footnote{The reduction of $U(1)$-valued field on torsion cycles and free cycles should be treated differently. More details are given in Section \ref{sec:torsiondiscretefreecycle} and \cite{Gukov:2020inprepare}.} 
Since $\int \phi^I_\gamma d\phi^J_{\gamma'}=-\int \phi^J_{\gamma'}d\phi^I_\gamma+\phi^I_\gamma\phi^J_{\gamma'}\big|_{\text{endpoints}}$ and $K^{IJ}$ is symmetric, this is consistent with the anti-symmetric $\eta^{\gamma,\gamma'}=-\eta^{\gamma',\gamma}$.
The operators $e^{i\phi^I_\gamma}$ obey fusion algebra given by $D$.
The braiding in the original 7d TQFT becomes the braiding between the operators $e^{i\phi}$: exchanging the order of insertions of the point operator produces a phase, which when summed over different $\gamma,\gamma'$ weighted with the intersection $\eta$ produces the 7d braiding for the corresponding three-dimensional operators.
This is in general a non-trivial theory, coupled to a relative 0-dim theory with ``$(-1)$-form symmetry'' $H^3(M_6,D)$ on the boundary, which, however, can be made absolute by gauging a subgroup of $H^3(M_6,D)$.

\subsubsection*{$(-1)$-form symmetries and their anomaly}

A $(-1)$-form symmetry in a quantum field theory is not a symmetry in the usual sense. Its existence implies that there are (continuous or discrete) theta angles that can be turned on to modify the theory. When such a theory is obtained from reduction of a theory in one higher dimension on $S^1$ with a usual (0-form) symmetry, a theta angle is the holonomy of the symmetry along the circle. Familiar examples include the $U(1)$ ``$(-1)$-form instanton symmetry'' in 4d gauge theories; this symmetry corresponds to the theta angle $\theta\int \mathrm{Tr}F^2$ with periodic continuous $\theta$ parameter. The discrete ``$(-1)$-form'' symmetry corresponds to discrete theta angle. The discrete theta angles are discrete topological actions of the gauge fields and can be thought of as coupling the gauge theory to SPT phases. They are classified by cobordism groups \cite{Kapustin:2014tfa,Kapustin:2014dxa,Freed:2016rqq,Yonekura:2018ufj}, and in some cases they can be described by group cohomology \cite{Chen:2011pg}.
Both continuous and discrete theta angles can lead to interesting consequences for other global symmetries \cite{Hsin:2020nts}.

The reason that $(-1)$-form symmetries are usually not referred to as a symmetry is because the symmetry action will in general transform the theory, as theories with different theta angles are in general distinct. Another way of saying this is that the symmetry defect of $(-1)$-form symmetry is space-filling.

In this paper, however, it is more convenient to refer to them as symmetries. One reason is that they can have anomalies which are direct generalizations of anomalies for usual symmetries \cite{Cordova:2019jnf,Cordova:2019uob,Hsin:2020cgg}.
 
In our present case, the $(-1)$-symmetry $H^3(M_6,D)$ will indeed have an anomaly coming from the inflow of a 1d TQFT \eqref{1dAction}. Namely, if we turn on a discrete theta angle $\alpha \in H^3(M_6,D)$, then it will break the symmetry group from $H^3(M_6,D)$ to a smaller one, composed of elements that pair trivially with $\alpha$. In other words, under a ``gauge transformation''\footnote{To see that this is the right notion of gauge transformation for $(-1)$-form symmetries, consider reducing a theory with 0-form symmetry on a circle. Then a large gauge transformation for the latter on the circle will leads to  transformation for ``gauge field'' $\alpha$ of the $(-1)$-form symmetries of this particular type.}
\begin{equation}
     \alpha \mapsto \alpha + N\beta,
\end{equation}
where $N\in \Z_{>0}$ is the smallest annihilator of $\beta\in H^3(M_6,D)$, the partition function will pick up a phase $\langle\alpha,\beta\rangle$ given by the pairing in $H^3(M_6,D)$. 
In the continuous notation, the anomaly can be described by the periodic scalars that are no longer periodic in the presence of a boundary: under a ``gauge transformation'' $\phi_\gamma^I\rightarrow \phi^I_\gamma +2\pi n_\gamma^I$ with integer $n^I_\gamma$, the action (\ref{1dAction}) transforms by the boundary term
\begin{equation}
\frac{1}{2}\sum K^{IJ}\eta^{\gamma,\gamma'} n^I_\gamma \phi^J_{\gamma '}(p)~,
\end{equation}
where there is a sum over the boundary (endpoints) $p$.

Another reason we refer to $(-1)$-form symmetries as symmetries is because one can consider gauging them {\it i.e.} summing over the (continuous or discrete) theta angles.
 
\subsubsection*{Gauging $(-1)$-form symmetry}
 
Just as with usual symmetries, a subgroup of $(-1)$-form symmetry can be gauged if it is anomaly-free. 
 
In the present example, we hope to start with a quite non-trivial 1d/0d coupled system and obtain an absolute theory on the boundary, coupled to an invertible TQFT in the bulk. To get closer to the goal, one can first consider a subgroup $\Lambda\subset H^3(M_6,D)$ that is anomaly-free. This is equivalent to requiring $\Lambda$ to be isotropic,
\begin{equation}
      \langle \alpha, \beta\rangle = 0, \quad\forall \alpha, \beta \in \L.
\end{equation}

Being anomaly-free also means that the action \eqref{1dAction} becomes trivial when restricted to $\Lambda$. 
The TQFT with trivial action is not necessarily invertible, as it can have different topological sectors given by non-interacting invertible TQFTs and labelled by $B\in \Lambda^\vee:=\text{Hom}(\Lambda,U(1))$.
One can project onto one of them. From the boundary point of view, this procedure corresponds to gauging $\Lambda$, and choosing a particular summand to project onto corresponds to choosing a background for the dual $(-1)$-form symmetry  $\Lambda^\vee\cong H^3(M_6)/\Lambda$ appearing after gauging. 
 
More explicitly, each $\alpha \in\ H^3(M_6,D)$ corresponds to an operator $\hat{\alpha}$ acting on the boundary Hilbert space, and
\begin{equation}\label{0dProjection}
     P_{\Lambda}:=\frac{1}{|\L|} \sum_{\alpha\in \L}\hat{\alpha}
\end{equation}
is a projection operator, $P_{\Lambda}^2= P_{\Lambda}$. Similarly, one can define
\begin{equation}
     P_{\Lambda,B}:=\frac{1}{|\L|} \sum_{\alpha\in \L}e^{-iB(\alpha)}\hat{\alpha} 
\end{equation}
where $B\in \L^\vee$ and $e^{2\pi iB(\alpha)}$ involves the pairing $\L\times \L^\vee \rightarrow U(1)$. These projection operators satisfy
\begin{equation}
     P_{\Lambda,B}P_{\Lambda,B'}=P_{\Lambda,B}\delta_{BB'},
\end{equation} 
and therefore give a decomposition of the Hilbert space.

Another effect of gauging $\L$ is breaking $H^3(M_6,D)$ to a subgroup $\L^\perp$ composed of those elements that pair trivially with elements in $\L$. Obviously, one has
\begin{equation}
     \L\subset \L^{\perp}.
\end{equation}
After gauging $\L$, we will have an invertible theory if and only if the remaining action on $\L^{\perp}/\L$ gives an invertible TQFT. The TQFT is similar to the original one but defined with the pairing  
 \begin{equation}
    \L^{\perp}/\L \times \L^{\perp}/\L \rightarrow U(1).
 \end{equation}
It is easy to check that this pairing is well-defined and is, in fact, non-degenerate. 
Requiring the theory to be invertible implies that $\L^{\perp}/\L$ has no further isotropic subgroups, which is equivalent to saying that $\L$ is a maximal isotropic subgroup.

When this is the case, $P_{\L,B}$'s give rise to a set of rank-1 projection operators that decompose the Hilbert space of TQFT into one-dimensional subspaces. This is not yet equivalent to choosing a basis in the Hilbert space on $M_6$, as there are still phase ambiguities, which can be readily fixed if the short exact sequence
 \begin{equation}
     0\rightarrow\L\rightarrow H^3(M_6,D)\rightarrow \L^\vee\rightarrow 0
 \end{equation}
splits. Because this gives a lift of $\L^\vee$ to $H^3(M_6,D)$, denoting the resulting subgroup $\bar{\L}$, we learn that the Hilbert space $\CH(M_6)$ can be canonically identified with $\C \bar{\L}$. Then there exists a canonical basis given by elements in $\bar{\L}$. We will denote these elements $|\L,B\rangle$, with $B\in\bar{\L}$.

In this case, we have 
\begin{equation}
    H^3(M_6,D) \; =\; \L\oplus \bar{\L}.
\end{equation}
Although $\bar{\L}\simeq \L^\vee$ as groups, there can be different lifts of $\L^\vee$ in $H^3(M_6,D)$. When a lift $\bar{\L}$ is changed to $\bar{\L}'$, the basis vectors are transformed by phases. Namely, for $B\in \L^\vee$
\begin{equation}\label{LBarChange}
|B\rangle \; =\; e^{2\pi i\varphi(B)}|B\rangle' 
\end{equation}
where $\varphi$ is a homomorphism from $\L^\vee$ to $U(1)$. Such basis changed will be studied in more general setting in Subsection \ref{sec:PolAndPar}.

If the extension 
\begin{equation}
    \L\rightarrow H^3(M_6,D) \rightarrow \L^\vee
\end{equation}
is non-trivial, then $\L^\vee$ cannot be lifted to a subgroup of $H^3(M_6,D)$. Although the projection operators are still well defined, there is no canonical choice of basis due to an anomaly that will be discussed further in later sections.

In the above, we presented a procedure to obtain absolute theories on the boundary, and 
one can now ask whether this is the only such procedure. As it is very natural ({\it i.e.}~functorial from the TQFT point of view) and has many nice properties (\textit{e.g.}~closed under mapping class group action which we will discuss later), we conjecture that other constructions will not yield any new theories, unless additional properties specific to the boundary theory (such as additional symmetries) are used. Then we can summarize the above discussion as follows:
\begin{equation}
    \Pol(M_6) \; \simeq \; \{\text{Absolute $0$-dim theory obtained via reduction on $M_6$}\}.
\end{equation}

Projections operators like $P_{\L,B}$ can be interpreted as domain walls between the non-trivial 1d TQFT and an invertible TQFT. This point of view will be explored further in later sections. Indeed, any such domain wall will look like a projection operator mapping states in the Hilbert space $\CH(M_6)$ to a one-dimensional vector space. As we will see later, the invertible TQFT will actually be trivial if $\L^\vee$ can be lifted to a subgroup of $H^3(M_6,D)$ (in other words, $\L^\vee$ is anomaly free). Otherwise, it will be a non-trivial SPT whose action can be written in terms of $B\in \L^\vee$.

\subsection{Polarizations and partition functions}\label{sec:PolAndPar}

As explained above, one purpose of choosing a polarization $\L$ is to obtain a state in the Hilbert space $\CH(M_6)$ of the 7d TQFT on $M_6$ and render the partition function of $T[M_6]$ unambiguous. In this subsection, we will study explicitly these partition functions. In order to be more explicit, though, one needs to choose a basis in $\CH(M_6)$. One way to achieve this is by choosing a decomposition of the abelian group $H^3(M_6,D)$ into a direct sum
\begin{equation}\label{eqn:decompose}
    H^3(M_6,D)=\L\oplus\bar{\L},
\end{equation}
where $\bar\L$ can also be chosen to be a maximal isotropic subgroup. Although such a decomposition always exists, only specific $\L$ can appear.\footnote{As an example, when $M_6=S^3\times S^3$ and $D=\Z_{mn}$, one cannot use the maximal isotropic subgroup $\Z_m\times \Z_n \subset H^3(M_6,\Z_{mn})=\Z_{mn}^2$.} 

This decomposition sometimes can be realized geometrically by choosing a 7-manifold $M_7^\L$, with $\partial M_7^\L=M_6$, such that the two maps in the following long exact sequence for the relative cohomology
\begin{equation}
    \cdots\rightarrow H^3(M_7,D)\xrightarrow{i^*} H^3(M_6,D)\xrightarrow{\delta} H^4(M_7,M_6;D) \rightarrow \cdots
\end{equation}
give a decomposition
\begin{equation}
    H^3(M_6;D)\simeq \mathrm{Ker}(\delta)\oplus \mathrm{Im}(\delta).
\end{equation}
In other words, in the discussion here we choose $M_7$ such that the following short exact sequence splits
\begin{equation}
    0\rightarrow \mathrm{Ker}(\delta)\rightarrow H^3(M_6,D)\rightarrow \mathrm{Im}(\delta)\rightarrow 0~,
\end{equation}
and we choose a splitting identifying $\mathrm{Im}(\delta)$ with a subgroup of $H^3(M_6;D)$. We will refer to this data as a choice of a ``framing'' on $M_7$. Later in Section \ref{sec:GeneralTheory} we will discuss more general cases where the sequence might not split.

Since $i^*$ is the pullback under the inclusion $i: M_6\rightarrow M_7$ and
\begin{equation}
    \L \; := \; \mathrm{Ker}(\delta) \; = \; \mathrm{Im}(i^*)
\end{equation}
it follows that $\L$ contains 3-cocycles on $M_6$ that can be obtained from restricting 3-cocycles in $M_7$.  On the other hand, we identify
\begin{equation}
    \bar{\L} \; := \; \mathrm{Im}(\delta)
\end{equation}
as a subgroup of $H^3(M_6,D)$. Using relative version of the Poincar\'e duality, its elements can be interpreted as dual of the $3$-cycles in $M_7$ that come from 3-cycles in $M_6$. The pairing between $\L=\mathrm{Ker}(\delta)$ and $\bar{\L}=\mathrm{Im}(\delta)$ can be computed using either the intersection pairing on $H^3(M_6,D)$, or using the natural pairing between $H^4(M_7,M_6;D)$ and $H^3(M_7,D)$. This unifies the boundary perspective and the bulk perspective.

Then, once such $M_7$ is chosen, one can consider a basis given by the TQFT states $|M_7;B\rangle$ with $B\in \bar{\L}$ prescribing a defect inserted along $H_3(M_7,D)$. These states yield a set of independent partition functions $Z^{\L}_B$ of the boundary theory.

For a given $\L$, the choice of $\bar{\L}$ may not be unique. Similarly, for the same $M_7$, there are different choices of framing, leading to different basis, with the state $|M_7;B\rangle$ modified by a phase proportional to $B$ as mentioned in the previous subsection. Only the state corresponding to $B=0$ can be specified without choosing a $\bar{\Lambda}$.  
Given a pair $(\L,\bar{\L})$, any other maximal isotropic subgroup $\L'$ can be decomposed as
\begin{equation}
    \Gamma':=\L'\cap \L\lhook\joinrel\longrightarrow\L'\longrightarrow \bar{\Gamma'}:=p_{\bar{\L}}(\L')
\end{equation}
where $p_{\bar{\L}}$ is the projection to $\bar{\L}$
\begin{equation}
  p_{\bar{\L}}:   H^3(M_6,D)\rightarrow \bar{\L}.
\end{equation}

With such choice of basis, for any polarization $\L'$ we have a partition function\footnote{In this paper, we mostly ignore the overall normalization of partition functions. In other words, we specify $\varphi$ up to a constant term, independent of $B$.}
\begin{equation}\label{6dPartition}
    Z^{\L'}=\sum_{\alpha\in \L'}e^{2\pi i\varphi(\alpha)}Z^\L_{B_\alpha}.
\end{equation}
Here $B_\alpha\in \bar{\Gamma'}$ is the image of $\alpha$ under the map $p_{\bar{\L}}$, up to a possible shift,\footnote{Technically, there might be a ``quantum effect'' modifying this condition, which is only exactly correct if $\varphi(\alpha)$ is 0 when restricted to $\L\cap \L'$. In general, this function can be also be a non-trivial one, $\L\cap \L'\rightarrow\{0,\frac12\}$. Then $B_\alpha$ will be shifted from $p_{\bar{\L}}(\L')$ by a constant amount. The right condition for the shift $B_{\varphi}$ is such that $\varphi(\alpha)+B_{\varphi}(\alpha)$ become the zero function when restricted to $\L\cap \L'$. We will address this subtlety in Section \ref{sec:PolAndQuad}, after giving the explicit formula for $\varphi$.\label{ShiftSubtlety}} and the sum of the phase factors $e^{2\pi i\varphi(\alpha)}$ --- whose explicit form will be given below --- over a fixed $B_\alpha$
\begin{equation}
    \sum_{\alpha\in p_{\bar{\L}}^{-1}(B_\alpha)}e^{2\pi i\varphi(\alpha)}
\end{equation}
can be identified with the inner product of states $|M_7;B_\alpha\rangle$ and $|M'_7\rangle$. Here $M'_7$ is a choice of a $7$-manifold such that the image of the restriction map $H^3(M'_7,D)\rightarrow H^3(M_6,D)$ is $\L'$. 

Alternatively, $e^{2\pi i\varphi(\alpha)}$ can be computed as the partition function of the 7d TQFT on $M_7\cup M'_7$ with insertion of defects along $B\in H_3(M_7,D)$. This calculation can be performed classically on-shell, by finding configurations of the $C$-field on the 7-manifold compatible with the defects and summing over such configurations. When no such configurations exist, $e^{\varphi(\alpha)}$ is set to zero. Up to a normalization factor, the partition function does not depend on a particular choice of manifolds $M_7$ and $M'_7$,\footnote{As the TQFT is invertible, replacing $M_7$ with another 7-manifold $\tilde{M}_7$ only results in an overall phase change, as long as it does not give 0 in $\CH(M_6)$.} and it usually makes sense to choose ones that are sufficiently simple. (The simplest ones are those on which there is a unique on-shell configuration for $C$ given a choice of a $B$.)

A special case is $\L'=\L$. Then, the configuration with $B\neq 0$ cannot extend to the entire 7-manifold on-shell, as the image of $B$ under
\begin{equation}
    H^3(M_6,D)\rightarrow  H^4(M_7\cup M_7,D)
\end{equation}
is non-zero. Therefore, only
$B=0$ contributes, with $e^{\varphi(\alpha)}=1$. This gives us a consistency check.

The reason why in \eqref{6dPartition} one only needs to consider $B$ in  $\bar{\Gamma'}\subset\bar{\L}$ is that for other elements of $\bar{\L}$ there are no on-shell configurations of the $C$ field. This can be understood in the following way. 
When a $B\in \bar{\L}$ is not in $\bar{\Gamma'}$, then by exactness of the Mayer--Vietoris sequence
\begin{equation}\label{MV}
    H^3(M_7,D)\oplus H^3(M'_7,D)\rightarrow H^3(M_6,D) \rightarrow H^4(M_7\cup M'_7,D),
\end{equation}
it determines a non-trivial class in $H^4(M_7\cup M'_7,D)$, and this is an obstruction to having a flat connection for the $C$-field. On the other hand, different flat connections are classified by elements in the pre-image of the first map over a $B$.

\begin{figure}[t]
  \centering
    \includegraphics[width=0.6\textwidth]{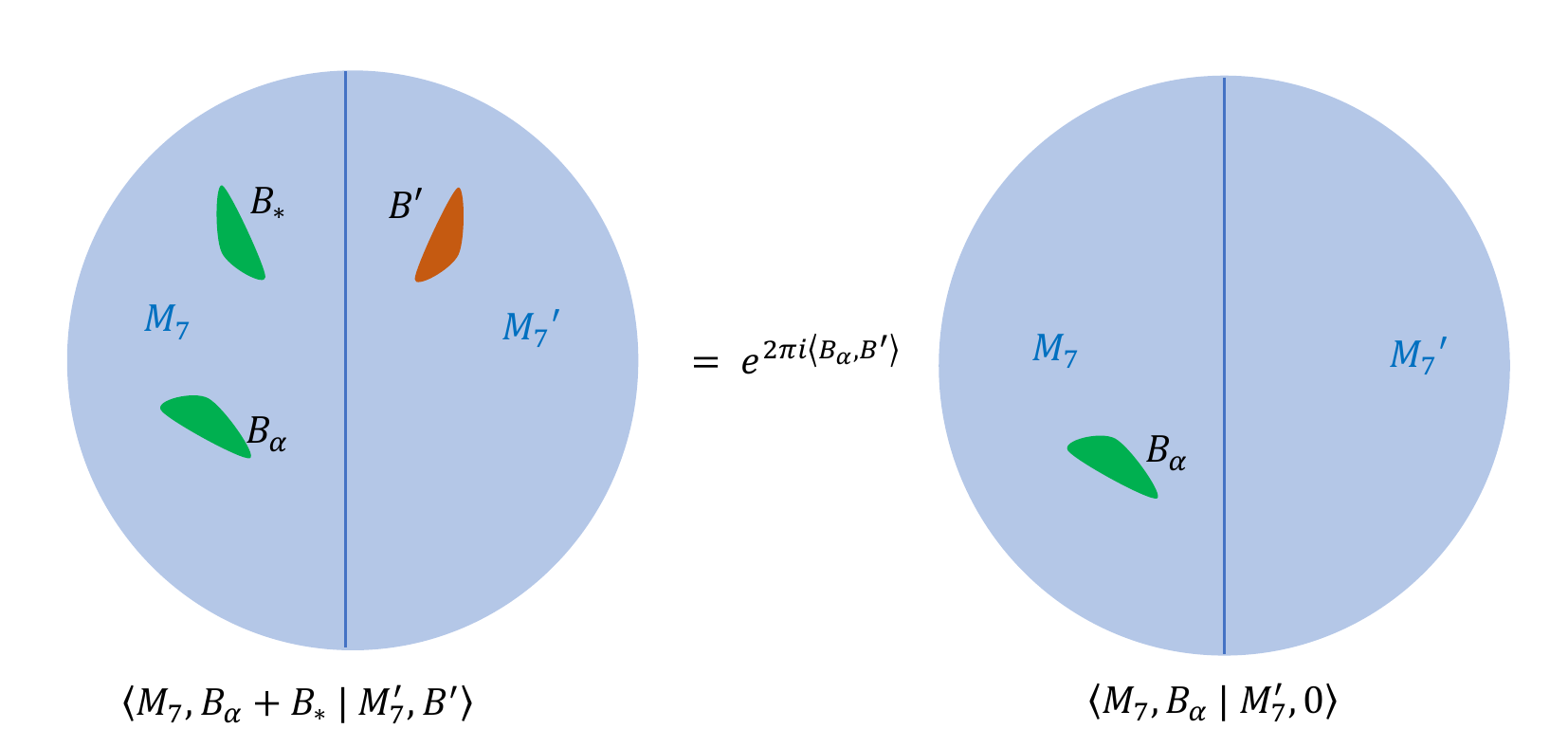}
      \caption{7d partition function from gluing $7$-manifolds with defects described by the Poincar\'e dual of $B$. Removing the pair $B',B_*$ together introduces a phase.}\label{fig:M7M7'}
\end{figure}

Next, we explore how the basis given by $|M_7,B\rangle$ changes when we change the decomposition of $H^3(M_6,D)$. Now we consider a different decomposition
\begin{equation}
    H^3(M_6,D)=\L'\oplus \bar{\L}'
\end{equation}
assumed to be realized geometrically by a 7-manifold $M'_7$,
and we should have a collection of TQFT states given by $|M'_7,B'\rangle$ with 
\begin{equation}
    B'\in \bar{\L}'\simeq H_3(M'_7,D).
\end{equation}
If we pair it with the state $|M_7,B\rangle$, the result is only non-zero if $B-B'$ is in the image of the first map of \eqref{MV}, as otherwise there is no on-shell solution. Let $B_*\in \bar{\L}$ denote the image of $B'\in \bar{\L}'$, then by inserting a complete basis, we have\footnote{Again, as in Footnote \ref{ShiftSubtlety}, there could be a shift $B_{\varphi}$ for $B_\alpha+B_*$. Since this is a constant shift, we will absorb it into the definition of $B_*$.} 
\begin{equation}\label{6dStateChange}
    |M'_7,{B}'\rangle=\sum_{\alpha\in \L'}e^{2\pi i\varphi(\alpha,{B}')} |M_7,B_\alpha+{B}_*\rangle
\end{equation}
where
\begin{equation}
   \sum_{\alpha\in p_{\bar{\L}}^{-1}(B_\alpha)}e^{2\pi i\varphi(\alpha,B')}=\langle M_7,B_\alpha +B_* |M'_7,B'\rangle
\end{equation}
is given by the partition function of the 7d TQFT on $M_7\cup M'_7$ with defects determined by $B=B_*+B_\alpha$ and $B'$. Here $\varphi(\alpha,B')$ is the action of the 7d TQFT for an extension of $C$-field over $M_7\cup M'_7$. In fact, it will be linear in $B'$, given by
\begin{equation}
    \varphi(\alpha,B')=\varphi(\alpha)+\langle B_\alpha, B'\rangle.
\end{equation}
Geometrically, this relation comes from the fact that the defects inserted on both sides can be combined and brought to the $M_7$ side, with $B_*$ cancelling $B'$, up to a change of framing. See Figure~\ref{fig:M7M7'} for an illustration. In other words,
\begin{equation}
    \langle M_7,B_\alpha +B_* |M'_7,B'\rangle=e^{2\pi i\langle B_\alpha,B'\rangle}\langle M_7,B_\alpha |M'_7,0\rangle.
\end{equation}
We will see this relation more clearly later in the next subsection.

At the level of partition functions, changing from old basis to the new basis is described by the relation
\begin{equation}\label{6dBasisChange}
    Z^{\L'}_{B'}=\sum_{\alpha\in \L'}e^{2\pi i\varphi(\alpha,B')} Z^{\L}_{B_\alpha+B_*},
\end{equation}
which reduces to \eqref{6dPartition} when $B'=0$. 

In the above, we used geometric realizations of TQFT states to describe $\varphi(\alpha)$. This is not necessary, though, and sometimes not practical in applications. We will find in the next subsection an alternative group-theoretic way of defining these states.

\subsection{Polarization and quadratic refinement}\label{sec:PolAndQuad}

Explicit formula for $\varphi(\alpha,B)$ can be obtained using the projection operators defined in \eqref{0dProjection}, by acting with $P_{\L'}$ (or $P_{\L',B'}$) on $|\L,0\rangle$ (respectively $|\L,B_*\rangle$, where $B_*\in\bar\L$ is the image of $B'$ under the projection to $\bar{\L}$) and rewriting it as a linear combination of $|\L,B\rangle$ with different $B$. To carry out this computation, we will first need to introduce some notation to describe the action of $\L'$ on $\CH(M_6)$. 

The function $\varphi$ will implicitly depend on the decomposition of $H^3(M_6,D)$ into $\L\oplus\bar{\L}$. We denote the two projections as $p_{\L}$ and $p_{\bar{\L}}$. Then, given an element $\alpha\in \L'$, one needs to define an operator $\hat\alpha$ acting on the Hilbert space $\CH(M_6)$.
There is a potential phase ambiguity, as neither
\begin{equation}\label{NormalOrdering}
    \alpha \mapsto \hat\alpha= \hat{p_{\bar{\L}}(\alpha)}\cdot \hat{p_{\L}(\alpha)},
\end{equation}
where $\hat{p_{\bar{\L}}(\alpha)}$ and  $\hat{p_{\L}(\alpha)}$ are operators corresponding to $p_{\bar{\L}}(\alpha)$ and $p_{\L}(\alpha)$, nor the opposite order $\hat\alpha=\hat{p_{\L}(\alpha)}\cdot  \hat{p_{\bar{\L}}(\alpha)}$ is canonical, and these two choices differ by a phase $\exp[2\pi i \langle p_{\bar{\L}}(\alpha),p_{\L}(\alpha)\rangle]$. More importantly, none of these two choices give a group action. Nevertheless, there is a choice of the phase factor, such that
\begin{equation}
    \alpha \mapsto\hat{p_{\bar{\L}}(\alpha)}\cdot \hat{p_{\L}(\alpha)}\cdot e^{2\pi i \frac{\langle p_{\bar{\L}}(\alpha),p_{\L}(\alpha)\rangle}{2}}= \hat{p_{\L}(\alpha)}\cdot \hat{p_{\bar{\L}}(\alpha)}\cdot e^{2\pi i \frac{\langle p_{\L}(\alpha),p_{\bar{\L}}(\alpha)\rangle}{2}}.
\end{equation}
In fact, this makes the operator $\hat{\alpha}$ well defined and free from ordering ambiguity. In this process, there is a sign choice, which can be interpreted as a choice of quadratic refinement of an inner product\footnote{It suffices to fix the sign for any $\alpha\in \L'$. In fact, the sign can't be fixed on the entire $H^3(M_6,D)$ to define an honest group action on the Hilbert space $\CH(M_6)$. 
This is related to the existence of a non-trivial associator in the category of defects. Analogous aspects in 3d abelian Chern--Simons theories have been discussed in \cite{Kapustin:2010hk}. See also \cite{Monnier:2016jlo} for a detailed discussion of this phenomenon on the free part of $H^3(M_6,D)$.} on $\L'$ that, in turn, can be defined by starting with the anti-symmetric pairing on $H^3(M_6,D)$.

\begin{figure}[t]
  \centering
      \includegraphics[width=0.5\textwidth]{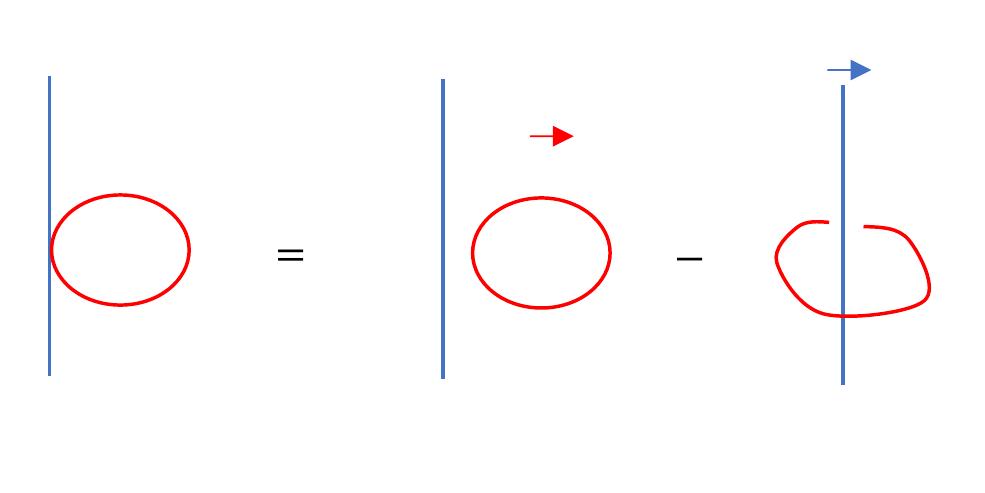}
  \caption{Linking in the bulk and intersection on the boundary.}\label{fig:linkresolve}
\end{figure}

The anti-symmetric intersection pairing on $M_6$ is related to the symmetric braiding pairing  $\br\langle\cdot,\cdot\rangle$ on $M_7$ by 
\begin{equation}
    e^{2\pi i\langle \alpha, \beta\rangle} = \frac{\left\langle \hat\alpha'\hat\beta \right\rangle_{M_7}}{\left\langle \hat\alpha\hat\beta' \right\rangle_{M_7}},
\end{equation}
where $\hat\alpha'$ and $\hat\beta'$ are obtained by resolving the intersection between the two operators by pushing either $\hat\alpha$ or $\hat\beta$, respectively, into the bulk of $M_7$, whose correlation functions in $M_7$ are denoted with a subscript $M_7$.
Thus the ratio of the two correlation functions is given by the braiding.
This is shown in Figure \ref{fig:linkresolve}. 
As elements in $\L$ correspond to 3-cycles in $M_6$ that are contractible in $M_7$, $\left\langle \hat{p_{\bar{\L}}(\alpha)}\, \hat{p_{\L}(\alpha)}' \right\rangle_{M_7}$ is an unlinked configuration. Therefore, one has
\begin{equation}
   \langle p_{\bar{\L}}(\alpha),p_{\L}(\alpha)\rangle =\br\langle \hat{p_{\bar{\L}}(\alpha)}', \hat{p_{\L}(\alpha)} \rangle.
\end{equation}
The braiding pairing is the linking pairing on $M_7$ combined with the pairing on $D$. The latter admits a quadratic refinement $\bf q$, and similarly, we can define the quadratic function
\begin{equation}
    2q_{\L,\bar{\L}}: \quad H^3(M_6,D) \rightarrow \R/\Z
\end{equation}
given by
\begin{equation}
   2q_{\L,\bar{\L}}(\alpha)=\langle p_{\bar{\L}}(\alpha), p_{\L}(\alpha)\rangle. 
\end{equation}
In fact, it is the quadratic refinement of the symmetric pairing 
 \begin{equation}
     2\langle \alpha,\beta\rangle_{\text{sym},\L,\bar{\L}}:=\langle p_{\bar\L}(\alpha),p_{\L}(\beta)\rangle +\langle p_{\bar{\L}}(\beta),p_{\L}(\alpha)\rangle.
 \end{equation}
 When restricted to an isotropic subgroup $\L'\subset H^3(M_6,D)$,  this pairing can be divided by 2, 
 \begin{equation}
      \langle \alpha,\beta\rangle_{\text{sym},\L,\bar{\L}}=\langle p_{\bar\L}(\alpha),p_{\L}(\beta)\rangle =\langle p_{\bar{\L}}(\beta),p_{\L}(\alpha)\rangle=q(\alpha+\beta)-q(\alpha)-q(\beta).
\end{equation}
Therefore, one can define a quadratic refinement of $\langle \cdot,\cdot\rangle_{\text{sym},\L,\bar{\L}}$ on $\L'$,\footnote{Both the  symmetric pairing $\langle \cdot,\cdot\rangle_{\text{sym},\L,\bar{\L}}$ and  $q_{\L,\bar{\L}}$ depend on the decomposition  $H^3(M_6,D)\simeq \L\oplus\bar{\L}$, and when the dependence is clear from the context we simply denote them by $\langle \cdot,\cdot\rangle_{\text{sym}}$ and $q$. 
}
\begin{equation}
    q_{\L,\bar{\L}}:\quad  \L' \rightarrow \R/\Z.
\end{equation}
There are choices to be made, parametrized by $\Z^n_2$ for some $n$; they form a torsor over the subgroup of $\L'^\vee$ composed of 2-torsion elements. To see this, consider any $\gamma\in \L'$ with $2\gamma=0$. Then it is easy to check that given a quadratic refinement, $q'=q+\gamma$ is another quadratic refinement. Conversely, the difference between any two quadratic refinements $q'-q$ automatically gives such a $\gamma$. They represent additional choices of background fields in the reduction of 7D TQFT.

More precisely, one can choose a collection of $\Z_2$-valued 3-form fields $X_i$, one for each $C_i$, and couple them to the 7D theory via
 \begin{equation}
     \sum_{i,j}\int \frac{K_{ij}}{4\pi}C_idC_j +  \frac{1}{2\pi}C_idX_i.
 \end{equation}
The effect is that although commutation relation is not changed, ``spin'' (in the 1d TQFT sense as the phase in the ordering in \eqref{NormalOrdering}) can get modified by a sign. 

For each $\mathbb{Z}_2\subset D$ (or $\mathbb{Z}_2\subset \tilde D$), the corresponding combination of the background fields can be related to a trivialization of the Wu structure. These are special backgrounds with ${dX\over 2\pi}=v_4$, the fourth Wu class, that makes them invariant under diffeomorphism symmetry of $M_6$ and allows a lift to the 7d TQFT. In other words, they can be used to shift the spin of  three-dimensional operators in the 7d TQFT by $\frac12$.
  See Appendix~\ref{sec:7ddualities} for more details.

The fact that $q_{\L,\bar{\L}}$ is a quadratic refinement ensures that the map
\begin{equation}
    \alpha \mapsto \hat{\alpha} \; = \; e^{2\pi i q(\alpha)} \,\hat{p_{\bar{\L}}(\alpha)}\cdot  \hat{p_{\L}(\alpha)}
\end{equation}
defines a group action of any isotropic subgroup $\L'$ on the Hilbert space $\CH(M_6)$ (again up to a sign). In other words, given $\alpha,\beta\in\L'$, the action of $\alpha+\beta$ is simply described by $\hat{\alpha}\hat{\beta} =\hat{\beta}\hat{\alpha}$, such that
$$
\hat{\alpha}\hat{\beta}=\left(e^{2\pi i q(\alpha)}\hat{p_{\bar{\L}}(\alpha)}\cdot  \hat{p_{\L}(\alpha)}\right)\cdot \left(e^{2\pi i q(\beta)}\hat{p_{\bar{\L}}(\beta)}\cdot  \hat{p_{\L}(\beta)}\right)=\left(e^{2\pi i q(\alpha+\beta)}\hat{p_{\bar{\L}}(\alpha+\beta)}\cdot  \hat{p_{\L}(\alpha+\beta)}\right).
$$
Since the projection operators depend on a choice of $q$, we will sometimes make this explicit by writing them as $P_{\L',q}$ and $P_{\L',B',q}$. Their eigenstates also depend on $q$ and can be written as $|\L',B',q\rangle$.\footnote{For $\L'=\L$ or $\L'=\bar{\L}$, there is a canonical choice of origin given by $q(\alpha)=0$. And when we write $|\L,B\rangle$, it is implicitly $|\L,B,q=0\rangle$.}

Furthermore, it makes sense to define the set of \emph{refined polarizations}:
\begin{equation}
    \tilde\Pol(M_6)=\left\{(\L',q)\,\middle\vert\,  \text{$\L'\in \Pol(M_6)$ and $q$ a quadratic refinement of $\langle \cdot,\cdot\rangle_{\text{sym},\L,\bar{\L}}$}  \right\}.
\end{equation}
The definition of $q$ depends on a choice of a decomposition of $H^3(M_3,D)\simeq \L\oplus \bar{\L}$. On the other hand, the set $\tilde\Pol(M_6)$ is independent of such choice because it can be identified with the image of the map
\begin{equation}\label{PoltoHilb}
    \tilde\Pol(M_6) \rightarrow \mathbb{P}\CH(M_6).
\end{equation}
In other words, they are represented by physical states in the Hilbert space $\CH(M_6)$. 
While both sides of \eqref{PoltoHilb} require a decomposition to be made explicit, the map is necessarily covariant under a change of basis. When there is a canonical choice of the background fields, one can identify the fiber of $\tilde\Pol(M_6)\rightarrow \Pol(M_6)$ as 2-torsion elements in $\L'^\vee$.

Such functions $\varphi=q$ can be divided into two types: the ones that are trivial when restricted to $\L\cap\L'$, and those which are not. The former have the property that $P_{\L',B',q}|\L,B_*\rangle$ is always non-zero, because
\begin{equation}
    \langle\L,B_*| P_{\L',B',q}|\L,B_*\rangle\neq 0.
\end{equation}
Then we have
 \begin{equation}
    P_{\L',B',q}|\L,B_*\rangle
    =\sum_{\alpha\in \L'} e^{2\pi i [-B'(\alpha)+q(\alpha)+B_*(\alpha)]}|\L,B_\alpha+B_*\rangle.
 \end{equation}
 This shows that only states $|\L,B\rangle$ with $B\in \bar{\Gamma'}+B_*$ can appear, and \begin{equation}
     \varphi(\alpha,B')=q(\alpha)-B'(\alpha) +B_*(\alpha)=q(\alpha)+\langle B_\alpha,B'\rangle.
\end{equation}
where
\begin{equation}
     B'(\alpha)=\langle B',\alpha \rangle=\langle B',B_\alpha \rangle+\langle B_*,\alpha \rangle .
\end{equation}
was used.

On the other hand, $\varphi$ of the second type defines a non-trivial homomorphism
\begin{equation}
    \Gamma'=\L\cap\L' \rightarrow \Z_2,
\end{equation}
such that
\begin{equation}
    P_{\L',B',\varphi}|\L,B_*\rangle = 0
\end{equation}
because $\sum_{\alpha\in\Gamma'}e^{2\pi i q(\alpha)}=0$. However, notice that
\begin{equation}
    P_{\L',B',\varphi}=P_{\L',B'+B_\varphi, q}
\end{equation}
for a certain $q$ of the first kind and $B_\varphi\in \bar{\L}$ satisfying
\begin{equation}
    B_\varphi(\beta)=\varphi(\beta),\quad \forall \beta\in \Gamma',
\end{equation}
and 
\begin{equation}
    \varphi(\alpha)-B_\varphi(\alpha)=q(\alpha),\quad \forall \alpha\in \L'.
\end{equation}
Such a $B_\varphi$ always exists because any one-dimensional representation of $\Gamma'$ can be extended to all of $\L$ and the pairing between $\bar{\L}$ and $\L$ is perfect. All choices for $B_\varphi$ live in a coset of $\bar{\Gamma'}$ in $\bar{\L}$. And then $\varphi(\alpha)-B_\varphi(\alpha)$ is obviously another quadratic refinement. Therefore, we can always view a quadratic refinement of the second type as a pair $(q,B_\varphi)$ where $q$ is of the first type and $B_\varphi\in \bar{\L}$. Then we have
\begin{equation}
    P_{\L',B',\varphi}|\L,B_*+B_{\varphi}\rangle=P_{\L',B'+B_\varphi, q}|\L,B_*+B_{\varphi}\rangle=\sum_{\alpha\in \L'} e^{2\pi i [q(\alpha)+\langle B_\alpha,B'\rangle]}|\L,B_\alpha+B_*+B_\varphi\rangle.
\end{equation}
By definition, the function $q(\alpha)$ in this formula is  a quadratic refinement of the first type. Therefore, the exponent is invariant under $\alpha\mapsto \alpha+\beta$ with $\beta \in\Gamma'$,
\begin{equation}
    q(\alpha+\beta)-q(\alpha)=q(\beta)+\langle \alpha,\beta \rangle_{\text{sym},\L,\bar{\L}}=q(\beta)+\langle \alpha,\beta \rangle=0.
\end{equation}
In other words, it only depends on $B_\alpha$, once the choice for $\L'$, $\L$ and $\bar{\L}$ is made. With a slight abuse of notation, we define
\begin{equation}
    q(B_\alpha;B'):=\left.q(\alpha)\right|_{\alpha\in p_{\bar{\L}}^{-1}(B_\alpha)} +\langle B_\alpha,B'\rangle.
\end{equation}
In fact, $q(B):=q(B_\alpha;B'=0)$ can be regarded as a quadratic refinement of the symmetric product on $\bar{\Gamma'}$, given by
\begin{equation}
    \langle B_1,B_2 \rangle'_{\text{sym}}=\langle\alpha_1, \alpha_2\rangle_{\text{sym},\L,\bar{\L}}, \quad \alpha_1\in p_{\bar{\L}}^{-1}(B_1), \quad \alpha_2 \in p_{\bar{\L}}^{-1}(B_2).
\end{equation}
This is well defined because $\langle\alpha_1, \beta\rangle_{\text{sym},\L,\bar{\L}}=\langle\alpha_2, \beta\rangle_{\text{sym},\L,\bar{\L}}=0$ for any $\beta\in \Gamma'.$

Then, up to an overall normalization factor, \eqref{6dBasisChange} can be rewritten as
\begin{equation}\label{6dBasisChange1}
    Z^{(\L',q,B_\varphi)}_{B'}=\sum_{B\in \bar{\Gamma'}}e^{2\pi i q(B;B')} Z^{\L}_{B+B_*+B_{\varphi}},
\end{equation}
where we made the dependence on the quadratic refinement $\varphi=(q,B_\varphi)$ explicit. When we set $B'=0$, this becomes
\begin{equation}\label{6dPartition1}
    Z^{(\L',q,B_\varphi)}=\sum_{B\in \bar{\Gamma'}}e^{2\pi i q(B)} Z^{\L}_{B+B_{\varphi}}.
\end{equation}
Notice that, although \eqref{6dPartition1} seems to be a special case of \eqref{6dBasisChange1}, it applies to any refined polarization $(\L',q)$ and does not require a decomposition of $H^3(M_6,D)$ into $\L'\oplus\bar{\L}'$ for some $\bar{\L}'$. One might wonder whether this would allow one to define a set of basis given by 
\begin{equation}
|\L',q,B'\in\L'^\vee\rangle :=P_{\L',q,B'}|\L,B_*\rangle
\end{equation} 
even when $\L'^\vee$ cannot be lifted to a subgroup $\bar{\L}'$ of $H^3(M_6,D)$. This is possible, but not in a canonical way because now $B_*$ is not determined by $B'$ and there are multiple choices, corresponding to different ways of lifting $B'$ to $H^3(M_6,D)$. Different choices differ by elements in $\L'$, which will lead to relative phases given by $
\langle B_\alpha,\beta \rangle$ with $\beta \in\L'$.

From the physics point of view, the obstruction to consistently choosing phases of the basis vectors is given by the anomaly of the $(-1)$-form symmetry $\L'^\vee$, which vanishes only if the following short exact sequence splits:
\begin{equation}
    \L'\rightarrow H^3(M_6,D) \rightarrow \L'^\vee\simeq H^3(M_6,D)/\L'.
\end{equation}
In other words, any $|\L',B'\in\L'^\vee\rangle$ can be obtained by action of operators on $|\L',B'\in\L'^\vee\rangle$, but different ways of getting the same state can differ by relative phases. Such symmetries and their anomalies will be the topic of the next subsection.

\subsection{Remaining symmetries and their anomalies}\label{sec:RemainingSymmetries}

In previous subsections, we studied interfaces between the 1d TQFT and invertible theories. In this subsection, We explore further the connection between the invertible theory and anomalies for the theory $T[M_6,\L]$. 

As we have seen, in the 1d TQFT, operators indexed by $\L\subset H^3(M_6,D)$ are mutually commutative. In this subsection, we explore what happens to the remaining symmetry $H^3(M_6,D)/\L$. One might also ask what happens to the dual symmetry obtained after gauging $\Lambda$. Indeed, it is a general phenomenon that, when one discrete symmetry (such as $\Lambda$) is gauged, a dual symmetry valued in $\Lambda^\vee:=\mathrm{Hom}(\Lambda,U(1))$ appears. In the present case, both are $(-1)$-form symmetries.

It turns out that $\L^\vee$ can be canonically identified with $H^3(M_6,D)/\L$, and acts in the same way on the Hilbert space. This is a general feature that we will also encounter in higher dimensions. One interesting phenomenon is that the $H^3(M_6,D)/\L$ symmetry can have an 't Hooft anomaly, which we have seen at the level of Hilbert space $\CH(M_6)$ using projection operators in the previous subsection. There are many different but equivalent ways of stating this. For example, at the level of partition functions, one can explicitly check that performing a gauge transformation in a non-trivial background specified by $B \in H^3(M_6,D)/\L$ leads to a phase that can not be absorbed by a redefinition if $\L^\vee$ is anomalous.

\begin{figure}[t]
  \centering
  \includegraphics[width=0.75\textwidth]{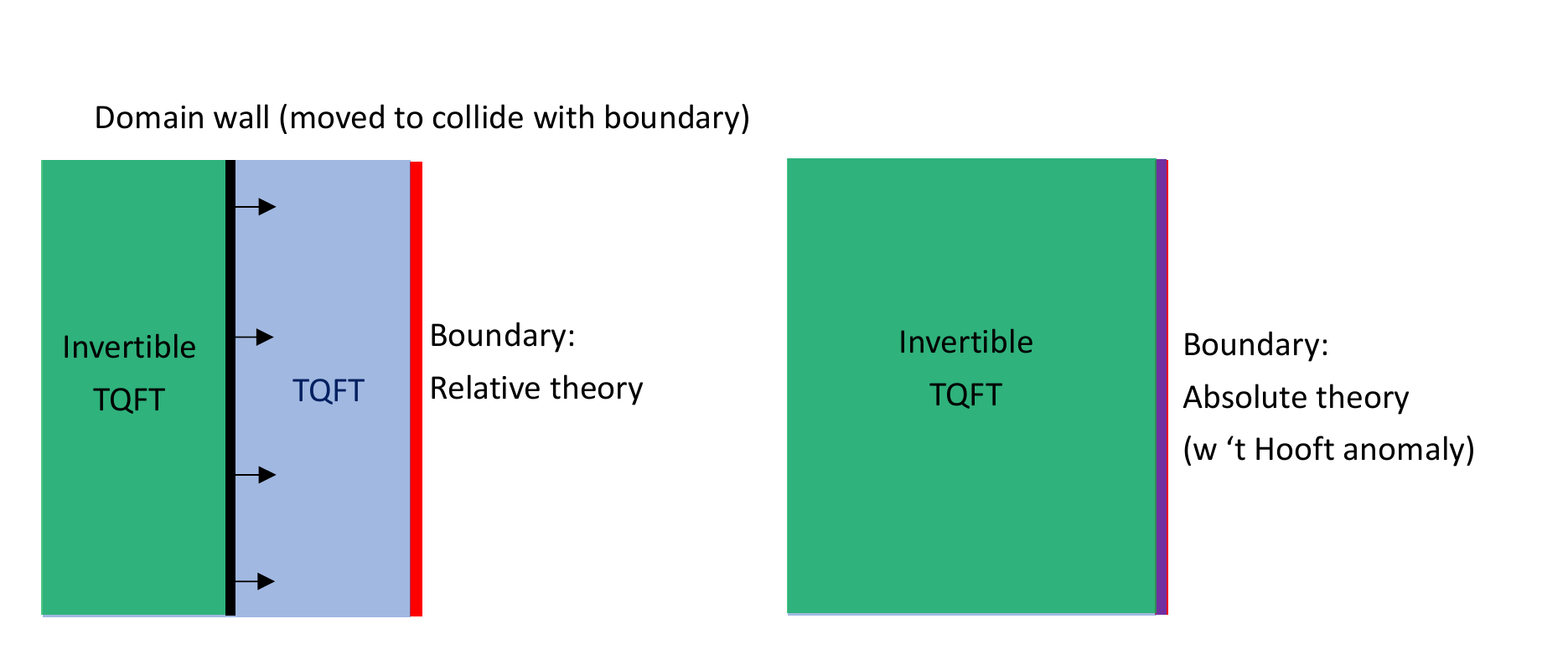}
    \caption{An absolute theory can be constructed from a domain wall separating the original bulk and an invertible TQFT, after colliding it with the boundary.}\label{fig:collidewall}
\end{figure}

As we have been focusing on the ``bulk perspective,'' it might be worthwhile to understand the 't Hooft anomaly of the $\L^\vee$ symmetry in the $T[M_6,\L]$ theory by identifying its anomaly field theory in 1d. This can be achieved in two steps: 
\begin{enumerate}
    \item Deformation. Each $B\in H^3(M_6,D)/\L$, after lifting to $H^3(M_6,\Z^r)$, determines a deformation of the action of the 1d TQFT,
    \begin{equation}
        \frac{1}{4\pi}\sum K^{IJ}\eta^{\gamma,\gamma'} \int \phi_\gamma^Id\phi_{\gamma'}^J\quad \leadsto\quad \frac{1}{4\pi}\sum K^{IJ}\eta^{\gamma,\gamma'} \int (\phi_\gamma^I-B_\gamma^I)d(\phi_{\gamma'}^J-B_{\gamma'}^J),
    \end{equation}
    and the boundary condition given by the projection operator $P_{\L}$ will be deformed to $P_{\L,B}$. The terms linear in $B$ in the deformed action combine to give a total derivative on-shell, and can be interpreted as a deformation of the boundary action. Then, what remains is the original action plus 
    \begin{equation}\label{1dAFT}
         \frac{1}{4\pi}\sum K^{IJ}\eta^{\gamma,\gamma'} \int B_\gamma^I dB_{\gamma'}^J.
    \end{equation}
    It turns out that this action, up to total derivatives and equations of motion, does not depend on the particular lift of $B$. Since this action interacts with the original one only via the boundary, we can perform the next step.
    \item ``Unfolding.'' We unfold the deformed theory coupled to the boundary to become two theories with an interface sandwiched between them. 
    As illustrated in Figure~\ref{fig:collidewall} and further explained in the next subsection, the 1d SPT \eqref{1dAFT} can be interpreted as the anomaly field theory for $T[M_6,\L]$.  
\end{enumerate}
This procedure is quite general and will be applied to TQFTs $\CT^{\text{bulk}}[M_d]$ of higher dimensions in later sections.

\subsubsection*{Gauging remaining symmetries}

Even if the remaining $\L^\vee$ symmetry has an anomaly, there can be an anomaly-free subgroup $G\subset\L^\vee$. Gauging this subgroup, one can obtain another theory, with a symmetry given by the extension of $\L^\vee/G$ by $G^\vee$. This new theory corresponds to a polarization $\L'$,
\begin{equation}
    \L^G \rightarrow \L' \rightarrow G,
\end{equation}
where $\L^G$ is the subgroup of $\L$ that pairs trivially with $G$. And, to realize $\L'$ as a subgroup of $H^3(M_6,D)$, one needs to choose a lift of $G$ to $H^3(M_6,D)$. This is possible because $G$ is anomaly-free. Furthermore, one can define states $|\L,\alpha\rangle$ with $\alpha\in G$. To fix the phase, a quadratic function on $G$ is needed.\footnote{When $G^\vee$ has 2-torsion, there can be different choices of the quadratic function on $G$. We will very soon encounter analogues of this ambiguity in higher dimensions in later sections. Also, if a quadratic function $q$ on $\L$ is given, the chosen quadratic function on $G$ will combine with $q|_{\L^G}$ to a quadratic function on $\L'$, and uniquely determine the partition function of the new theory in a consistent way.} At the level of the TQFT states, gauging $G$ leads to 
\begin{equation}
    |\L'\rangle = \sum_{\alpha\in G} |\L,\alpha\rangle.
\end{equation}
The analysis in higher dimensions is very similar (see also \cite{Tachikawa:2017gyf} for a systematic study on gauging finite group symmetries), one interesting feature is that the choice of a lift of $G$ generalizes to a choice of a $G$-SPT in higher dimensions.  

Another source of symmetries comes from isometries of $M_6$. The discrete ones coming from $\MCG(M_6)$ in general act non-trivially on $\Pol(M_6)$, and will be discussed separately, later in this section. The continuous ones are also interesting, and will be discussed when we move on to the higher-dimensional $T[M_d]$ theories.

Next, we will discuss an alternative way of looking at polarizations, which will be very useful later when we generalize it to other dimensions.

\subsection{Polarizations and topological boundary conditions}

The projection operator $P_{\L,B}$ can be interpreted as a topological domain wall between the original 1d TQFT and the invertible one discussed above. When the remaining symmetry $\L^\vee$ is anomaly-free, the invertible theory is trivial, and the domain wall becomes a topological boundary condition.

Then, the operation of gauging the symmetry $\L$ of $T[M_6]$ can be interpreted geometrically as moving the topological domain wall to collide with $T[M_6]$, creating an absolute (or projective) theory coupled to an invertible TQFT, as illustrated in Figure~\ref{fig:collidewall}.

Shrinking the interval where the 1d TQFT lives produces a theory $T[M_6,\L]$ which is now coupled to an invertible theory. The theory $T[M_6,\L]$ is now absolute, with symmetry $\L^\vee$, whose anomaly is captured by the invertible TQFT.

We remark that the colliding picture of the topological domain wall with the boundary provides a correspondence between the polarizations and the topological boundary conditions of the TQFT. This  holds true in any space-time dimension. For instance, in $3\d$ Abelian TQFT where the line operators ({\it i.e.}~the worldlines of the anyons) form a fusion algebra ${\cal A}$, the polarization is given by a Lagrangian subgroup $\Lambda$ of
\begin{equation}
H^1(M_2,{\cal A})=\Lambda\oplus\bar\Lambda~,
\end{equation}
where $M_2$ is a Riemann surface, and $H^1(M_2,{\cal A})$ has the pairing given by the composition of the intersection form on the Riemann surface $M_2$ with the braiding of the line operators in ${\cal A}$. The possible polarizations are given by the Lagrangian subalgebra ${\cal A}_\text{Lag}$ of the fusion algebra ${\cal A}$ of the line operators with respect to the braiding,
\begin{equation}
    \Lambda=H^1(M_2,{\cal A}_\text{Lag})~.
\end{equation}
On the other hand, the topological boundary conditions in the $3\d$ Abelian TQFTs are also known to be classified by the Lagrangian subalgebra \cite{Kapustin:2010hk,Kapustin:2010if,Levin:2013gaa,Barkeshli:2013yta,Kapustin:2013nva}.
This is in agreement with the correspondence between the polarizations and the topological boundary conditions explained above using the colliding topological domain wall picture.

\subsection{Example: $M_6=S^3\times S^3$}

We will now illustrate the general framework introduced above with a concrete example of $M_6=S^3\times S^3$. This example will be closely related to many examples that will appear in later parts of the paper.

In this case 
\begin{equation}
    H^3(M_6,D)=D\oplus D,
\end{equation}
and there is a decomposition of $D\oplus D$ into direct sums of abelian groups of the form $\Z_N\oplus \Z_N$, with a pairing determined by a number $\kappa\in\Z_N$. Without loss of generality, we can assume that $D\oplus D$ contains a single such copy. The non-degeneracy requires that $\kappa$ is invertible in $\Z_N$. When $N$ is even, we will also need a refinement of $\kappa$, which we denote by $\tilde{\kappa}\in \Z_{2N}$. Because $\tilde \kappa$ equals $\kappa$ when modded by $N$, there are two such refinements, namely $\tilde \kappa$ and $N+\tilde\kappa$.

In order to distinguish the two $S^3$'s in $M_6$, we denote them as $S^3_E$ and $S^3_M$, respectively, for ``electric'' and ``magnetic.'' They have non-trivial intersection pairing but have no self-intersection. Similarly, we will write $H^3(M_6,D)=\L_E\oplus \L_M$, with both $\L_E$ and $\Lambda_M$ isomorphic to $D$. Elements in $\L_E$ are the cocycles dual to cycles along $S^3_E$ and, therefore, have non-trivial periods over $S^3_M$ but trivial period over $S^3_E$, and vice versa.

In this example, there is a canonical electric decomposition of $H^3(M_6,D)$ into $\L\oplus\bar{\L}$, given by 
\begin{equation}
    \L=\L_E,\quad \bar{\L}=\L_M.
\end{equation}
Geometrically, this corresponds to choosing  $M_7=D^4_E\times S^3_M$ with $\partial D^4_E=S^3_E$. Then, we can identify
\begin{equation}
    \L_E=H^3(D^4_E\times S^3_M)\hookrightarrow H^3(S^3_E\times S^3_M).
\end{equation}
A choice of framing can be realized geometrically as a trivialization of the tangent bundle of $D^4_E$ in $D^4_E\times S^3_M$. There is a unique one (up to homotopy) that is compatible with the product structure. This choice enables us to find a section of the quotient map 
\begin{equation}
    H^3(S^3_E\times S^3_M)= H_3(S^3_E\times S^3_M) \rightarrow H_3(D^4_E\times S^3_M)=  
    \L_E^\vee,
\end{equation}
by pushing a 3-cycle in $M_7$ to the boundary $M_6$ using a vector field given by the framing of the normal bundle, similar to pushing a framed loop lying along the core of the solid torus to the boundary two-torus.\footnote{Notice that in either case the framing consist of a collection of vector fields, but using any non-degenerate linear combination of them will lead to the same element in the homology group of the boundary. }
This identifies $\L_E^\vee\simeq \bar{\L}={\L}_M$ thus canonically splitting the short exact sequence
\begin{equation}
    \L_E\rightarrow \L_E\oplus \L_M \rightarrow \L_E^\vee.
\end{equation}

There are other ways of decomposing $H^3(M_6,D)$, such as the magnetic one given by $\L'=\L_M$ and $\bar{\L}'=\L_E$. The basis change is given by
\begin{equation}
    |\L_M,B_M\rangle=\frac{1}{\sqrt{N}}\sum_{B_E\in \Z_N} e^{\frac{2\pi i \kappa}{N} B_{E}B_M} |\L_E,B_E\rangle,
\end{equation}
whose inverse transform is 
\begin{equation}
    |\L_E,B_E\rangle=\frac{1}{\sqrt{N}}\sum_{B_M\in \Z_N} e^{-\frac{2\pi i \kappa}{N} B_{E}B_M} |\L_M,B_M\rangle.
\end{equation}

In the special case of $\kappa=1$, the TQFT has action 
\begin{equation}
    \frac{N}{4\pi}\int\alpha d \beta,
\end{equation}
for $U(1)$-valued $\alpha$ and $\beta$,
and on shell we have $\alpha \in \L_E$ and $\beta \in \L_M$. For more general $\kappa$, the 1d TQFT can be described by the same action using the relabelling $B_E\rightarrow B_E \kappa^{-1}$ where $\kappa^{-1}$ is the inverse of $\kappa$ in $\mathbb{Z}_N$.

The projective action of $\L_{E}\times \L_M$ on the Hilbert space is carried by  a discrete Heisenberg group, a central extension of $\Z_N\times \Z_N$. More explicitly, in the electric basis, the action of $\alpha$ is diagonalized
\begin{equation}
    \hat\alpha|\L_E,B_E\rangle=e^{\frac{2\pi i \kappa}{N} B_E\alpha}|\L_E,B_E\rangle
\end{equation}
while $\beta$ acts by shifting,
\begin{equation}
    \hat\beta|\L_E,B_E\rangle=|\L_E,B_E+\beta\rangle.
\end{equation}
These two actions do not commute:
\begin{equation}
    \hat\beta\hat\alpha=e^{\frac{2\pi i \kappa}{N}\alpha\beta}\hat\alpha \hat\beta .
\end{equation}
In the magnetic basis, we have instead 
\begin{equation}
    \hat\beta|\L_M,B_M\rangle=e^{-\frac{2\pi i \kappa}{N} B_M\beta}|\L_M,B_M\rangle
\end{equation}
and
\begin{equation}
    \hat\alpha|\L_M,B_M\rangle=|\L_M,B_M+\alpha\rangle.
\end{equation}

We now study more general polarizations. For any subgroup $\L$ of $\L_E\times \L_M$, it is an extension 
\begin{equation}
    \Gamma_E\rightarrow\L\rightarrow\Gamma_M
\end{equation}
where $\Gamma_E=\L\cap\L_E$, and $\Gamma_M\subset \L_M$ is the image of $\L$ under the projection to $\L_M$. Notice that the roles of $\Gamma_E$ and $\Gamma_M$ are asymmetric in this definition, reflecting a preference to work with the ``electric basis.'' If one is working in the magnetic basis, the roles of $\Gamma_E$ and $\Gamma_M$ should be reversed.

Because all subgroups of $\Z_N$ are cyclic, we assume $\Gamma_E\subset \L_E $ is an index-$k$ subgroup of $\Z_N$, generated by $(k,0)$ after embedding it into $\Z_N\times \Z_N$. Then, $\Gamma_M\simeq \Z_{\frac Nk}$. Its generator can be lifted to  $(p,\frac{N}{k})\in \Z_N\times \Z_N$ for some $p$. Using the freedom to add copies of $(k,0)$ to this representative, we can assume $p\in \Z_k$. Now, it is easy to check that these two elements, $(p,\frac{N}{k})$ and $(k,0)$, generate a subgroup $\Z_l\times \Z_{\frac{N}{l}}$, where $l:=\gcd(\frac Nk, p)$, and it is already maximal isotropic. To summarize: 

\begin{quote}
 \emph{Maximal isotropic subgroups $\L$ of $\Z_N\times \Z_N$ are uniquely specified by a pair of integers $(k,p)$ with $k|N$ and $p\in\Z_k$. Given such a choice, $\L=\Z_l\times \Z_{\frac{N}{l}}$ while $\Gamma_E=\Z_{\frac Nk}$ and $\Gamma_M=\Z_{k}$.}

\end{quote}

\noindent
Denote this maximal isotropic subgroup by $\L_{k,p}$. It gives a state $|\L_{k,p}\rangle$ in the TQFT Hilbert space:
\begin{equation}\label{kpState}
    |\L_{k,p}\rangle=\frac{1}{\sqrt{|\Gamma_M|}}\sum_{B_E\in\Gamma_M}e^{2\pi i\cdot\frac{ B_E^2 k p  }{N}\cdot\frac{\tilde \kappa}{2N}}|\L_E,B_E\rangle.
\end{equation}
Here, because $B_E\in\Gamma_M$ is a multiple of $N/k$,  $B_E^2 k p /N$ is a well-defined element in $\Z_N$. When multiplied further by $\tilde{\kappa}\in \Z_{2N}$, it gives an element in $\Z_{2N}$, 
 which can be further exponentiated to obtain a phase. If $N$ is odd, then $\tilde{\kappa}/2\in\Z_N$ is interpreted as multiplying $\kappa$ by the inverse of 2.

When $N$ is even, there are two quadratic refinements which correspond to lifting $\kappa$ to either $\tilde\kappa$ or $N+\tilde\kappa$ in \eqref{kpState}. We denote the latter state by $|\L_{k,p},-\rangle$. If we have both $2|l$ and $2|\frac Nl$, then there is another $\Z_2$ choice, associated with a choice of the homomorphism $\Gamma_E\rightarrow \Z_2$. If this is non-trivial, we will have a ``polarization of the second type.'' Activating this choice corresponds to shifting every $|\L,B_E\rangle$ in the sum \eqref{kpState}, which in general is a phase. In the simplest case of $\L_{l,0}$, the $\Z_2\times \Z_2$ choice of quadratic refinement leads to the following four states:
\begin{align}
    |\L_{l,0},B_E,++\rangle&=\sum_{j=0}^{N/l}\left\vert\L,B_E=\frac{Nj}{k}\right\rangle\\
    |\L_{l,0},B_E,-+\rangle&=\sum_{j=0}^{N/l}(-)^{j}\left\vert\L,B_E=\frac{Nj}{k}\right\rangle\\
    |\L_{l,0},B_E,+-\rangle'&=\sum_{j=0}^{N/l}\left\vert\L,B_E=\frac{Nj}{k}+\frac{N}{2k}\right\rangle\\
    |\L_{l,0},B_E,--\rangle'&=\sum_{j=0}^{N/l}(-)^{j}\left\vert\L,B_E=\frac{Nj}{k}+\frac{N}{2k}\right\rangle.
\end{align}
The explicit expression in the case of more general $\L_{k,p}$ can be obtained by performing an $\SL(2,\Z)$ transformation, which we discuss in the next subsection.

There is a special class of theories obtained by choosing a $\L$ with $\Gamma_E=0$ and $\L\cap \L_M=0$. They are labeled by $(N,p)$ with $p$ coprime with $N$, generated by $(p,1)\in \Z_N\times \Z_N$. Then,
\begin{equation}
    |\L_{N,p}\rangle=\frac{1}{\sqrt{N}}\sum_{B_E\in\L_M}e^{2\pi i\cdot B_E^2  p  \cdot\frac{\tilde \kappa}{2N}}|\L_E,B_E\rangle.
\end{equation}
When expanded in the magnetic basis, one similarly has
\begin{equation}
    |\L_{N,p}\rangle=\frac{1}{\sqrt{N}}\sum_{B_M\in\L_E}e^{-2\pi i\cdot B_M^2  p^*  \cdot\frac{\tilde \kappa}{2N}}|\L_M,B_M\rangle,
\end{equation}
where $p^*$ is the inverse of $p\in \Z_N$. Now it is easy to check that this is compatible with the basis change between the electric and magnetic basis.\footnote{In fact, there is an overall phase given by Gauss quadratic sum. It can be incorporated if one suitably normalizes the expression \eqref{6dPartition} by including a constant term for $\varphi$. We will not deal with this phenomenon here. In higher dimension, the analogue for such a phase is a decoupled invertible TQFT. See {\it e.g.} Appendix B of \cite{Gaiotto:2014kfa} for more details.}

We now look at symmetries that act on $|\L_{k,p}\rangle$. They are given by $\L^\vee$, isomorphic to $H^3(M_6,D)/\L$ as a group. The anomaly for this symmetry is captured by the failure of lifting it to a subgroup of $H^3(M_6,D)$. 

An example where the anomaly is non-vanishing is $p=0$ and $k\neq 1$ a proper divisor of $N$. Then, $\L^\vee\simeq \Z_{k}\times \Z_{N/k}$, and the anomaly is given by
\begin{equation}
    \kappa \int B_1dB_2.
\end{equation}
This can be explicitly checked using
\begin{equation}
    |\Lambda_{k,p}\rangle = \sum_{j=0}^{N/k}|\L_E,jk\rangle.
\end{equation}
Now, let $\alpha$ and $\beta$ be the generator for the two factors of $\Z_N\times \Z_N$, then the action of $\alpha^{N/k}$ and $\beta^{k}$ on $|\Lambda_{k,p}\rangle$ is trivial, and the remaining symmetry is generated by $\alpha^i$ and $\beta^j$, with $0\leq i < N/k$ and $0\leq j < k$, subject to
\begin{equation}
    \hat\beta\hat\alpha=e^{\frac{2\pi i \kappa}{N}}\hat\alpha \hat\beta.
\end{equation}
This action is anomalous, however, because
\begin{equation}
    \hat\beta^k \hat\alpha=e^{\frac{2\pi i \kappa k}{N}}\hat\alpha \hat\beta^k.
\end{equation}
This shows that, in the presence of a $\Z_{N/k}$ $(-1)$-form symmetry background, the gauge transformation of the $\Z_{k}$ symmetry given by $\beta^k$ is broken by a phase.

\subsection{Mapping class group action on polarizations}\label{sec:MCG6d}

In this section, we will discuss how the mapping class group $\MCG(M_6)$ acts on $\Pol(M_6)$.

Let $\frak{M}(M_6):=\mathrm{Aut}(H^3(M_6,D))$
be the group of automorphisms of $H^3(M_6,D)$ that preserves the symplectic structure. Since the action of $\frak{M}(M_6)$ preserves the maximal isotropic condition, it also acts on $\Pol(M_6)$.
As any orientation-preserving diffeomorphism of $M_6$ preserves the symplectic form, we have a homomorphism
\begin{equation}
    \MCG(M_6)\rightarrow \frak{M}(M_6).
\end{equation}
This can be made very explicit when $M_6$ is simple enough.

Choosing a decomposition of $H^3(M_6,D)$ into $\L\oplus \bar{\L}$ enables one to identify $\CH(M_6)\simeq \C\bar{\L}$. The action of $\frak{M}$ is then represented by unitary matrices acting on $\C\bar{\L}$, as the inner product is preserved. 

Let $\frak{M}_\L$ be the stabilizer of $\L$. Although it fixes $\L$, there can be a non-trivial action on $\L^\vee$. As explained before, such action is diagonalized in the basis of $|\L,B\rangle$. On the other hand, the coset space $\frak{M}/\frak{M}_\L$ is isomorphic to the orbit of $\L$ under the action of $\frak{M}$. 
One might view different polarizations related by $\frak{M}$ (or by $\MCG(M_6)$) as giving the same theory in different duality frames. From this point of view, different theories are labeled by cosets of $\frak{M}_\L$. Although this is a valid perspective, in the present paper we find it more convenient (linguistically) to refer to different duality frames as different theories. Notice that they will necessarily have the same symmetries and anomalies. 

\subsubsection*{Anomaly of mapping class group action}

Since $\MCG(M_6)$ acts on both $\Pol(M_6)$ and $\CH(M_6)$, one could have assumed that the map
\begin{equation}
    \Pol(M_6) \rightarrow \CH(M_6)
\end{equation}
obtained by sending $\L$ to $|\L,0\rangle$ discussed previously is $\MCG(M_6)$-equivariant. However, this is not always the case due to a possible anomaly of the $\frak{M}$ action on $\CH(M_6)$, coming from the phases factor $\varphi(\alpha)$ in \eqref{6dPartition} defined through a quadratic refinement. On the other hand, the map taking into account this additional choice,
\begin{equation}
        \tilde{\Pol}(M_6) \rightarrow \CH(M_6)
\end{equation}
is expected to be equivariant under the action of $\MCG(M_6)$. 
More precisely, the action of $\frak{M}$ on $\CH(M_6)$ is only projective, given by an extension $\tilde{\frak{M}}$ of $\frak{M}$. In other words, for an order $N$ element $A$ in $\frak{M}$, it can have order $2N$ when acting on $\CH(M_6)$, with the action of $A^N$ being central.

Let us illustrate general principles outlined here with a concrete example of $S^3\times S^3$ and $D=\Z_N$.
In this case, the action of the mapping class group $\MCG(S^3\times S^3)$ on $H^3(M_6,D)$ factors through the quotient\footnote{The whole mapping class group was determined in \cite{Krylov:2003,crowley2009mapping}. 
Modulo $\Theta_7$, it is given by $\Z^2\oplus \SL(2,\Z)$.} 
\begin{equation}
    \SL(2,\Z)\rightarrow \frak{M}(S^3\times S^3)=\SL(2,\Z_N)=\SL(2,\Z)/\Gamma(N),
\end{equation}
and the $\Z^2$ part of $\MCG(S^3\times S^3)$ acts trivially on homology.
The group $\SL(2,\Z_N)$ is generated by two elements, $T$ and $S$, acting on $\Z_N\oplus\Z_N$ by
\begin{equation}
    S: \quad (a,b)\mapsto (b,-a),
    \end{equation}
and 
    \begin{equation}
    T:\quad (a,b)\mapsto (a+b,b),
\end{equation}
that has order $N$. This group is isomorphic to a product $\prod_{p_i}\SL(2,\Z_{p_i^{r_i}})$, where $p_i$ are prime factors of $N$ and $r_i$ is the maximal number such that $p_i^{r_i}|N$. 

First, let us classify orbits of the action of $\MCG$ on $\Pol(M_6)$. When $N$ is a prime, $\SL(2,\Z_N)$ acts transitively on $\Pol(S^3\times S^3)$ as \begin{equation}
    T^pS:\quad \L_{1,0}\mapsto \L_{N,p}
\end{equation}
sending $\L_{1,0}$ to any other maximal isotropic subgroup.
The stabilizer of $\L_{1,0}$ is given by matrices of the form 
\begin{equation}
    \left(\begin{matrix}
        m &n\\
        0 & m^*
    \end{matrix}\right) 
\end{equation}
with $m^*$ being the inverse of $m$ in $\Z^{\times}_N$ and $n\in \Z_N$. This stabilizer group has order $N(N-1)$, and contains a subgroup $\Z_N$ generated by powers of $T$. The coset space  is then parametrized by $\{1,S,TS,\ldots T^{N-1}S\}$. It has cardinality $N+1$, consistent with the fact that the order of $\SL(2,\Z_N)$ is $N(N-1)(N+1)$ when $N$ is prime.

When $N$ is not a prime, we know that there must be multiple orbits for the action of the mapping class group on $\Pol(M_6)$, at least one for each $k|N$ with $k\leq\sqrt{N}$, as the action does not change $\L$ and, therefore, cannot send $\L_{k,0}$ to $\L_{k',0}$ unless $kk'=N$.

One can actually show that such orbits are in bijection with the isomorphism classes of $\L$. In other word, if $\L_{k,p}$ and $\L_{k',p'}$ are isomorphic as groups, then they are in the same $\SL(2,\Z_N)$ orbit. A group element relating them can be explicitly constructed as follows. Without loss of generality, let $(k',p')=(\frac Nl,0)$, where  $l:=\mathrm{gcd}(\frac Nk,p)$. Because $\frac pl$ and $\frac{N}{kl}$ are coprime, there exist $a,b\in\Z_N$ such that \begin{equation}
    a\cdot \frac pl+b\cdot\frac{N}{kl} \equiv 1 \pmod N.
\end{equation} 
Then 
\begin{equation}
    \left(\begin{matrix}
        \frac{N}{kl} & -\frac{p}{l}\\
        a & b
    \end{matrix}\right) \in \SL(2,\Z_N)
\end{equation}
will send $(p,\frac{N}{k})$ to $(0,l)$, hence mapping $\L_{k,p}$ to $\L_{N/l,0}$.
Therefore, there is an orbit for each divisor $l$ of $N$ with $l\leq \sqrt{N}$. The stabilizer for the action of $\SL(2,\Z_N)$ on the polarization $\L_{p,k}$ is isomorphic to the stabilizer for $\L_{\frac Nl,0}$, which is generated by  $T^{\frac Nl}$ and $U^l$.

As an example, consider $N=4$. There are two types of orbits: those with $\L$ isomorphic to $\Z_4$, and the other one consisting of a unique $\L_{2,0}\simeq \Z_2\times \Z_2$. The action of $\SL(2,\Z)$ on the first orbit is given by 
\begin{center}
    \begin{tabular}{rcl}
    & $\L_{4,1}$&\\
         \rotatebox[origin=c]{45}{$\xrightarrow{\phantom{s}T\phantom{s}}$} & & \rotatebox[origin=c]{-45}{$\xrightarrow{\phantom{s}T\phantom{s}}$} \\
   $\L_{1,0}$   $\xrightarrow{\phantom{s}S\phantom{f}}$    $\L_{4,0}$ \quad\quad&   &\quad $\L_{4,2}$  $\xrightarrow{\phantom{s}S\phantom{f}}$ $\L_{2,1}$ \\
       \rotatebox[origin=c]{135}{$\xrightarrow{\phantom{s}T\phantom{s}}$}& & \rotatebox[origin=c]{-135}{$\xrightarrow{\phantom{s}T\phantom{s}}$} \\
       & $\L_{4,3}$&
    \end{tabular}
\end{center}

Using \eqref{6dBasisChange}, it is easy to work out the action of $S$ and $T$ on $\CH(M_6)\simeq \C^N$. We have
\begin{equation}
    S|\L_E,B\rangle=|\L_M,B\rangle=\frac{1}{\sqrt{N}}\sum_{B'\in \L_E}e^{2\pi i \kappa BB'/N}|\L_E,B'\rangle
\end{equation}
and $T$ shifts $\bar{\L}$ to be the diagonal of $\Z_N\times \Z_N$,
\begin{equation}
    T|\L_E,B\rangle=e^{2\pi i\cdot \frac{\tilde\kappa B^2}{2N}}|\L_E,B'\rangle,
\end{equation}
where we have restored an overall phase to ensure this is a representation of $\SL(2,\Z)$. When $\kappa=1$, this is the same as the action of $\SL(2,\Z)$ on the torus conformal blocks of $U(1)_{N}$ Chern--Simons theory, while for higher $\kappa$ it is the minimal Abelian theory ${\cal A}^{N,\kappa}$ discussed in \cite{Moore:1988qv,Hsin:2018vcg}.

In general, the action of $\SL(2,\Z_N)$ is projective because $(ST)^3$ usually acts by a non-trivial phase. Such overall phases typically will not concern us. In fact, it can be corrected by modifying the action of $T$ by a ``central charge'' correction $e^{2\pi i c/24}$ (with $c=1$ for $\kappa=1$, and see {\it e.g.}~\cite{Hsin:2018vcg} for other values of $\kappa$).
However, the action also has a more interesting anomaly if $N$ is even, in the sense that it leads to a relative phase. When $N$ is even, due to the quadratic refinement, only $T^{2N}=1$, while the action of $T^{N}$ is given by 
\begin{equation}
       T^N|\L_E,B\rangle=(-1)^{B}|\L_E,B\rangle.
\end{equation}
Then, if one starts with the states $|\L\rangle$ and acts by $\SL(2,\Z)$, one finds more than $\Pol(S^3\times S^3)$ in the orbit if $N$ is even. In fact, the ${PSL}(2,\Z_N)$ action on  $\Pol(S^3\times S^3)$, will be lifted to an action of ${PSL}(2,\Z_{2N})$ on $\tilde\Pol(S^3\times S^3)$. The fiber of $\L$ is the union of $|\L,q\rangle$, with all possible quadratic refinements $q$ (a total of four if $\L$ is a product of two cyclic groups of even order, which can only happen when $4|N$, and two otherwise). The kernel $\{1, T^N, ST^NS, ST^NS T^N\}$ of the map ${PSL}(2,\Z_{2N})\rightarrow {PSL}(2,\Z_{N})$ acts transitively on this fiber.

This is the analogue of the metaplectic correction for $\SL(2,\R)$, but one difference is that the extension is no longer central. 

\subsubsection*{$D=\Z_2$ and the ``$\frak{su}(2)$ theories''}

As an example, consider $D=\Z_2$. Then, we have three choices of $\L$ generated by $(1,0)$, $(0,1)$ or $(1,1)$ in $\Z_2 \oplus \Z_2$. They correspond to the states
\begin{equation}
    |0\rangle:=|\L_E,0\rangle, \quad S|0\rangle=\frac{1}{\sqrt{2}}(|0\rangle+|1\rangle),\quad \text{and }\quad TS|0\rangle=\frac{1}{\sqrt{2}}(|0\rangle+i|1\rangle).
\end{equation}
On the other hand, the orbit of the mapping class group action contains three additional states
\begin{equation}
     T^2S|0\rangle=\frac{1}{\sqrt{2}}(|0\rangle-|1\rangle), \quad T^3S|0\rangle=\frac{1}{\sqrt{2}}(|0\rangle-i|1\rangle)\quad\text{and }\quad ST^2S|0\rangle=|1\rangle.
\end{equation}
The four states $T^iS|0\rangle$ correspond to theories that are analogues of the four $SO(3)$ theories in 4d distinguished by different discrete theta angles, while $|0\rangle$ and $|1\rangle$ are the analogues of the $SU(2)$ and Spin-$SU(2)$ theory. For example, the theory given by $\L'=\L_E$ with the non-trivial quadratic refinement $q(1)=2\in \Z_4$ leads to a projection operator
\begin{equation}
    P_{\L',q}=1-\hat\alpha,
\end{equation}
which indeed projects onto a one-dimensional space generated by $|1\rangle$.

We can summarize the $\SL(2,\Z)$ action in the following diagram:
\begin{center}
    \begin{tabular}{rcl}
    & $SO(3)_1$&\\
         \rotatebox[origin=c]{45}{$\xrightarrow{\phantom{s}T\phantom{s}}$} & & \rotatebox[origin=c]{-45}{$\xrightarrow{\phantom{s}T\phantom{s}}$} \\
   $SU(2)   \xrightarrow{\phantom{s}S\phantom{f}}$    $SO(3)_{0}$ &   & $SO(3)_2$  $\xrightarrow{\phantom{s}S\phantom{f}}$ Spin-$SU(2)$ \\
       \rotatebox[origin=c]{135}{$\xrightarrow{\phantom{s}T\phantom{s}}$}& & \rotatebox[origin=c]{-135}{$\xrightarrow{\phantom{s}T\phantom{s}}$} \\
       & $SO(3)_3$&
    \end{tabular}
\end{center}
When lifted to 4d theories, this confirms the $\SL(2,\Z)$ action on $SU(2)$ gauge theories on non-spin manifolds conjectured in \cite{Ang:2019txy}, and it equally applies to the reduction of any relative 6d theory whose bulk is the same TQFT. For example, we also have
\begin{center}
    \begin{tabular}{rcl}
    & ($E_7/\Z_2)_1$&\\
         \rotatebox[origin=c]{45}{$\xrightarrow{\phantom{s}T\phantom{s}}$} & & \rotatebox[origin=c]{-45}{$\xrightarrow{\phantom{s}T\phantom{s}}$} \\
   $E_7$   $\xrightarrow{\phantom{s}S\phantom{f}}$   ($E_7/\Z_2)_0$
    &   & ($E_7/\Z_2)_2$  $\xrightarrow{\phantom{s}S\phantom{f}}$ Spin-$E_7$ \\
       \rotatebox[origin=c]{135}{$\xrightarrow{\phantom{s}T\phantom{s}}$}& & \rotatebox[origin=c]{-135}{$\xrightarrow{\phantom{s}T\phantom{s}}$} \\
       & ($E_7/\Z_2)_3$ &
    \end{tabular}
\end{center}
when we start with the 6d $(2,0)$ theory labeled by $E_7$.

\begin{figure}[t]
  \centering
    \includegraphics[width=0.7\textwidth]{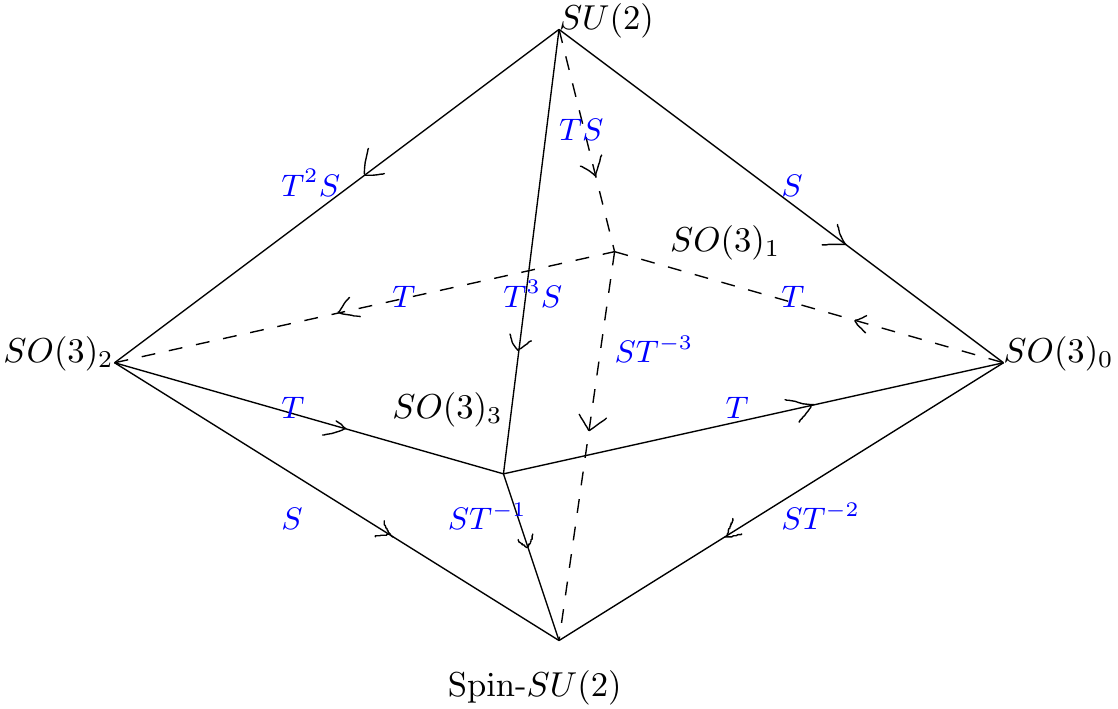}
    \caption{Six theories permuted by $PSL(2,\mathbb{Z}_4)$ arranged as vertices of an octahedron.  $PSL(2,\mathbb{Z}_4)$ then acts as orientation preserving isometries of the octahedron.}
    \label{fig:octahedron}
\end{figure}

The action of $\SL(2,\Z)$ factors through the quotient ${PSL}(2,\Z_4)=S_4$, which is an extension of ${PSL}(2,\Z_2)=S_3$ by $\Z_2\times \Z_2$ generated by $T^2$ and $ST^2S$,
\begin{equation}
    \Z_2\times \Z_2\rightarrow  {PSL}(2,\Z_4) \rightarrow {PSL}(2,\Z_2).
\end{equation}
A good way to visualize this is to put the six theories at the six vertices of a regular octahedron, see figure \ref{fig:octahedron}. Then, ${PSL}(2,\Z_4)$ is isomorphic to the group of orientation preserving isometries of the octahedron ($S_4$ acts by permuting the four pairs of opposing faces), and the subgroup $\Z_2\times\Z_2$ contains $\pi$ rotations along the three diagonals. The regular octahedron maps to a regular triangle by collapsing the three diagonals, and this $\Z_2\times \Z_2$ subgroup is exactly the kernel of the quotient map  $S_4\rightarrow S_3$.   

Next, let us consider an example where $D$ is no longer a single copy of $\Z_N$.

\subsubsection*{$D=\Z_2\times \Z_2$ and the ``$\frak{spin}(8)$ theories''}

To give another example --- relevant to compactification of 6d $(2,0)$ theory of Cartan type $D_4$ --- consider $D=\Z_2\times \Z_2$ with the pairing between $(a,b)$ and $(a',b')$ given by $ab'+ba'$. The quadratic refinement of $D$ is given by 
\begin{equation}
    \tilde\kappa=(2,2,1)\in \Z_4\times\Z_4\times\Z_2=\Gamma(D).
\end{equation}

Since all maximal isotropic subgroups in $H^3(S^3\times S^3,D)=\Z_2^4$ are isomorphic to $\Z_2\times \Z_2$, it is enough to specify two generators. Each generator is specified by four binary digits, equal to 0 or 1. We use an abbreviated notation $v=(1,0)$, $s=(0,1)$ and $c=(1,1)$ to write $(v,s)=(1,0,0,1)$ and $(c,0)=(1,1,0,0)$, and so on. We also assemble two such strings into a $2\times 2$ matrix. Then, up to permutation and addition of rows, each matrix uniquely specifies a subgroup in $\Z_2^4$. To make sure it is isotropic, one needs to check the inner product between rows is zero. For example, a good choice is
\begin{equation}
    \text{``Spin(8) theory''}: 
    \left(\begin{matrix}
    v & 0\\
    s & 0
    \end{matrix}\right).
\end{equation}
We refer to it as the Spin(8) theory since this is the direct analogue of the 4d theory with gauge group Spin(8). Furthermore, we have
\begin{equation}
    \text{``$SO(8)_+$ theory''}: 
    \left(\begin{matrix}
    v & 0\\
    0 & v
    \end{matrix}\right), \quad \text{ and\,\,\, ``$SO(8)_-$ theory''}: \left(\begin{matrix}
    v & 0\\
    s & v
    \end{matrix}\right).
\end{equation}
There are two more ``$Ss(8)$ theories'' and two more ``$Sc(8)$ theories'' obtained by replacing $v$ above with $s$ or $c$ and replacing $s$ with $v$. There are 8 more theories given by 
\begin{align}
     \text{``$PSO(8)_{0,0,0}$ theory''}: 
    \left(\begin{matrix}
    0 & v\\
    0 & s
    \end{matrix}\right),& \quad \text{ and\,\,\, ``$PSO(8)_{v,s,c}$ theory''}: \left(\begin{matrix}
    v & v\\
    s & s
    \end{matrix}\right),\\
     \text{``$PSO(8)_{0,v,v}$ theory''}: 
    \left(\begin{matrix}
    0 & v\\
    v & s
    \end{matrix}\right),& \quad \text{ and\,\,\, ``$PSO(8)_{v,c,s}$ theory''}: \left(\begin{matrix}
    v & v\\
    c & s
    \end{matrix}\right),\\
     \text{``$PSO(8)_{s,0,s}$ theory''}: 
    \left(\begin{matrix}
    s & v\\
    0 & s
    \end{matrix}\right),& \quad \text{ and\,\,\, ``$PSO(8)_{c,s,v}$ theory''}: \left(\begin{matrix}
    c & v\\
    s & s
    \end{matrix}\right),\\
     \text{``$PSO(8)_{c,c,0}$ theory''}: 
    \left(\begin{matrix}
    c & v\\
    c & s
    \end{matrix}\right),& \quad \text{ and\,\,\, ``$PSO(8)_{s,v,c}$ theory''}: \left(\begin{matrix}
    s & v\\
    v & s
    \end{matrix}\right).
\end{align}
Here we use a convention that the $PSO(8)_{x,y,z}$ theory corresponds to the isotropic subgroup containing $0$, $(x,v)$, $(y,s)$ and $(z,c)$. 

It is easy to study the action of $\SL(2,\Z_2)$ on $\Pol(M_6)$ by directly applying $S$ and $T$ to the above matrices as column operations. In the end, one finds seven orbits of the form

\begin{center}
\begin{tabular}{rccc}
     & Spin(8)\quad& $\xrightarrow{\phantom{asadfd}S\phantom{asddsf}}$& $PSO(8)_{0,0,0}$\\
  \textcircled{1} : \quad&\quad \rotatebox[origin=c]{135}{$\xrightarrow{\phantom{as}ST\phantom{as}}$} & &\rotatebox[origin=c]{225}{$\xrightarrow{\phantom{asd}T\phantom{asd}}$}\\
    & & $PSO(8)_{v,s,c}$  
\end{tabular}
\end{center}

and three orbits that are related by the triality of Spin(8):
\begin{center}
\begin{tabular}{rccc}
     & $SO(8)_-$\quad& $\xrightarrow{\phantom{asadfd}S\phantom{asddsf}}$& $PSO(8)_{0,v,v}$\\
  \textcircled{2} : \quad&\quad \rotatebox[origin=c]{135}{$\xrightarrow{\phantom{as}ST\phantom{as}}$} & &\rotatebox[origin=c]{225}{$\xrightarrow{\phantom{asd}T\phantom{asd}}$}\\
    & & $PSO(8)_{v,c,s}$  
\end{tabular}
\end{center}

\begin{center}
\begin{tabular}{rccc}
     & $Ss(8)_-$\quad& $\xrightarrow{\phantom{asadfd}S\phantom{asddsf}}$& $PSO(8)_{s,0,s}$\\
  \textcircled{3} : \quad&\quad \rotatebox[origin=c]{135}{$\xrightarrow{\phantom{as}ST\phantom{as}}$} & &\rotatebox[origin=c]{225}{$\xrightarrow{\phantom{asd}T\phantom{asd}}$}\\
    & & $PSO(8)_{c,s,v}$  
\end{tabular}
\end{center}

\begin{center}
\begin{tabular}{rccc}
     & $Sc(8)_-$\quad& $\xrightarrow{\phantom{asadfd}S\phantom{asddsf}}$& $PSO(8)_{c,c,0}$\\
  \textcircled{4} : \quad&\quad \rotatebox[origin=c]{135}{$\xrightarrow{\phantom{as}ST\phantom{as}}$} & &\rotatebox[origin=c]{225}{$\xrightarrow{\phantom{asd}T\phantom{asd}}$}\\
    & & $PSO(8)_{s,v,c}$  
\end{tabular}
\end{center}

and, finally, three orbits --- again, related by triality --- containing a single theory,

\begin{center}
\begin{tabular}{lcr}
\textcircled{5} : S\: \rotatebox[origin=c]{275}{$\circlearrowleft$}\: $SO(8)_+$ \rotatebox[origin=c]{90}{$\circlearrowleft$}\; T, &\textcircled{6} : S\: \rotatebox[origin=c]{275}{$\circlearrowleft$}\: $Ss(8)_+$ \rotatebox[origin=c]{90}{$\circlearrowleft$}\; T, &\textcircled{7} : S\: \rotatebox[origin=c]{275}{$\circlearrowleft$}\: $Sc(8)_+$ \rotatebox[origin=c]{90}{$\circlearrowleft$}\; T.
\end{tabular}
\end{center}

One can check that (after lifting to 4d theories) this indeed agrees with the result of \cite{Aharony:2013hda}.\footnote{
The label in \cite{Aharony:2013hda} for $PSO(8)$ theories is $\left(SO(8)/\mathbb{Z}_2\right)^{n_1,n_2}_{n_3,n_1}$ with $n_i=0,1$ ($+,-$ in the notation of \cite{Aharony:2013hda}), which corresponds to the discrete theta angle $n_1\pi\int w_2^{(1)}\cup w_2^{(2)}+\frac{n_2\pi}{2}\int {\cal P}(w_2^{(1)})+{n_3\pi\over 2}\int {\cal P}(w_2^{(2)})$. The $T$-transformation relates $n_1\rightarrow n_1+1$ while leaving $n_2,n_3$ invariant on spin manifolds. The relation between these labels and the ones used here is as follows ($(\alpha,\beta,\gamma):(n_1,n_2,n
_3)$ with $\alpha,\beta,\gamma=0,v,s,c$)
\begin{align}
    &(0,0,0):(0,0,0),\quad
    (v,s,c):(1,0,0);\quad (0,v,v):(0,1,1),\quad
    (v,c,s):(1,1,1);\cr
    &(s,0,s):(0,0,1),\quad
    (c,s,v):(1,0,1);\quad
    (c,c,0):(0,1,0),\quad
    (s,v,c):(1,1,0)~.
\end{align}
} We now proceed to classify the refined polarizations $(\L',q)$, and label them simply by the state in the Hilbert space $\mathbb{P}\CH(S^3\times S^3)$. As before, we choose the ``electric basis'' for $\CH\simeq\C\L_M$ spanned by $|0\rangle:=|00\rangle$, $|v\rangle:=|10\rangle$, $|s\rangle:=|01\rangle$ and $|c\rangle:=|11\rangle$.

As all $\L'$ are of the form $\Z_2\times \Z_2$, there are always four quadratic refinements. A short computation leads to the following states for each theory.

First, we have
\begin{equation}
    \text{Spin(8) theory}: \{|0\rangle, |v\rangle, |s\rangle, |c\rangle\}.
\end{equation}
In this case, as $\L'=\L_E$, quadratic refinements are labeled by 2-torsion elements in $\L_M$, which are all 4 of them, namely $0$, $v$, $s$ and $c$. All of these are quadratic refinements of the second type; therefore, they correspond to these four states with shifted background fields. We refer to these theories as Spin(8)$^{++}$, Spin(8)$^{-+}$, Spin(8)$^{+-}$ and Spin(8)$^{--}$, and adopt similar conventions for all theories below.

Next, 
\begin{equation}
    \text{$SO(8)_+$ theory}: \{|0\rangle+|v\rangle, |0\rangle-|v\rangle, |s\rangle +|c\rangle, |s\rangle -|c\rangle\}.
\end{equation}
In this case, $\Gamma'\simeq\bar{\Gamma'}=\Z_2$, and there are two $q$ of the first type, whose effect is to flip a sign of $|v\rangle$, and two $q$'s of the second type, which further shift the background fields by $v\in\bar{\L}$. The cases of $Ss(8)_{+}$ and $Sc(8)_{+}$ are similar, with $s$ and $c$ taking the special role played by $v$ in the $SO(8)_+$ theory. 

Then, the $SO(8)_-$ theory is different from the previous ones because the vector $(s,v)$ has non-trivial inner product under $\langle\cdot,\cdot\rangle_{\text{sym},\L,\bar{\L}}$, and $q$ sends $(s,v)$ to either 1 or 3 mod 4. As a consequence, $i$ can now appear in the coefficients,
\begin{equation}
    \text{$SO(8)_-$ theory}: \{|0\rangle+i|v\rangle, |0\rangle-i|v\rangle, |s\rangle +i|c\rangle, |s\rangle -i|c\rangle\}.
\end{equation}
The $Ss(8)_-$ and $Sc(8)_-$ theories are, again, very similar, obtained by permuting $v$, $s$ and $c$.

For $PSO(8)_{0,0,0}$, as $\L'=\bar{\L}$, a quadratic refinement corresponds to choosing a different homomorphism from $\L'$ to $\Z_2$. So, we have
\begin{equation}
    \text{$PSO(8)_{0,0,0}$ theory}: \left\{\begin{matrix}
    |0\rangle+|v\rangle+|s\rangle+|c\rangle, &|0\rangle+|v\rangle-|s\rangle-|c\rangle,\\ |0\rangle-|v\rangle+|s\rangle-|c\rangle,&|0\rangle-|v\rangle-|s\rangle+|c\rangle
    \end{matrix} \right\}.
\end{equation}
For $PSO_{v,s,c}$, it is given by 
\begin{equation}
    \text{$PSO(8)_{v,s,c}$ theory}: \left\{\begin{matrix}
    |0\rangle-|v\rangle-|s\rangle-|c\rangle, &|0\rangle-|v\rangle+|s\rangle+|c\rangle,\\ |0\rangle+|v\rangle-|s\rangle+|c\rangle,&
    |0\rangle+|v\rangle+|s\rangle-|c\rangle
    \end{matrix} \right\}.
\end{equation}
Next, 
\begin{equation}
    \text{$PSO(8)_{0,v,v}$ theory}: \left\{\begin{matrix}|0\rangle+|v\rangle-i|s\rangle-i|c\rangle,& |0\rangle+|v\rangle+i|s\rangle+i|c\rangle,\\ |0\rangle-|v\rangle-i|s\rangle+i|c\rangle,&
    |0\rangle-|v\rangle+i|s\rangle-i|c\rangle \end{matrix} \right\},
\end{equation}
and similarly for $PSO(8)_{s,0,s}$ and $PSO(8)_{c,c,0}$ theories. Lastly, we have the triple $PSO(8)_{v,c,s}$, $PSO(8)_{c,s,v}$ and $PSO(8)_{s,v,c}$. The first is given by
\begin{equation}
    \text{$PSO(8)_{v,c,s}$ theory}: \left\{\begin{matrix}
    |0\rangle-|v\rangle+i|s\rangle+i|c\rangle,& |0\rangle-|v\rangle-i|s\rangle-i|c\rangle,\\ |0\rangle+|v\rangle+i|s\rangle-i|c\rangle,&
    |0\rangle+|v\rangle-i|s\rangle+i|c\rangle \end{matrix} \right\}.
\end{equation}
with the rest related by permutations.

We now study the action of the mapping class group. In the basis $\{|0\rangle,|v\rangle,|s\rangle,|c\rangle\}$ the action of $S$ and $T$ generators look like
\begin{equation}
    S=\frac{1}{2}\left(\begin{matrix}
    1 & 1 &1 &1\\
    1 & 1 & -1 & -1\\
    1 & -1 & 1 & -1\\
    1 & -1 & -1 & 1
    \end{matrix}\right) \quad \text{and}\quad T=\left(\begin{matrix}
    1 & 0 &0 &0\\
    0 & -1 & 0 & 0\\
    0 & 0 & -1 & 0\\
    0 & 0 & 0 & -1
    \end{matrix}\right).
\end{equation}
Therefore, the group $\SL(2,\Z_2)$ acts genuinely on $\tilde\Pol(S^3\times S^3)$ without the need to be extended. Some of the orbits in $\Pol(S^3\times S^3)$ become larger, although this doesn't happen for \textcircled{1}, which has four orbits: one of the form
\begin{center}
\begin{tabular}{rccc}
     & Spin(8)$^{++}$\quad& $\xrightarrow{\phantom{asadfd}S\phantom{asddsf}}$& $PSO(8)_{0,0,0}^{++}$\\
  \textcircled{1}$^{++}$ : \quad&\quad \rotatebox[origin=c]{135}{$\xrightarrow{\phantom{as}ST\phantom{as}}$} & &\rotatebox[origin=c]{225}{$\xrightarrow{\phantom{asd}T\phantom{asd}}$}\\
    & & $PSO(8)_{v,s,c}^{++}$  
\end{tabular}
\end{center}
and three other obtained by replacing $(++)$ with other sign combinations. As for \textcircled{2}, \textcircled{3}, and \textcircled{4}, each turns into three orbits of cardinality 6, 3 and 3. For example, \textcircled{2} splits into
\begin{center}
    \begin{tabular}{lccccc}
    & $SO(8)_-^{++}$    &$\xrightarrow{\phantom{sf}S\phantom{sf}}$ & $PSO(8)_{0,v,v}^{++}$ & $\xrightarrow{\phantom{sf}T\phantom{sf}}$ & $PSO(8)_{v,s,c}^{++}$ \\
   \textcircled{2}$^+$:&   \rotatebox[origin=c]{90}{$\xrightarrow{\phantom{as}T\phantom{sf}}$}   & & & & \rotatebox[origin=c]{270}{$\xrightarrow{\phantom{as}S\phantom{sf}}$}\\
     & $SO(8)_-^{-+}$    &$\xleftarrow{\phantom{sf}S\phantom{sf}}$ & $PSO(8)_{0,v,v}^{-+}$ & $\xleftarrow{\phantom{sf}T\phantom{sf}}$ & $PSO(8)_{v,s,c}^{-+}$
    \end{tabular}
\end{center}
in the regular representation of ${PSL}(2,\Z_2)=S_3$ and two orbits composed of three theories
\begin{center}
\begin{tabular}{rccc}
     & $SO(8)_-^{+-}$ \quad& $\xrightarrow{\phantom{asadfd}S\phantom{asddsf}}$&$PSO(8)_{0,v,v}^{+-}$\\
  \textcircled{2}$^{+-}$ : \quad&\quad \rotatebox[origin=c]{135}{$\xrightarrow{\phantom{as}ST\phantom{as}}$} & &\rotatebox[origin=c]{225}{$\xrightarrow{\phantom{asd}T\phantom{asd}}$}\\
    & &$PSO(8)_{v,s,c}^{+-}$  
\end{tabular}
\end{center}
and
\begin{center}
\begin{tabular}{rccc}
     & $SO(8)_-^{--}$  \quad& $\xrightarrow{\phantom{asadfd}S\phantom{asddsf}}$&$PSO(8)_{0,v,v}^{--}$\\
  \textcircled{2}$^{+-}$ : \quad&\quad \rotatebox[origin=c]{135}{$\xrightarrow{\phantom{as}ST\phantom{as}}$} & &\rotatebox[origin=c]{225}{$\xrightarrow{\phantom{asd}T\phantom{asd}}$}\\
    & &$PSO(8)_{v,s,c}^{--}$
\end{tabular}
\end{center}

As for \textcircled{5}, \textcircled{6}, \textcircled{7}, each becomes two orbits, one containing only a single object, such as 
\begin{center}
\begin{tabular}{c}
\textcircled{5}$^{--}$ : \quad S\: \rotatebox[origin=c]{0}{$\acts$}\: $SO(8)_{+}^{--}$ \rotatebox[origin=c]{180}{$\acts$}\; T, 
\end{tabular}
\end{center}
while the other containing three theories,
\begin{center}
\begin{tabular}{rccc}
     & $SO(8)_{+}^{-+}$\quad& $\xrightarrow{\phantom{asadfd}S\phantom{asddsf}}$& $SO(8)_{+}^{+-}$\\
  \textcircled{5}$^{++}$ : \quad&\quad \rotatebox[origin=c]{135}{$\xrightarrow{\phantom{as}ST\phantom{as}}$} & &\rotatebox[origin=c]{225}{$\xrightarrow{\phantom{asd}T\phantom{asd}}$}\\
    & &$SO(8)_{+}^{++}$  
\end{tabular}.
\end{center}

Again, the results above universally apply to any 6d theory coupled to the same 7D TQFT, such as 6d $(2,0)$ theories labeled by $\frak{so}(2N)$ for any even $N\geq2$.

 \section{General aspects of compactifications of 6d theories}\label{sec:GeneralTheory}
 
We now study compactification of 7d/6d coupled systems on a manifold $M_d$ of dimension $d<6$, generalizing the discussion in the previous section. The goal is to define and study the notion of ``polarizations on $M_d$'' --- choices that one can make when reducing the coupled system to obtain absolute theories in $(6-d)$ dimensions. 

We expect it to enjoy the following list of properties.
\begin{itemize}

\item Similar to the $d=6$ case, the set of polarizations on $M_d$ should capture all reductions of the 7d/6d system that are ``bulk universal'' ({\it i.e.}~not involve additional choices specific to the boundary theory and, therefore, are robust under deformation of the coupled system):
\begin{equation}\label{PoldAb}
    \tilde\Pol(M_d)\simeq \{\text{Absolute $(6-d)$-dim theory obtained by reduction on $M_d$}\}.
\end{equation}
One can identify theories that differ by a choice of the quadratic refinement (whose meaning will become clear shortly), leading to 
\begin{equation}
    \tilde\Pol(M_d)\rightarrow \Pol(M_d).
\end{equation}
In practice, it is usually easier to first obtain the latter and then classify compatible quadratic refinements.

\item Just as in the case of $d=6$, one can construct absolute theories using topological domain walls between the $(7-d)$-dimensional TQFT obtained by reducing the 7d TQFT on $M_d$ and an invertible TQFT. For simplicity, we refer to such domain walls as ``topological boundary conditions'' even when the invertible theory is non-trivial. Then, each such boundary condition gives rise to an absolute theory, as shown in Figure \ref{fig:collidewall}. The converse is also expected to be true --- the set $\tilde\Pol(M_d)$ is expected to be isomorphic to the topological boundary conditions for the $(7-d)$-dimensional TQFT (with two boundary conditions deemed  equivalent if they differ by a $(6-d)$-dimensional TQFT). Alternatively, one can use this as a definition of $\tilde\Pol(M_d)$.\footnote{We thank Dan Freed for discussions on this interpretation.}

\item Since the absolute theory has a well-defined reduction on a given manifold $M_{d'}$ with $d'\leq 6-d$, we have a map
\begin{equation}
    \tilde\Pol(M_d)\rightarrow \tilde\Pol(M_d \times M_{d'}),
\end{equation}
which in general is neither injective nor surjective, as we shall see later.

\item For a given polarization $\CP \in \tilde\Pol(M_d)$, the theory $T[M_d,\CP]$ has different symmetries, and their 't Hooft anomolies are determined by $\CP$. And the spectrum of charged operators is constrained by $\CP$.

\end{itemize}

We now proceed to define and study this generalized version of polarizations.

\subsection{Polarization and compactification}\label{sec:PolandComp}

When the 7d/6d coupled system is reduced on a $d$-dimensional manifold $M_d$, on the boundary we have a $(6-d)$-dimensional theory on a manifold $M_{6-d}$. In the $(7-d)$-dimensional bulk, we have a TQFT with the action 
\begin{equation}\label{GeneralBulk}
    \sum_{I,J,i,j,r} {K^{IJ}\eta^{ij}_{r}\over 4\pi}\int b^{I,i}_{r}db^{J,j}_{6-d-r}
\end{equation}
where $b^{I,i}_r$ is the $r$-form from reducing $C^I$ and $i$ labels the $(3-r)$-cycle on $M_{d}$ along which $C^I$ is reduced. $\eta_r^{ij}$ is the intersection pairing between $H^{3-r}(M_d)$ and $H^{d-3+r}(M_d)$.
For simplicity, in the above action we only write the fields from the reduction on free cycles; it can be easily generalized to the reduction on torsion or discrete cycles using the method in Section \ref{sec:torsiondiscretefreecycle}, which we will explicitly elaborate in the companion paper \cite{Gukov:2020inprepare}.

The action can be written compactly as
\begin{equation}\label{1dActionp}
    \int \langle \alpha,d\alpha\rangle, \quad \text{with} \quad  \alpha \in \bigoplus_{i=0,1,2,3}H^i(M_{6-d},H^{3-i}(M_d,D))\quad \text{on shell,}
\end{equation}
where the pairing $\langle\cdot,\cdot\rangle$, when restricted to on-shell configurations,
\begin{equation}
    H^i(M_{6-d},H^{3-i}(M_d,D))\otimes H^{6-d-i}(M_{6-d},H^{d-3+i}(M_d,D)) \rightarrow U(1),
\end{equation}
comes directly from the pairing on 
\begin{equation}
    H^3(M_d\times M_{6-d}, D )\simeq \bigoplus_{i=0,1,2,3}H^i(M_{6-d},H^{3-i}(M_d,D)).
\end{equation}

Again, one can argue that the boundary theory can be made absolute by gauging a subgroup $\Lambda$ that is maximal isotropic. However, one important difference now is that one wants not just a single choice for a particular $M_{6-d}$, but a consistent family of $\Lambda$ for all possible $M_{6-d}$. How can one achieve this? 

 One option is to consider families of maximal isotropic groups of the form $H^*(M_{6-d},L)\{3\}$,
where $L$ is a maximal subgroup of $H^{d-3\leq*\leq 3}(M_d,D)$ trivial under the pairing
\begin{equation}
    H^*(M_d,D) \otimes H^*(M_d,D) \rightarrow U(1),
\end{equation}
and $H^*(M_{6-d},L)\{3\}$ is the degree 3 piece of the this cohomology group. If we assume that $L$ has a decomposition into graded pieces
\begin{equation}
    L=L^{(0)}\oplus L^{(1)}\oplus L^{(2)}\oplus L^{(3)},
\end{equation}
then
\begin{equation}
    H^*(M_{6-d},L)\{3\}=\bigoplus_{i=0}^{3}H^i(M_{6-d},L^{(3-i)}).
\end{equation}

The image of the map
\begin{equation}
    H^*(M_{6-d},L)\{3\}\rightarrow  H^3(M_{6-d}\times M_6,D)
\end{equation}
is always isotropic. If it is also maximal for all $M_{6-d}$, then this defines a polarization on $M_d$, denoted as $\CP_L$.

Physically, $H^{d-3\leq*\leq 3}(M_d,D)$ is the symmetry in $6-d$ dimensions coming from the reduction of the 2-form symmetry $D$ in 6d. And the above corresponds to gauging an anomaly-free subgroup $L$ of it. The fact that $L$ is maximal guarantees that after gauging, the boundary theory is coupled to an invertible $(7-d)$-dimensional TQFT. One can again think of this process using topological interfaces between $\CT^{\text{bulk}}[M_d]$ and an invertible theory as illustrated in Figure~\ref{fig:collidewall}.

This motivates a classification of polarizations on $M_d$ by looking at the union of images of $\L\subset H^3(M_{6-d}\times M_d,D)$ under the map
\begin{equation}
   H_*(M_{6-d},\Z) \times H^*(M_{6-d},H^*(M_d,D))\{3\} \rightarrow H^*(M_d,D)
\end{equation}
as we scan over all cycles in $H_*(M_{6-d},\Z)$ for all $M_{6-d}$. We will refer to it as the \emph{spectrum group}, and denote it as $\CS({\CP})$. Then $\CS({\CP}_L)=L$. 

Then, we can classify polarizations as follows
\begin{enumerate}
    \item We say a polarization $\CP$ is \emph{pure} if $\CS(\CP)$ has trivial pairing with itself in $H^*(M_d,D)$.
    \item $\CP$ is said to be a \emph{mixed} polarization if $\CS(\CP)$ has non-trivial pairing with itself.
\end{enumerate}
These two classes are on equal footing from the viewpoint of the topological boundary conditions in the $(7-d)$-dimensional TQFT. Pure polarizations, however, are easier to classify since they are in bijection with choices of $L$. Therefore, one of the main goals of this section will be to develop tools for understanding mixed polarizations. 

Many pure polarizations can be obtained geometrically as $(d+1)$-manifolds $W_{d+1}$ bounding $M_d$. Then, $L$ can be interpreted as the image of the first map
\begin{equation}
   H^{i}(W_{d+1},D)\rightarrow H^{i}(M_d,D)\rightarrow H^{i+1}(W_{d+1},M_d;D)
\end{equation}
in degree $d-3\leq i\leq 3$. 
Then, given any $M_{6-d}$, this determines the maximal isotropic subgroup $\L\subset H^3(M_d\times M_{6-d},D)$ as the image of the map from $H^{i}(W_{d+1}\times M_{6-d},D)$. Mixed polarizations could also admit geometric constructions, though they require using 7-manifolds that are not products like $W_{d+1}\times M_{6-d}$.

For some $L$ there can be special choices of $\bar{L}\subset H^*(M_d,D)$ such that 
\begin{equation}
    H^{d-3\leq*\leq 3}(M_d,D)\simeq L\oplus\bar{L},
\end{equation}
also trivializing the pairing.
Then, for any $M_{6-d}$, it gives a decomposition
\begin{equation}\label{H3Split}
    H^3(M_{6-d}\times M_d,D)=H^*(M_{6-d},L)\{3\}\oplus H^*(M_{6-d},
    \bar{L})\{3\}=\L\oplus\bar{\L}.
\end{equation}
Such special polarization is said to be \emph{splittable}, with $\bar{L}$ being a splitting of it. In this case, there is a canonical choice of a quadratic refinement given by $q=0$ on $\L$.

As an example, for $M_d=S^1$, we have $H^*(M_d,D)=D^{(0)}\oplus D^{(1)}$ in degree 0 and 1. Then, $L=D^{(0)}$ or $D^{(1)}$ give pure polarizations, both of which are splittable. A non-splittable example is $D=\Z_4$ and $L=\Z^{(0)}_2\times\Z^{(1)}_2$. In general, there can be other polarizations, including some mixed ones with their spectrum group being the entire $D^{(0)}\oplus D^{(1)}$.

Often it is not hard to find a splittable polarization.\footnote{Note, however, that for some $M_d$ and $D$ it could be the case that no polarizations exist. Another way of saying this is that there might not be any topological boundary condition for the $(7-d)$-dimensional TQFT. An example is $M_d=\C \mathbf{P}^2$ and $D=\Z_p$. The 3d TQFT in the bulk of $T[\C \mathbf{P}^2]$ is an abelian Chern--Simons theory with no topological boundary conditions.} For example, when $d=3$ and $M_d$ is an arbitrary 3-manifold, one can take $L=H^0(M_3,D)\oplus H^1(M_3,D)$ and $\bar{L}=H^2(M_3,D)\oplus H^3(M_3,D)$. When $d=2$, and $M_d$ is a genus-$g$ Riemann surface, one can take $L$ to be $H^0(M_2,D)\oplus L^{(1)}$, where $L^{(1)}$ is a maximal isotropic subgroup of $H^1(M_2,D)$. There are many such choices, acted upon by ${Sp}(2g,\Z)$.

In later sections and the companion paper \cite{Gukov:2020inprepare}
, when we study reductions on various $M_d$, we will always start by first classifying pure polarizations and finding among them one that is splittable. Just as in the 0-dimensional case ($d=6$), they provide a basis for us to study partition functions in other polarizations, especially the mixed ones. 

Furthermore, given a polarization, one would also wish to understand how to explicitly construct a theory, not just its partition function on a given $M_{6-d}$. This question can be better understood by relating polarizations with SPT phases in $6-d$ dimensions, which we turn to next.

\subsection{Polarizations and discrete theta angles}

A refined polarization on $M_d$ determines a maximal isotropic subgroup $\L$ of $H^3(M_d\times M_{6-d},D)$, for any $M_{6-d}$, together with a quadratic refinement.
Then, one can consider projection operators \eqref{0dProjection} from the previous section acting on the TQFT Hilbert space $\CH(M_d\times M_{6-d})$, such as
\begin{equation}
    P_{\CP}=\sum_{\alpha\in \L}\hat{\alpha}
\end{equation}
as well as their cousins with non-trivial phase factors with background fields given by $B\in H^3(M_d\times M_{6-d},D)/\L$,
\begin{equation}
    P_{\CP,B}=\sum_{\alpha\in \L}\hat{\alpha}e^{-2\pi i B(\alpha)}.
\end{equation}

If $\CP=\CP_L$ is a pure polarization, then action of these projection operators on the boundary can be interpreted as gauging the subgroup $L$ of the symmetry $H^*(M_d,D)$ and then turning on a background field for the dual $L^\vee$ symmetry. 

Just as in the 0d case, where the set of projection operators defines a basis in the Hilbert space up to phase factors, in higher dimensions too it defines a partition function of the $6-d$ theory on all manifolds $M_{6-d}$  up to a $(6-d)$-dimensional SPT phase. There is usually not a canonical choice for this SPT phase. For most part, one can ignore this ambiguity. However, when we gauge an anomaly-free subgroup $G$ of the remaining symmetry $L^\vee$, the resulting theory can depend on the SPT phase chosen. This additional piece of data fixes a lift of $G$ to $H^*(M_d,D)$. We have already encountered this phenomenon in the case of $d=6$ in Section~\ref{sec:RemainingSymmetries}, which we will see again for $d=0$ and 1.

For mixed polarizations, they can be often obtained by turning on certain discrete theta angles when gauging $L$ as we will see in detail later.

Now we turn to the question of what characterizes a nice family of polarizations on $M_d\times M_{6-d}$ that give a polarization on $M_d$.
Because we can fully characterize polarizations on a 6-manifold, and expect an ``embedding''\footnote{For a given $M_{6-d}$, this map may not be injective, but it is natural to expect that any two polarizations can always be distinguished by a suitable choice of $M_{6-d}$.
}
\begin{equation}
    \tilde\Pol(M_d)\rightarrow \tilde\Pol(M_d \times M_{6-d}),
\end{equation}
it is natural to start with partition functions of the theory in $6-d$ dimensions.
From this viewpoint, different choices of polarization correspond to different ways of summing the components of the partition vector into combinations that satisfy certain TQFT functoriality conditions. 

However, to make this explicit, one needs to choose a basis, and we assume that there is a splittable polarization $\CP_{L}$ given by $(L,\bar{L})$. Then, given an $M_{6-d}$, there is decomposition as in \eqref{H3Split}, while another polarization $\CP'$ gives a maximal isotropic subgroup $(\L',q)$ of $H^3(M_{6-d}\times M_d,D)$ together with a quadratic refinement $q$, such that 
\begin{equation}\label{dPartition}
Z^{\L',q}(M_{6-d})=\sum_{\alpha\in \L'} e^{2\pi i\varphi(\alpha)} Z^{\L}_{B_\alpha},
\end{equation}
where again $B_\alpha$ is the image of $\alpha$ under projection to $\bar{\L}$. Notice that the phase $e^{2\pi i\varphi(\alpha)}$ only depends on $B_\alpha$ since, if $
\alpha+\beta\in \L'$ also projects onto $B_\alpha$, then a short computation shows that
\begin{equation}\label{BAlphaComputation}
    \varphi(\alpha+\beta)-\varphi(\alpha)=\langle B_\alpha, \beta\rangle +\langle \alpha, p_{\bar{\L}}(\beta)\rangle + \varphi(\beta) =\langle \alpha, \beta\rangle=0.
\end{equation}
When we need to emphasize this point, we shall also write it as $\varphi(B_\alpha)$, with a slight abuse of notation. For the functionality, this phase factor itself must be a $(6-d)$-dimensional SPT phase. 

\begin{figure}[t]
  \centering
    \includegraphics[width=0.6\textwidth]{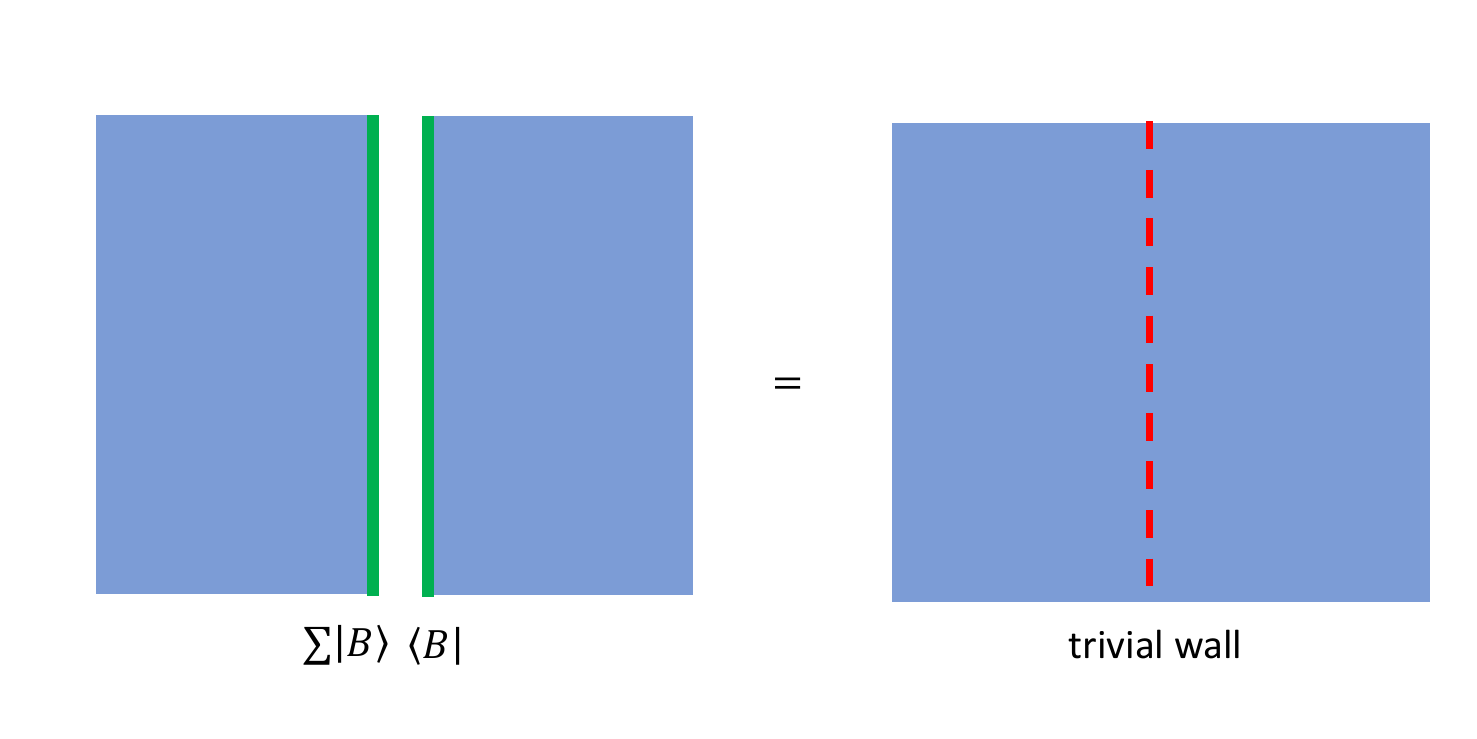}
    \caption{Cutting the manifold and summing over the boundary conditions $\sum_B|B\rangle \langle B|=1$ is equivalent to inserting the identity interface.}\label{fig:cutandglue}
\end{figure}

This can be better understood from another point of view as follows. Projection operators labeled by $\CP_L$ and $B\in \bar\L$ define topological boundary conditions $\CB_B$ for the $(7-d)$-dimensional theory $\CT^{\mathrm{bulk}}[M_d]$ on $M_{6-d}$. One can create an identity interface by summing over $B$. This is illustrated in Figure \ref{fig:cutandglue}. An interface for a fixed $B$ determines the value of the bulk field on the interface. On the other hand, summing over all of possible values makes the interface transparent; therefore, such an interface is expected to be the identity interface.

\begin{figure}[t]
  \centering
    \includegraphics[width=0.7\textwidth]{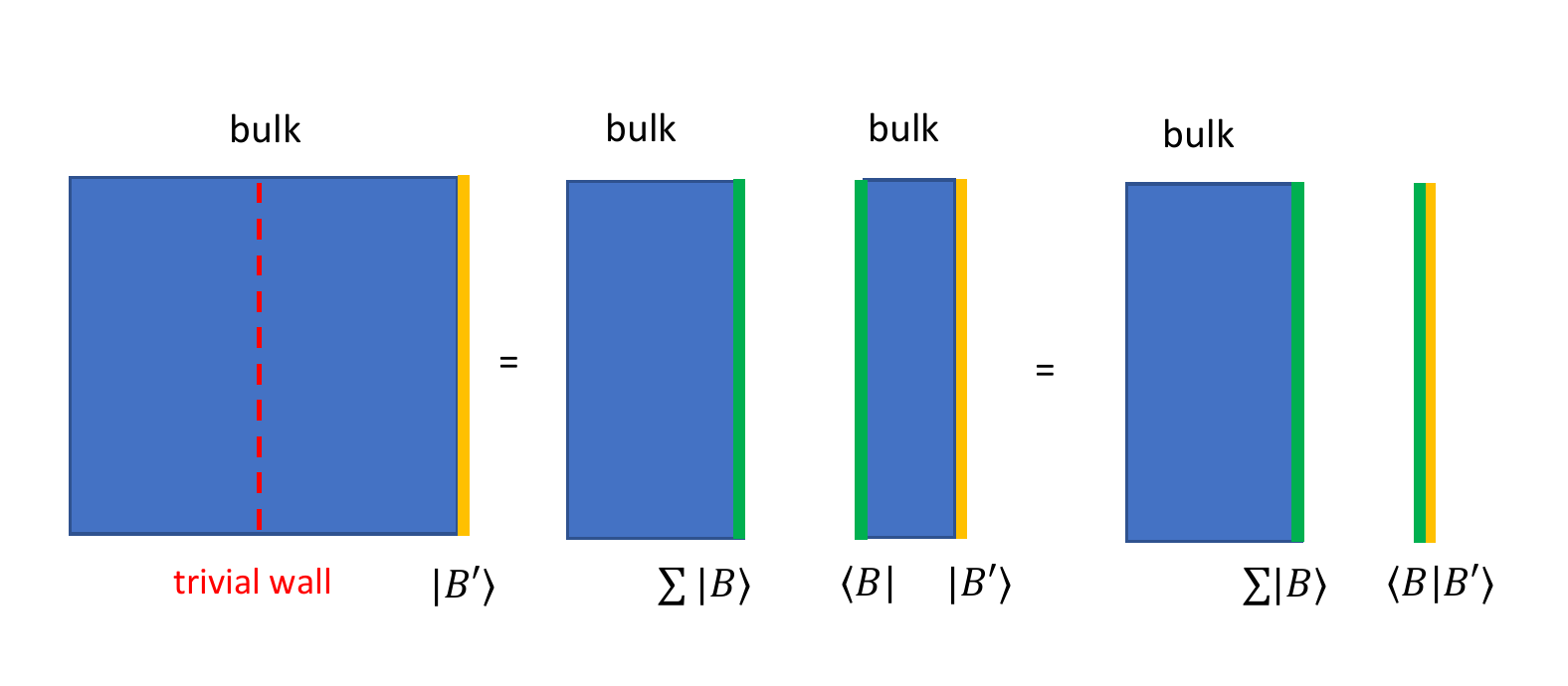}
    \caption{Computation by inserting the trivial wall.}\label{fig:trivialwall}
\end{figure}

In order to compute the partition function given by any other topological boundary condition $\CB'$, it is convenient to insert this identity interface. Then, one only needs to compute the partition function of the theory $\CT^{\mathrm{bulk}}[M_d]$ sandwiched between $\CB_B$ and $\CB'$, illustrated in Figure~\ref{fig:trivialwall}. This gives a TQFT with $\bar L$ symmetry in $6-d$ dimensions. For the class of 7d theories that we are interested in, the partition functions turn out to be, up to an overall normalization, an SPT phase with $\bar L$ symmetry in $6-d$ dimensions due to the identity \eqref{BAlphaComputation}. 

Then, the task of classifying polarizations on $M_d$ becomes closely related to the problem of classifying SPTs. In the present case, they are conjectured to be classified by the dual of the torsion part of\footnote{Here, $s$ stand for structures on the $M_{6-d}$ manifolds that we are imposing. The default choice is $s=$SO for most of the paper, or equivalently only requiring the manifold to be oriented. Depending on the application, one can require that it is spin, or even relax it to be un-oriented. } 
\begin{equation}
    \Omega^{s}_{6-d}(\mathbf{B}^3L^{(0)}\times\mathbf{B}^2L^{(1)}\times\mathbf{B}L^{(2)}\times L^{(3)}),
\end{equation}
and in practice one can search for these SPTs by writing actions for $B_\alpha$ fields of degrees 0, 1, 2 and 3, and characteristic classes of the tangent bundle of $M_d$. 

However, there is an important caveat: the sum in \eqref{dPartition}, when viewed as a sum over $B_\alpha \in \bar{\L}$, may not run over all elements in $\bar{\L}$. Therefore, one needs to first restrict to a subgroup of $\bar{\L}$. For a given $M_{6-d}$, there can be many such subgroups, but we are only interested in the ``universal'' ones in the sense that they can be defined without referring to the details of $M_{6-d}$. For instance, it has to be invariant under the action of the mapping class group of $M_{6-d}$. Examples of such subgroups include those obtained by choosing a subgroup of $\bar L$, and those obtained by imposing conditions on $B_\alpha$ (such as $\int_{M_4}B^2_\alpha=0$ for a $\mathbb{Z}_2$-valued $2$-form $B_\alpha$). 

On the other hand, given an SPT, or a function $\varphi(B_\alpha)$ on all $M_{6-d}$, it is often simple to obtain the corresponding refined polarizations, 
as we shall see in examples below. It is not guaranteed that any SPT can be realized by a polarization. For example, as the phase factors ultimately come from the $CdC$ theory, the resulting SPT phase has an effective action quadratic in the background fields of the symmetry.

In case $T[M_d,\CP_L]$ is realized by a gauge theory with $\bar{\L}$ classifying topological class of the (twisted) gauge bundle, restricting to a subgroup of $\bar{L}$ restricts the set of gauge bundles to be summed over in the path integral, while the phase $\varphi$ represents discrete theta angles of the gauge theory. These are specified by the polarization $\CP'$ and often lead to an explicit description of the theory $T[M_d,\CP']$ in terms of either path integral or Lagrangian.

\subsection{Remaining symmetries and their anomalies}\label{sec:RemainingSymmetry}

A choice of pure polarization $\CP_L$ leads to a theory with global symmetry 
\begin{equation}
L^\vee=H^{d-3\leq *\leq 3}(M_d,D)/L.
\end{equation} 
The background fields for these symmetries are valued in $L^\vee$, while the charged objects are labeled by the dual of $L^\vee$, which is $(L^\vee)^\vee=L$ itself. Geometrically, such polarizations can sometimes be obtained by choosing a $W_7=W_{d+1}\times M_{6-d}$ with $\partial W_{d+1}=M_{d}$,\footnote{Note that in general there are also many pure polarizations that are not ``geometric.'' In the special case of $d=0$, $M_0$ being a point is not null-cobordant. Therefore, none of the polarizations in $\Pol(\mathrm{pt})$ are geometric. On the other hand, recall that all oriented 1, 2 and 3-manifolds are null-cobordant, which lead to many geometric pure polarizations in $\Pol(M_d)$.}  such that $L^{(d-n)}$ is the image of the first map in 
\begin{equation}
\begin{matrix}
    H^{d-n}(W_{d+1},D) &   \rightarrow & H^{d-n}(M_d,D)   &\rightarrow& H^{d-n+1}(W_{d+1},M_d;D)\\\nonumber
    \rotatebox[origin=c]{270}{$\simeq$}& & \rotatebox[origin=c]{270}{$\simeq$} & & \rotatebox[origin=c]{270}{$\simeq$}  \\
    H_{n+1}(W_{d+1},M_d;D)&   \rightarrow & H_{n}(M_d,D) &\rightarrow& H_{n}(W_{d+1},D)
\end{matrix}
\end{equation}
when restricted to the relevant degrees, while $L^{\vee(d-n)}$ can be identified with the image of the second map. 

\begin{figure}
  \centering
    \includegraphics[width=0.5\textwidth]{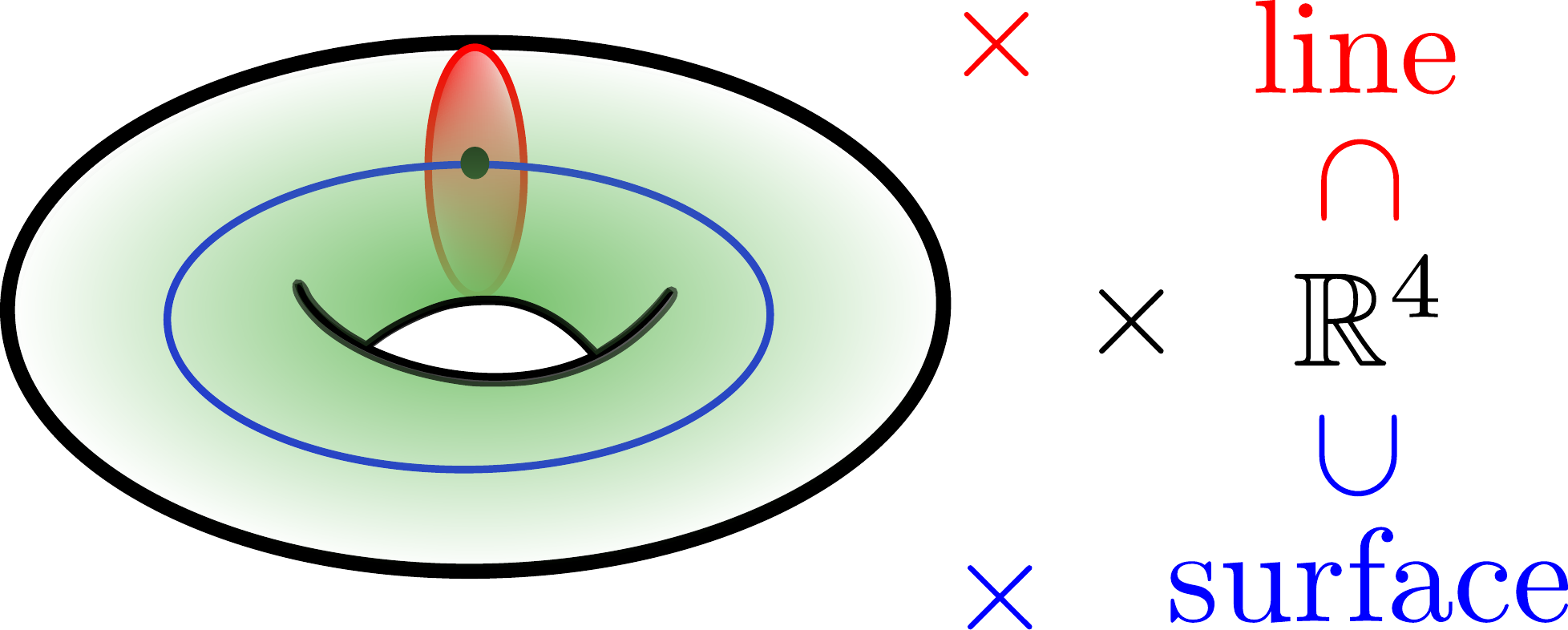}
    \caption{When $d=2$ and $M_2=T^2$ is a two-torus, the solid torus $D^2\times S^1$ gives a pure polarization. A three-dimensional operator wrapping $D^2$ (red in the figure) becomes a line operator in $\R^4$, while those along $S^1$ (blue in the figure) give rise to topological surface operators that generate a 1-form symmetry. Since they intersect in the solid torus, the line operator is charged under the 1-form symmetry.}\label{fig:SolidTorus}
\end{figure}

The group $L^{(d-n)}$ labels objects charged under the $L^{\vee(n)}$ piece in the remaining $L^\vee$ symmetry. These come from elements in $H_{n+1}(W_{d+1},M_d;D)$, which can be interpreted as 3-dimensional defect operators in 7d wrapping $(n+1)$-dimensional cycles with boundary represented by $n$-cycles in $M_{6-d}$. In the theory $T[M_d]$ that lives on $M_{6-d}$, these are $(2-n)$-dimensional charged objects coupled to the $(2-n)$-form symmetry $L^{\vee (n)}$. On the other hand, the symmetry defects for $L^{\vee (n)}$ come from those $(d-n)$-cycles in $M_{d}$ that are non-contractible in $W_{d+1}$. In $T[M_d]$, they have dimension $3-d+n$. Figure~\ref{fig:SolidTorus} is an illustration of this for $d=2$, $M_2=T^2$ and $W_3=D^2\times S^1$ with $n=1$. 

In the geometric setup, it is clear that both charged objects and symmetry defects are robust for topological reasons.

The remaining symmetry in general has anomalies, if $L^\vee$ cannot be lifted to be a subgroup of $H^{*}(M_d,D)$, similar to what we have observed in the case of 0d theories. In other words, the anomaly is captured by the extension class of
\begin{equation}
    L\rightarrow H^{d-3\leq *\leq 3}(M_d,D)\rightarrow L^\vee.
\end{equation}
The invertible TQFT for this anomaly can again be obtained via the ``deformation--unfolding'' trick introduced for $d=6$. Namely, one simply needs to replace fields in \eqref{GeneralBulk} by discrete-valued background fields to obtain the anomaly field theory for the $L^\vee$ symmetry.

\subsubsection*{Symmetries of $T[M_d]$ from mixed polarization}

The story is more interesting when we consider mixed polarizations. The quotient \begin{equation}
    \tilde{\CS}(\CP):=H^{3-d\leq * \leq 3}/\CS(\CP)
\end{equation} 
is no longer isomorphic to $\CS(\CP)^\vee$, but it  still corresponds to symmetries of the remaining theory. What is different in the mixed case is that there are more symmetries coming from the dual of a subgroup in $\CS(\CP)$. To discuss this, we first need to better interpret $\CS(\CP)$ in the mixed case. Naively, it becomes larger than it should be, but not all elements in $\CS(\CP)$ corresponds to independent charged objects. In fact, some will always live on the boundary of other objects. Therefore, the fact that $\CS(\CP)$ is non-trivial under the pairing in $H^*(M_d,D)$ doesn't imply inconsistency, as charged objects on the boundary of other charged objects can be improperly quantized, as long as the latter carry the right amount of flux for the former. We will see this more concretely when we discuss examples.

When looking for symmetries, one should first look at the subgroup $\CS(\CP)_{\text{ind}}$ of $\CS(\CP)$ that corresponds to charged objects that can exist independently on their own. Then, the symmetry of the theory is given by the dual $\CS(\CP)_{\text{ind}}^\vee$.

The subgroup $\CS(\CP)_{\text{ind}}$ trivializes the pairing on $H^*(M_d,D)$, and it  is also maximal. Therefore, there is a pure polarization associated with it, defining a theory with the same symmetry $\CS(\CP)_{\text{ind}}^\vee$ (and in fact the same anomaly for $\CS(\CP)_{\text{ind}}^\vee$) as in the theory given by the mixed polarization.

\subsection{Mapping class group action on $\Pol(M_d)$}

As in the 0d case, the action of the mapping class group $\MCG(M_d)$ on $\Pol(M_d)$ factors through the automorphism group of $H^*(M_d,D)$. And, just like in the 0d case, this action has an anomaly in the sense that the map from $\Pol(M_d)$ to the partition functions of $T[M_d]$ is not equivariant, and it is more natural to consider the action on the space of refined polarizations 
\begin{equation}
    \Z_2^{\#} \rightarrow \tilde\Pol (M_d) \rightarrow \Pol (M_d),
\end{equation}
where the number of $\Z_2$ factors can vary and essentially counts the number of possible quadratic refinements that are ``functorial.'' Examples include the Pontryagin square of the middle cohomology of $M_{6-d}=M_{4}$ for $d=2$ and the quadratic refinement for the $\Z_2$-valued intersection pairing on the middle cohomology of $M_{2}$ or $M_6$ for $d=4$ and $d=0$ respectively.

Another property of this action is that pure polarizations do not transform into mixed polarizations and vice versa. Furthermore, it doesn't change the global symmetry and their 't Hooft anomalies.

A generic choice of polarization does not respect the full diffeomorphism invariance.\footnote{For instance, if the 6d geometry is a product $M\times N$ and the polarization respects the diffeomorphism invariance on $N$, the corresponding symmetry may be regarded as a ``subsystem symmetry'' in 6d.} But, given a refined polarization $\CP$, there can be a subgroup $\mathrm{Stab}_{\CP}$ of $\MCG(M_d)$ that acts trivially.  It is potentially a symmetry of $T[M_d,\CP]$. $\MCG(M_d)$ can be viewed as the group of components of $\mathrm{Diff}(M_d)$ which also includes infinitesimal diffeomorphism of $M_d$. They also have an interesting interplay with the symmetries of $T[M_d]$ that come from reduction of the 2-form symmetry in 6d. We will analyze this problem next.

\subsection{Higher-group symmetry from isometries of $M_d$}

Besides a choice of polarization $\CP$, the theory $T[M_d,\CP]$ can in principle also depend on choices of additional structures on $M_d$, such as a metric, spin structure, fluxes and holonomies for background fields, etc. For simplicity, we shall refer to the subgroup of $\mathrm{Diff}(M_d)$ that preserves all such structures as the group of ``isometries'' $\mathrm{Iso}(M_d,\CP)$ with the understanding that it takes into account not only the metric, but also other structures used in the construction of $T[M_d,\CP]$. Defined in this way, $\mathrm{Iso}(M_d,\CP)$ will be a symmetry of the theory $T[M_d,\CP]$. One can also consider the group of isometries $\mathrm{Iso}(M_d)$ without specifying a polarization. And this group acts on the relative theory $T[M_d]$ as 0-form symmetry. Notice that the identity component $\mathrm{Iso}_0(M_d)$ consists of continuous isometries, and they preserve any $\CP$. We will focus on this part below.\footnote{
There are also interesting consequences of the discrete isometries, which will be further discussed in \cite{Gukov:2020inprepare}.}

To study the interplay between this symmetry and the other ones, we can turn on a non-trivial background field $A$ for the principal $I=\mathrm{Iso}_0(M_d)$-bundle. This changes the gauge transformations for other symmetries. Indeed, consider symmetries of various degree in $T[M_d]$ that are reductions of the 2-form symmetry in 6d along cycles $\gamma_n^{i_n}\in H_n(M_d,D)$. The background fields $B^{i_n}_{3-n}$ are related to $C$ via
\begin{equation}\label{CReduction}
    C=B_3+\sum_{i_1}\omega^{i_1}_1\wedge B^{i_1}_2+\sum_{i_2}\omega^{i_2}_2\wedge B^{i_1}_1+\sum_{i_3}\omega^{i_3}_3\wedge B^{i_1}_0.
\end{equation}
Here $\omega^{i_n}_n$ is a basis of $H^n(M_d,D)$ dual to the basis of cycles in $H_n(M_d)$,
\begin{equation}
    \omega^{i}_n(\gamma_n^{j})=\delta_{ij}.
\end{equation}
Notice that \eqref{CReduction} is only valid when $A=0$. When a background $A$ is turned on for isometry of the internal manifold $M_d$, the $\omega_n^{i_n}$ are modified to an $I$-equivariant form $\tilde\omega_n^{i_n}(A)$ via minimal coupling\footnote{In the Cartan model for equivariant cohomology, this map is obtained by applying the operator 
\begin{equation}
    P=\prod_a(1-A_a\iota_a)
\end{equation}
with $\iota_a$ is the contraction with the Killing vector (see {\it e.g.} \cite{Figueroa-OFarrill:1994uwr}) .}
\begin{equation}
    \Omega^*(M_d)\rightarrow \Omega^*_I(M_d).
\end{equation}
This form is gauge invariant but in general not closed.\footnote{Such behavior of $\omega(A)$ is an obstruction for gauging the WZW term given by $\omega$, as explained in \cite{Hull:1990ms}. However, in our setting, it is precisely this obstruction that leads to interesting higher-group structure.} As a consequence, demanding $dC$ to be gauge invariant requires modification of the gauge transformations for different $B$'s. In fact, there is always a canonical modification that makes $dC$ gauge invariant. To see this, one first turns on $A$ in \eqref{CReduction} and then applies the differential rearranging the result to look like
\begin{multline}
    dC=(dB_3+f_3(B_2,B_1,B_0,A))+\sum_{i_1} (dB^{i_1}_2+f_{2}^{i_1}(B_1,B_0,A))\wedge\tilde\omega^{i_1}_1\\+\sum_{i_2} (dB^{i_1}_1+f_{1}^{i_1}(B_0,A))\wedge\tilde\omega^{i_2}_2+\sum_{i_3}dB^{i_1}_0\wedge\tilde\omega^{i_3}_3 ,
\end{multline}
for certain functions $f_{1,2,3}$ that come from the non-vanishing $d\tilde\omega$'s. Then, one can recursively solve for gauge transformations of $B_1, B_2$ and $B_3$ by demanding that $dB_n+f_n$ is gauge invariant, while the gauge transformation of $B_0$'s remains unchanged.
After this procedure, $dB_n$ themselves are not gauge invariant, whereas the combinations
\begin{equation}
    H_{n+1}^{i_n}:= dB_n^{i_n}+f_n^{i_n}
\end{equation}
are gauge invariant. 
This is the global symmetry analogue of the Green--Schwarz mechanism \cite{Green:1984sg} and is called $n$-group symmetry \cite{Kapustin:2013uxa,Sharpe:2015mja,Cordova:2018cvg,Benini:2018reh}. Discussions and examples of higher-group symmetry in physics can be found in \cite{Kapustin:2013uxa,Barkeshli:2017rzd,Cordova:2018cvg,Benini:2018reh,Hsin:2019fhf,Hsin:2020nts,Cordova:2020tij}. 

To generalize to discrete symmetries, instead of demanding the gauge invariance of $dC$, we can demand the gauge invariance of its holonomy $e^{i\oint C}$. Since the $I$-equivariant form $\tilde\omega_n^{i_n}(A)$ does not have integral period in contrast to $\omega_n^{i_n}$, the usual gauge transformations of $B_n$ would change the holonomy $e^{i\oint C}$. Demanding the gauge invariance of $e^{i\oint C}$ then modifies the gauge transformations in the same way as discussed before and we again conclude the compactified theory has higher-group symmetry. We will illustrate both arguments in an example below in Section \ref{sec:5d3-groupreduction}.

\subsubsection{Example: $M_d=S^1$, 5d theory with 3-group symmetry }\label{sec:5d3-groupreduction}

Parametrizing the circle with coordinate $\varphi\in[0,2\pi)$, the $S^1$ isometry acts by shifting $\varphi$.\footnote{
It is also the Kaluza-Klein momentum symmetry: for the momentum-$n$ Kaluza-Klein mode in any expansion
\begin{equation}
f_n(x^\mu) e^{in\varphi}~,
\end{equation}
the $U(1)$ shift symmetry given by the diffeomorphism $\varphi\rightarrow\varphi +\lambda(x^\mu)$ transforms the momentum-$n$ mode as
\begin{equation}
f_n(x^\mu)\rightarrow f_n(x^\mu) e^{in\lambda}~.
\end{equation}
The gauge field for the isometry can be identified with the graviphoton gauge field.
} Then,
\begin{equation}
    C=B_3+{d\varphi\over 2\pi} \wedge B_2
\end{equation}
after turning on the background $A$ for the $S^1$ isometry becomes
\begin{equation}
    C=B_3+\tilde\omega_1 \wedge B_2,\quad 
    \tilde\omega_1:={d\varphi-A\over 2\pi}~.
\end{equation}

Then gauge invariance of
\begin{equation}
    dC=(dB_3-\frac{1}{2\pi} dA\wedge B_2)+dB_2\wedge \tilde\omega_1
\end{equation}
requires that, if $A$ and $B_2$ are transformed as
\begin{equation}
    A\mapsto A+d\lambda_0,\quad B_2\mapsto B_2+d\lambda_1,
\end{equation}
we should have 
\begin{equation}\label{eqn:red3grouptransformB3}
    B_3\mapsto B_3+ \frac{1}{2\pi} dA\wedge \lambda_1.
\end{equation}
Therefore, $dB_3$ is not gauge invariant, while the combination
\begin{equation}
    H_4:=dB_3-\frac{1}{2\pi} dA\wedge B_2
\end{equation}
is invariant.

There are several other ways to derive the three-group symmetry.
Instead of demanding the invariance of the field strength of $C$, we can also demand the invariance of the holonomy $e^{i\oint C}$, which can generalize the argument to discrete gauge fields. We can also expand the three-form gauge field $C$ in different ways:
\begin{enumerate}
    \item[(1)] Expand the three-form gauge field $C$ as before
    \begin{equation}
        C=B_3+\tilde \omega_1 B_2,\quad \tilde\omega_1=\frac{d\varphi-A}{2\pi}~.
    \end{equation}
 The holonomy of $C$ is
 \begin{equation}
         \oint C=\oint B_3+\oint \tilde \omega_1 B_2\quad \text{mod }2\pi\mathbb{Z}~.
 \end{equation}
    Under the transformation
\begin{equation}\label{eqn:redgaugetransf}
    A\rightarrow A+d\lambda_0,\quad B_2\rightarrow B_2+d\lambda_1~
\end{equation}
which also transforms $\varphi\rightarrow \varphi+\lambda_0$ and leaves $\tilde\omega_1=(d\varphi-A)/(2\pi)$ invariant, the holonomy $\oint\tilde \omega_1 B_2$ changes by
\begin{equation}
\oint \tilde\omega_1 d\lambda_1=-{1\over 2\pi}\oint A d\lambda_1=-{1\over 2\pi}\oint dA\lambda_1 \quad \text{mod }2\pi\mathbb{Z}~.
\end{equation}
Thus the gauge invariance of the holonomy  $e^{i\oint C}$ requires that $B_3$ must transform as
\begin{equation}
B_3\rightarrow B_3+{1\over 2\pi} dA \lambda_1~.
\end{equation}
This reproduces the transformation (\ref{eqn:red3grouptransformB3}).
    
    \item[(2)]
Instead of modifying $d\varphi$ to be $d\varphi-A$ in the expansion of the three-form gauge field $C$, we can also use the original expansion without including the isometry gauge field $A$:
\begin{equation}
    C = B_3+\frac{d\varphi}{2\pi}B_2~.
\end{equation}
The holonomy of $C$ is
\begin{equation}
    \oint C=\oint B_3+\oint \frac{d\varphi}{2\pi}B_2\quad\text{mod }2\pi\mathbb{Z}~.
\end{equation}
Then under the gauge transformation (\ref{eqn:redgaugetransf}), the holonomy $\oint \frac{d\varphi }{2\pi}B_2$ changes by
\begin{equation}
\oint\left( {d\lambda_0\over 2\pi}B_2+ \frac{d\varphi }{2\pi}d\lambda_1+{d\lambda_0\over 2\pi}d\lambda_1\right)=\oint \frac{d\lambda_0}{2\pi}B_2\quad\text{mod }2\pi\mathbb{Z}~.
\end{equation}
Thus, the gauge invariance of the holonomy $e^{i\oint C}$ requires that $B_3$ must transform as
\begin{equation}
B_3\rightarrow B_3-{1\over 2\pi} d\lambda_0 B_2~.
\end{equation}
After the redefinition $B_3\rightarrow B_3+\frac{1}{2\pi}AB_2$, this gives the same three-group symmetry as the previous case.
\end{enumerate}

We can investigate the anomaly of the higher-group symmetry by reducing the 7d TQFT.
For instance, consider reducing the three-form Chern--Simons theory on a circle 
\begin{equation}
    \frac{N}{4\pi}\int_{7\d} CdC~.
\end{equation}
Using the decomposition
\begin{equation}
    C=B_3+\left(\frac{d\varphi-A}{2\pi}\right)B_2~.
\end{equation}
We find the $6\d$ bulk theory that describes the anomaly of the three-group symmetry in $5\d$
\begin{equation}
\frac{N}{4\pi}\int_{7\d} CdC
=
\frac{N}{2\pi}\int_{6\d}B_3dB_2-\frac{N}{4\pi}\int_{6\d} B_2^2\frac{dA}{2\pi}~.    
\end{equation}

Later in Section \ref{sec:5d3-groupfield} we will provide a field theory explanation of the three-group symmetry and its anomaly in the 5d gauge theory.

When $M_d$ is $T^2$ or $T^3$, which we shall encounter in later parts, the analysis is very similar. A more interesting example is $M_d=S^2$, for which the isometry group is non-abelian and there is an interesting discrete higher-group that requires extending the analysis above using the Borel construction for equivariant cohomology. This example and its physics will be discussed in detail in the companion paper \cite{Gukov:2020inprepare}.

\subsection{Symmetry fractionalization in compactification}

The extended operators can carry anomalous quantum numbers under a global symmetry. This phenomenon is familiar from the fractional quantum hall effect, where line operators that describe anyons carry projective representations of the $U(1)$ 0-form symmetry. For general $(2+1)$-dimensional TQFT this is discussed in \cite{EtingofNOM2009,Barkeshli:2014cna}. Different symmetry fractionalizations represent different ways of coupling the theory to the background gauge field and can be classified by higher-form (higher-group) symmetries \cite{Benini:2018reh,Hsin:2019fhf,Hsin:2019gvb}. 

Different symmetry fractionlizations in compactified theories can be described by turning on different backgrounds for the two-form symmetry in 6d, which reduce to various background fields for higher-form symmetries in lower dimension. More generally, we can first consider the 7d Chern--Simons theory enriched by the symmetry {\it i.e.} coupled to the background gauge field for the symmetry and then study the 6d theory. In some cases different backgrounds --- {\it i.e.} different symmetry fractionalizations --- correspond to different choices of polarization.

We will see examples of symmetry fractionalization for Lorentz symmetry \cite{Wang:2016cto,Hsin:2019fhf,Hsin:2019gvb,Ang:2019txy} in the compactified theory that come from choices of polarization.
These different symmetry fractionlizations correspond to gauge theory with different spins for line operators, namely different consistent possibilities for line operators to be either bosons or fermions.
These choices can be distinguished on a non-spin manifold; otherwise, on a spin manifold, there is a gravitational fermion line operator that can fuse with line operators in the gauge theory to change their spins. This requires defining the 6d theory on a non-spin manifold. As discussed in Appendix \ref{sec:7dCSlevel}, we can define the theory on more general spin$^c$ manifolds (theories with Dirac fermions can always be defined on a spin$^c$ manifold by coupling to background spin$^c$ connection; examples can be found in \cite{Seiberg:2016rsg,Seiberg:2016gmd,Hsin:2016blu}). Thus, we can study these different Lorentz symmetry fractionalizations from compactification.

\subsection{Reduction on free, torsion and discrete cycles}
\label{sec:torsiondiscretefreecycle}

Suppose the internal manifold $M_d$ has torsion cycles and discrete cycles, where a discrete $\mathbb{Z}_n$ $r$-cycle has boundary that consists of $n$ copies of a $(r-1)$-cycle (which is then an $n$-torsion cycle, since $n$ copies of it equals a boundary), and thus  the boundary of the $\mathbb{Z}_n$ discrete cycles is trivial with the $\mathbb{Z}_n$ coefficient. 
Their Poincar\'e duals are
\begin{equation}
n^{(2)}\tau_2=d \hat\tau_1,\quad
n^{(3)}\tau_3=d \hat\tau_2,\quad
n^{(4)}\tau_4=d \hat\tau_3,\quad
n^{(5)}\tau_5=d \hat\tau_4~.
\end{equation}
Namely, the Poincar\'e dual $\text{PD}(\hat\tau_i)$ is a  $(d-i)$-cycle with $\mathbb{Z}_{n^{(i+1)}}$ coefficients (it is an open $\mathbb{Z}$ chain which is closed only mod $n^{(i+1)}$), while $\text{PD}(\tau_i)$ is a $n^{(i)}$-torsion $(d-i)$-cycle. In general, they carry index labelling different cycles, so that $n$'s are integer matrices. We have the following isomorphisms
\begin{equation}
\text{Tor }H_i(M_d)\cong \text{Tor }H^{i+1}(M_d)
\cong \text{Tor }H_{d-i-1}(M_d)\cong \text{Tor }H^{d-i}(M_d)~.
\end{equation}
Thus, $n^{(i+1)}=n^{(d-i-1)}$. The matrix $n$ can be symmetric or anti-symmetric depending on the dimension of the corresponding cycles.

The integer forms $\tau$ and $\hat\tau$ obey the orthonormality condition
\begin{equation}
    \int_{M_d} \tau_i\wedge \hat \tau_{d-i}=\mathbf{1},
\end{equation}
where $\mathbf{1}$ is the identity matrix with indices labelling different pairs of cycles that appear on the left-hand side of this relation.
We note that, for a free cycle $\alpha$,
\begin{equation}
    \int_{M_d}\tau_i\wedge \alpha_{d-i}=0=\int_{M_d}\tau_i\wedge \tau'_{d-i}~.
\end{equation}
Similarly, since $\hat\tau$ is defined up to a closed integral cycle, we can take
\begin{equation}
    \int_{M_d}\alpha_i\wedge \hat  \tau_{d-i}=0=\int_{M_d}\hat \tau_i\wedge \hat \tau'_{d-i}~.
\end{equation}

Using the orthonormality condition we can decompose $C_3^I$ as follows:
\begin{equation}
C_3=\sum \alpha_i B^F_{3-i} + \tau_i \hat B_{3-1}+\hat \tau_i B_{3-i}~, 
\end{equation}
where
\begin{equation}
B^F_{3-i}=\int_{\text{PD}(\alpha_i)} C_3,\quad
\hat B_{3-i}=\int_{\text{PD}(\hat\tau_i)} C_3,\quad
 B_{3-i}=\int_{\text{PD}(\tau_i)} C_3~.
\end{equation}
The superscript $F$ denotes the free part.

Next, we use the Dirac quantization of the three-form gauge field, $\int dC_3\in 2\pi\mathbb{Z}$, to impose constraint on $B_i^F,\hat B_i,B_i$, where $dC_3$ is expressed as
\begin{equation}
dC_3=\sum (-1)^i\alpha_i d B^F_{3-i} + \tau_i \left((-1)^i d\hat B_{3-i}+n^{(i)}B_{4-i}\right)+(-1)^i\hat \tau_i d B_{3-i}~.
\end{equation}
Thus, for properly quantized $\int dB^F_i,\int d\hat B_i,\int dB_i\in 2\pi\mathbb{Z}$, the gauge fields $B_i$ that come from reduction of $C_3$ on torsion cycles $\tau_{3-i}$ must take discrete values with holonomies valued in $2\pi{\cal A}=\mathbb{Z}^r/n^{(4-i)}\mathbb{Z}$, where $r=\text{dim}(n^{(4-i)})$.
We note this is an off-shell condition independent of the dynamics, and it follows directly from the Dirac quantization condition of $C_3$.

Moreover, since $dC_3$ is gauge invariant, the gauge transformation of $B_i$ also transforms $\hat B_{i-1}$:
\begin{equation}
B_{i}\rightarrow B_{i}+d\lambda_{i-1},\quad
\hat B_{i-1}\rightarrow \hat B_{i-1}-(-1)^{i}n^{(4-i)}\lambda_{i-1}~,
\end{equation}
and thus one can think of $\hat B_i$ as the analogue of Stueckelberg field.

Let us compare this discussion with \cite{Camara:2011jg,BerasaluceGonzalez:2012vb}. The difference is that in \cite{Camara:2011jg,BerasaluceGonzalez:2012vb} the discrete holonomy condition of $B_i$ is derived by taking the low energy limit of the kinetic term coupling for $C_3$ that contains $\frac{1}{g^2} \big|((-1)^id\hat B_{3-i}+n^{(i)}B_{3-i})\big|^2$, by sending the coupling $g^2\rightarrow 0$. Here we emphasize that the condition of discrete holonomy follows directly from the Dirac quantization of $C_3$, and the gauge transformation of $\hat B_i$ follows from the gauge invariance of $dC_3$. Unlike the discussion in \cite{Camara:2011jg,BerasaluceGonzalez:2012vb}, this does not require any information about the dynamics, such as the value of the coupling constants.

\section{$\Pol(\mathrm{pt})$ and 6d
topological boundary conditions}\label{sec:compactification6d}

In Section \ref{sec:compactification0d}, we considered the extreme case of $d=6$ and $M_{6-d}$ being zero-dimensional. In this subsection, we will consider another extreme case of $d=0$ and $M_d=\mathrm{pt}$, a point. In this case, the elements of $\Pol(\mathrm{pt})$ corresponds to topological boundary conditions of the 7d TQFT. 

\subsection{Pure and mixed polarizations}

The pure polarizations are classified by maximal subgroups $L$ of $H^*(\mathrm{pt},D)\simeq D$ that trivialize the pairing on $D$. Here, being maximal means that 
\begin{equation}
    |L|=|L^\vee|
\end{equation}
where $L^\vee=D/L$. Then, the non-degeneracy of the pairing on $D$ implies that $L^\vee=\mathrm{Hom}(L,U(1))$ is the Pontryagin dual of $L$. 

In the context of 6d $(2,0)$ theories labeled by Lie algebras $\frak{g}$, searching for such $L$ is equivalent to searching for a self-dual sub-lattice of the weight lattice ${\frak g}$. Every such sub-lattice corresponds to some group $G=\tilde G/L$ that acts faithfully on any representation labeled by a non-zero dominant weight of this lattice,  where $\tilde G$ is the simply connected Lie group with the Lie algebra ${\frak g}$. Such 6d theory is often said to admit a standard field theory description with a two-form symmetry $Z(\tilde G/L)$ \cite{Gaiotto:2014kfa}.

Given any $\CP\in \Pol(\mathrm{pt})$ and any 6-manifold $M_6$, we expect to get a maximal isotropic subgroup $\L\subset H^3(M_6,D)$. In the present case, it is constructed as follows. For a subgroup $L\subset D$ and any $M_6$, we have a long exact sequence of cohomology groups:
\begin{equation}
    \ldots\xrightarrow[]{} H^2(M_6,L^\vee)\xrightarrow[]{\beta_2} H^3(M_6,L)\xrightarrow[]{\iota_*} H^3(M_6,D) \xrightarrow[]{\pi_*} H^3(M_6,L^\vee)\xrightarrow[]{\beta_3} H^3(M_6,L)\rightarrow \ldots
\end{equation}
Here $\iota_*$ and $\pi_*$ are induced from the inclusion $\iota:\; L\rightarrow D$ and the quotient map $\pi:\; D\rightarrow L^\vee$, respectively, while $\beta_2$ and $\beta_3$ are Bockstein homomorphisms.
Denote by $\L=\iota_*(H^3(M_6,L))\subset  H^3(M_6,D)$ the image of $\iota_*$. Then, it is easy to see that it is isotropic. It is, in fact, also coisotropic (and hence maximal), as 
\begin{equation}
    \L^\vee := H^3(M_6,D)/\L
\end{equation}
can be identified with the image of $\pi_*$ in $H^3(M_6,L^\vee)$ by exactness.

We should emphasize that, in general, $\beta_2$ is non-trivial, $\iota_*$ is not injective, and $H^3(M_6,L)$ is not a subgroup of $H^3(M_6,D)$. Nonetheless, the image $\L$ is always maximal isotropic. To better illustrate this point, let us consider the example of $M_6=\mathbb{R}\mathbb{P}^3\times \mathbb{R}\mathbb{P}^3$ and $D=\mathbb{Z}_4$ (relevant {\it e.g.} to~$(2,0)$ theories labeled by ${\frak g}=\mathfrak{so}(4n+2)$).
We have
\begin{equation}
H^3(M_6,\mathbb{Z}_4)=\bigoplus_{i=0}^3 H^i(\mathbb{R}\mathbb{P}^3,H^{3-i}(\mathbb{R}\mathbb{P}^3,\mathbb{Z}_4))= \mathbb{Z}_4\times \mathbb{Z}_2\times \mathbb{Z}_2\times \mathbb{Z}_4~,
\end{equation}
which has a Lagrangian subgroup
\begin{equation}
\Lambda=\iota(H^3(M_6,\mathbb{Z}_2))=
\iota
\left(\mathbb{Z}_2\times \mathbb{Z}_2\times \mathbb{Z}_2\times \mathbb{Z}_2\right)=\mathbb{Z}_2\times \mathbb{Z}_2\times\bf{0}\times \mathbb{Z}_2~,
\end{equation}
where $\iota:\mathbb{Z}_2\rightarrow\mathbb{Z}_4$ is the inclusion, ${\bf0}=\Z_1$ is the trivial group, and $H^1(\mathbb{R}\mathbb{P}^3,M)=\text{2-Tor}(M)$, $H^2(\mathbb{R}\mathbb{P}^3,M)=M/2M$ for $M=\mathbb{Z}_m$.
Note, this is different from (4.14) in \cite{Eckhard:2019jgg}, which states that the polarization for the standard field theory is $\Lambda=H^3(M_6,L)=\mathbb{Z}_2^4$ but does not have the required order to be Lagrangian for $H^3(M_6,\CZ(\tilde G))=\mathbb{Z}_2^2\times \mathbb{Z}_4^2$ (in this case $\tilde G=\mathrm{Spin}(4n+2)$, $\CZ(\tilde G)=\mathbb{Z}_4$ and $L=\mathbb{Z}_2$).

\subsubsection*{More general background fields}

Another caveat is that background fields for the the remaining $L^\vee$ symmetry are now labeled by elements in $H^3(M_6,L^\vee)$ and not the subgroup $\L^\vee$. Thisis not a problem for splittable polarizations, for which $D=L\oplus \bar{L}$ with some $\bar{L}$ and the Bockstein homomorphisms are all trivial. For a polarization 
$L'$ that is not necessarily splitable, it leads to the following partition function on a 6-manifold,\footnote{Here, for simplicity, we suppressed the dependence on the choice of quadratic refinement, which enters through $\varphi$ and an additional shift of $B_\alpha$ by $B_\varphi$ (cf.~\eqref{dPartition}), and can be easily restored.}
\begin{equation}
    Z^{L'}_{B'}(M_6)=\sum_{\alpha \in \L'} e^{2\pi i [\varphi(\alpha)+\langle B_\alpha,\tilde{B'}\rangle]}\cdot Z^{L}_{B_\alpha+B_*}
\end{equation}
where $\L'$ is again the image of $H^3(M_6,L')$ in $H^3(M_6, D)$, $B'\in \L'^\vee$ with a chosen lift $\tilde{B'}$ in $H^3(M_6, D)$, and $B_*$ is the projection of $\tilde{B'}$ to $H^3(M_6,\bar{L})$. Two choices of the lift of $B'$ can differ by an overall phase in the partition function which cannot be canceled by changing the normalization of partition functions (or, at the level of the physical theory, by adding local counter terms). This means that the remaining symmetry $\L'^\vee$ can have an 't Hooft anomaly. 

To incorporate more general backgrounds for non-splittable polarizations  at the level of partition functions, one needs to generalize the equation above to $B'\in H^3(M_6,L'^\vee)$ that is not in $\L'^\vee$. The right procedure is to modify $B_*$ by including a magnetic defect in the theories given by the Poincar\'e dual of $\beta_3(B')$.\footnote{
We remark that a similar consideration applies in general, when the trivialization is by a dynamical field such as a ``dynamical spin structure'' ({\it i.e.}~a dynamical $\mathbb{Z}_2$ gauge field $a$ that satisfies $\delta a=w_2$). Since a Wilson line for a dynamical field (or, a Wilson $n$-surface for a field of degree $n$) has non-trivial correlation functions (by ``remote-detectablility'' an non-trivial operator must have a non-trivial correlation function), the trivialization is not well-defined everywhere. For instance, moving a magnetic observable through $\oint_{\gamma} a$ picks up a sign and, therefore, the spin structure, which only depends on $\gamma$, is not well-defined. On a manifold with non-trivial $w_2$ there are insertions of magnetic objects that link with the cycles where the spin structure is ill-defined.
}

\subsubsection*{Non-existence of mixed polarizations in 6d}

We end this section with a comment about mixed polarizations. It turns out that, although there can be many of them when $d>0$, they can not exist for $d=0$ in $\Pol(\mathrm{pt})$. One can see this in the following way.

Assume $\CP$ is a mixed polarization. Then, it determines a family of maximal isotropic subgroups $\L_{M_6}$ for each $M_6$, and the union of the image of
\begin{equation}
   H_3(M_6,\Z)\times \L_{M_6}\rightarrow D, 
\end{equation}
denoted as $\CS(\CP)$ is not isotropic. 
However, this is not possible as $\CS(\CP)$ labels charges of strings in the theory $T[\mathrm{pt},\CP]$ and all such objects are independent. In other words, $\CS(\CP)=\CS(\CP)_{\text{ind}}$ have to be isotropic. For mixed polarizations to exist, $\CS(\CP)$ has to have at least two pieces in different cohomological degree, which is simply not possible for $d=0$. Therefore, in the remainer of this section, we will focus on pure polarizations.

\subsection{Classification}\label{sec:ClassificationL}

Analysis in the previous subsection  allows us to classify pure polarizations in $\Pol(\mathrm{pt})$ when $D$ is simple. 

Recall that any finite abelian group is a product of cyclic groups with order being a prime power, $\Z_{p^k}$, and any subgroup is given by a product of subgroups of each $\Z_{p^k}$. Furthermore, any bilinear pairing on $D$ vanishes between different $p$-groups. Therefore, we can concentrate on the case where $D$ is a $p$-group ({\it i.e.}~$D$ is a product of $\Z_{p^k}$ with fixed $p$).
It splits into be a sum of homogeneous parts, each being a sum of $\Z_{p^k}$ with fixed $k$, and a Lemma of Wall \cite{WALL1963281} ensures that the pairing also splits. However, for classification of subgroups $L$, we cannot assume that $D$ is  homogeneous, as there could  still be interesting maximal isotropic subgroups $L$ of $D$ that does not respect the decomposition into homogeneous pieces. Examples include $D=\Z_3\times \Z_{27}$. For any non-degenerate pairing on this $D$, one can always find an $L=\Z_9$ isotropic subgroup which is generated by a non-homogeneous element of $D$. This illustrate that the classification of $L$'s is in general more complicated compared to the classification of non-degenerate bilinear forms. Because of this, we focus on those cases that are most relevant to 6d $(2,0)$ theories.

We start with the case when $D$ is a cyclic $p$-group.

\subsubsection*{$D=\Z_{p^k}$ with a pairing given by $\kappa\in \Z_{p^k}^{\times}$.}

There are only two isomorphism classes of $\kappa$, that correspond to $\kappa$ being either a square or a non-square. In both cases, the sought-after maximal isotropic subgroup $L$ exists only when $k$ is even, in which case $L=\Z_{p^{k/2}}$ is generated by $p^{k/2}\in\Z$.

This generalizes to  $D=\Z_N$. In this case it is easy to see that such $L$ exists if and only if $N=n^2$ is a perfect square, in which case $L=\Z_n$ is generated by $n\in \Z_N$. An example of this type is the ``$SU(n^2)/\Z_n$ theory'' in 6d. The corresponding topological boundary condition will be discussed in the next subsection.

Another remark is that the quotient $L^\vee=\Z_n$  does not have a canonical lift to $D$. This translates into the fact that the corresponding topological boundary condition and the 6d theory have an 't Hooft anomaly for their $L^\vee$ symmetry.

\subsubsection*{$D=\Z_{p^k} \times \Z_{p^k}$ with split pairing.}

When $D=\Z_{p^k}^m$ for some $m\in \Z_+$, the pairing can always be made to split among the $m$ copies if $p$ is an odd prime. This is not always possible for $p=2$, which will be discussed separately. 

After the bilinear form is made diagonal, it can be characterized by a pair $(\kappa,\kappa')$ of square or non-square numbers in $\Z_{p^k}^{\times}$. However, having both of these numbers non-square is equivalent to having both square \cite{WALL1963281}. So, there are really only two choices. 

When $\kappa=\kappa'=1$, any element $(a,b)$ in $L$ needs to satisfy $a^2+b^2=0 \pmod{p^k}$. Assume $p^\ell | a$ and $p^\ell|b$ for an integer $\ell$. If $\ell=1$, then $L$ contains an element of the form $(1,b_{k})$ with $b_{k}^2\equiv -1\pmod{p^k}$. Therefore, such subgroup only exists if $-1$ is a quadratic residue mod $p^k$. On the other hand, if any elements in $L$ has $\ell=k/2$, then $L$ is $\Z_{p^{k/2}}\times \Z_{p^{k/2}}\subset \Z_{p^k}\times \Z_{p^k}$, generated by $(p^{k/2},0)$ and $(0,p^{k/2})$.

Furthermore, if $-1$ is a quadratic residue mod $p^{k-2\ell}$, for $1<\ell<k/2$, then there exists an $L=\Z_{k-\ell}\times \Z_{\ell}$ generated by $(p^\ell,p^\ell b_{k-2\ell})$ and $(0,p^{k-\ell})$, where $ b_{k-2\ell}$ is a square root of $-1$ mod $p^{k-2\ell}$. 

An interesting fact is that the square roots of $-1$ mod $p^m$ for any $m\in Z_+$ and $p$ odd are in bijection with square roots of $-1$ mod $p$, which exist if and only if $p\equiv 1 \pmod 4$. And when $p\equiv 1 \pmod 4$, there are two choices. Summing them up (and adding the extra one for $\ell=k/2$ when $k$ is even) leads to $k+1$ choices for $L$. 

On the other hand, if $p=2$, then only for $m=1$ one can have $-1$ as a quadratic residue. In that case, there is only one choice of $L$ in $\Z_{2^k}\times\Z_{2^k}$, generated by either $(2^n ,0)$ and $(0,2^n)$ when $k=2n$, or $(2^n,2^n)$ and $(0,2^{n+1})$ when $k=2n+1$.

Therefore, the total number of the desired subgroups $L$ is given by 
\begin{equation}
    F(p^k,p^k)=\left\{\begin{array}{ll}1, &\quad p=2\\
    k+1, &\quad p\equiv 1 \pmod 4\\ k+1 \bmod 2, &\quad p\equiv 3 \pmod 4.
    \end{array}\right.
\end{equation}

The case when one of $\kappa$ or $\kappa'$ is non-square can be analyzed in a similar way, with the role played by the residue-1 and residue-3 primes switched. This is because the condition for existence of $L$ is now linked with the condition of $-\kappa'$ being a quadratic residue, and as $\kappa'$ is a non-square, this implies that $-1$ is also a non-square. This is only true mod $p^m$ with $m\in\Z_+$ when $p\equiv 3 \pmod 4$.

\subsubsection*{Generalizing to $D=\Z_N \times \Z_N$}

Again, we first assume that the pairing is given by $\kappa=\kappa'=1$ as other choices will be completely analogous. 

To find the desired $L$, one only needs to apply the result to $\Z_{p^k}\times\Z_{p^k}$ for all prime factors of $N$. Assuming $N=\prod_{i}p_i^{r_i}$, the total number of choices of $L$ is given by
\begin{equation}
    F(N,N)=\prod_{i=1}^{r}F(p_i^{r_i},p_i^{r_i}).
\end{equation}
This is non-zero only if $r_i$ for residue-3 primes are all even. Then,
\begin{equation}
    F(N,N)=\sum_{p_i\equiv1\pmod4} r_i+1.
\end{equation}
One might also be interested in the number of choices modulo outer-automorphisms of $\Z_N\times\Z_N$ by $(\Z_2\times \Z_2)\rtimes \Z_2$, generated by multiplying with $-1$ each factor and switching the two factors. This counting is given by
\begin{equation}
    G(N,N)=\sum_{p_i\equiv1\pmod4} \left\lfloor\frac{r_i+1}{2}\right\rfloor.
\end{equation}
The table below summarizes the number of choices of $L$ in $\Z_N\times \Z_N$ for small $N$, where we skipped the entries with $F(N,N)=0$.

\begin{center}
\begin{tabular}{c|c|c|c|c|c|c|c|c|c|c|c|c|c|c|c|c|c}
    $N$ & 2 & 4 & 5  & 8 & 9 & 10 & 13 & 16 & 17 & 18 & 20 & 25 & 26 & 29 & 32 & 34 &\ldots\\
    \hline
     $F(N,N)$& 1 & 1 & 2 & 1 & 1 & 2 & 2 & 1 & 2 & 1 & 2 & 3 & 2 & 2 & 1 & 2 &\ldots \\\hline
      $G(N,N)$& 1 & 1 & 1 & 1 & 1 & 1 & 1 & 1 & 1 & 1 & 1 & 2 & 1 & 1 & 1 & 1 &\ldots
\end{tabular}

\end{center}

\subsubsection*{$D=\Z_{p^k}\times \Z_{p^{k'}}$ and $D=\Z_{M}\times \Z_{N}$}

We now consider the ``inhomogeneous'' case of $D=\Z_{p^k}\times \Z_{p^{k'}}$, again assuming without loss of generality $\kappa=\kappa'=1$ and $k'>k$. Let $2s=k'-k$; obviously, $s$ has to be integer for $L$ to exist.

Let $p^\ell$ be the generator of the image of $L$ under the projection to $\Z_{p^k}$. Then, the pre-image is $(p^\ell,b)$ with a $b$ such that $p^{2s+2\ell}+b^2\equiv 0\pmod{p^{k'}}$. So $p^{s+\ell}|b$, and $b_{k-2\ell}:=b/p^{s+\ell}$ is a square root of $-1$ mod $p^{k-2\ell}$. 
Therefore, the story is completely similar to the case of $k'=k$, and the number of $L$ is given by
\begin{equation}\label{Ffunctionkk'}
    F(p^k,p^{k'})=\left\{\begin{array}{ll}
    F(p^k,p^k), &\text{if $k\le k'$ and $k'-k$ is even}\\
    F(p^{k'},p^{k'}),& \text{if $k\ge k'$ and $k'-k$ is even}\\
    0,& \text{if $k'-k$ is odd}.\end{array}\right.
\end{equation}
As for the function $G$, although the $\Z_2$ automorphism of the swapping the two factors is lost, one still has the exact same relation as in \eqref{Ffunctionkk'} between $G(p^k,p^{k'})$ and $G(p^k,p^k).$ 

This immedietly generalizes to the case of $D=\Z_M\times \Z_N$. If $N=\prod_{i}p_i^{r_i}$ and $M=\prod_{i}p_i^{r'_i}$, then the number of choices for the desired subgroup $L$ is given by
\begin{equation}\label{FfunctionMN}
    F(M,N)=\prod_{p_i}F(p^{r'_i},p^{r_i}).
\end{equation}

\subsubsection*{$D=\Z_2\times\Z_2\times\Z_2\times \Z_2$ with non-split pairing}

One difference between $p=2$ and odd $p$ is that the bilinear form sometimes cannot be made diagonal along cyclic subgroups. A minimal example is $D=\Z_2\times\Z_2$ with pairing given by multiplication mod 2. In this case, there are three $\Z_2$ subgroups and they all are maximal isotropic. 

The next case relevant for us is $D=\Z_2\times\Z_2\times\Z_2\times \Z_2$. We have actually already encountered the problem of finding maximal isotropic subgroups of it in Section~\ref{sec:MCG6d}. There, we found that there are 15 choices, 9 of which are generated by $(a,0)$ and $(0,b)$ where $a$ and $b$ can each be $(10)$, $(01)$ or $(11)$, while the remaining 6 are generated by $(01,b)$ and $(10,b')$ with $b\neq b'$.

\subsubsection*{Classification of absolute 6d $(2,0)$ theories}

The discussion above allows us to identify absolute 6d $(2,0)$ theories. We will assume the gauge algebra $\frak{g}$ of the theory does not contain abelian factors, and is decomposed into irreducible pieces 
\begin{equation}
    \frak{g}=\frak{g}_1\oplus \frak{g}_2\oplus\ldots\oplus \frak{g}_r,
\end{equation}
where each summand is of type ADE.

We first consider the case of $r=1$. There are four infinite families parametrized by $n\ge 2$ and two special ones. 
\begin{center}
\begin{tabular}{c|c|c|c|c|c}

    $\frak{g}$ & $D$ & $L$ & \# of choices of $L$ & \# mod Aut($\frak{g}$) & Theories \\
     
\hline\hline
$A_{n^2-1}$ & $\Z_{n^2}$ & $\Z_{n}$ & 1 & 1 & $SU(n^2)/\Z_{n}$ \\
\hline
$D_{2n+1}$ & $\Z_{4}$ & $\Z_{2}$ & 1 & 1  & $SO(4n+2)$\\\hline
$D_{4n}$ & $\Z_{2}\times \Z_2$ & $\Z_{2}$ & 3 & 2 & $SO(8n)$, $Ss(8n)$, $Sc(8n)$ \\\hline
$D_{4n-2}$ & $\Z_{2}\times \Z_2$ & $\Z_{2}$ & 1 & 1  & $SO(8n-4)$\\\hline\hline
$D_{4}$ & $\Z_{2}\times \Z_2$ & $\Z_{2}$ & 3 & 1  & $SO(8)$, $Ss(8)$, $Sc(8)$ \\
\hline
$E_{8}$ & 0 & 0 & 1 & 1 & $E_{8}$
\end{tabular}
\end{center}
In the above table, we labeled a theory by a Lie group $G$ when the charge lattice for strings $\L_{\mathrm{string}}$ coincides with the character lattice of $G$, while $Ss$ and $Sc$ (and $SO$ for $D_4$) are used to distinguish between different but isomorphic sub-lattices of the weight lattice of $D_{4n}$. The theory $D_4$ (and more generally $D_{4n}$) will be discussed in greater details in Section~\ref{sec:SO(8)}.

The case of $r=2$ is more interesting. The theories are listed in  Table~\ref{tab:Tabler2}. We assumed that there are no $E_8$ factors, as otherwise it would reduce to the $r=1$ case. 
\begin{table}\centering
\begin{tabular}{c|c|c|c|c|c}

    $\frak{g}$ & $D$ & $L$ & \# of choices of $L$ & \# mod Aut($\frak{g}$) & \# irreducible \\
     \hline\hline
$A_{m-1}\,\oplus A_{n-1}$ & $\Z_{m}\times\Z_n$ & \multicolumn{4}{c}{See analysis in Section~\ref{sec:ClassificationL}}   \\
\hline
$A_{m^2-1} \,\oplus$ $D_{4n-2}$ & $\Z_{m^2}\times \Z_{2}^2$ & $\Z_{m}\times \Z_{2}$ & 1 & 1 & 0\\\hline $A_{m^2-1} \,\oplus$ $D_{2n+1}$ & $\Z_{m^2}\times \Z_{4}$ & $\Z_{m}\times \Z_{2}$ & 1 & 1 & 0\\\hline
$A_{m^2-1}\, \oplus$ $D_{4n}$ & $\Z_{m^2}\times \Z_{2}^2$ & $\Z_m \times \Z_2$  &  3 & 2 & 0\\\hline
$A_{m^2-1}\, \oplus$ $D_{4}$ & $\Z_{m^2}\times \Z_{2}^2$ & $\Z_m \times \Z_2$  &  3 & 1 & 0\\
\hline
$A_{4m^2-1}\,\oplus$ $D_{4n-3}$  & $\Z_{4m^2}\times\Z_{4} $ & $\Z_{4m}$ & 2 & 1 & 1\\
\hline
$A_{3m^2-1}\,\oplus$ $E_{6}$  & $\Z_{3m^2}\times\Z_{3} $ & $\Z_{3m}$ & 2 & 1 & 1 \\
\hline
$A_{2m^2-1}\,\oplus$ $E_{7}$  & $\Z_{2m^2}\times\Z_{2} $ & $\Z_{2m}$ & 1 & 1 & 1 \\\hline
\hline $D_{4m-2}\, \oplus $ $D_{4n-2}$ & $\Z_{2}^4$ & $\Z_{2}\times \Z_2$ & 3 & 2  & 1\\
\hline $D_{4m-2}\,\oplus$ $D_{4}$ & $\Z_{2}^4$ & $\Z_2\times \Z_2$ & 3 & 1 & 0  \\
\hline $D_{4m-2}\,\oplus$ $D_{4n}$ & $\Z_{2}^4$ & $\Z_2\times \Z_2$ & 3 & 2 & 0  \\\hline\hline $D_{2m+1} \,\oplus$ $D_{2n+1}$ & $\Z_{4}\times \Z_{4}$ & $\Z_{2}\times \Z_{2}$ & 1 & 1 & 0\\\hline $D_{2m+1} \,\oplus$ $D_{4n-2}$ & $\Z_{4}\times \Z_{2}^2$ & $\Z_{2}\times \Z_{2}$ & 1 & 1 & 0\\\hline $D_{2m+1} \,\oplus$ $D_{4n}$ & $\Z_{4}\times \Z_{2}^2$ & $\Z_{2}\times \Z_{2}$ & 3 & 2 & 0\\\hline $D_{2m+1} \,\oplus$ $D_{4}$ & $\Z_{4}\times \Z_{2}^2$ & $\Z_{2}\times \Z_{2}$ & 3 & 1 & 0\\\hline $D_{4m-3} \,\oplus$ $D_{4n-1}$ & $\Z_{4}\times \Z_{4}$ & $\Z_{4}$ & 2 & 1 & 1\\
\hline\hline
$D_{4}\, \oplus$ $D_{4}$ &  &  &  & 2 &  1 \\\cline{1-1} \cline{5-6}
$D_{4} \,\oplus$ $D_{4m}$ &  &  &  & 3 & 1\\\cline{1-1} \cline{5-6}
2$D_{4m}$ & $\Z_{2}^4$ & $\Z_{2}\times \Z_{2}$ & 15 & 5  & 2\\\cline{1-1} \cline{5-6}
$D_{4m}\,\oplus$ $D_{4n}$ &  &  &  & 6 & 2 \\\hline\hline
\end{tabular}
\caption{Classification of absolute 6d $(2,0)$ theories labeled by $\frak{g}=\frak{g}_1\oplus\frak{g}_2$ a sum of two pieces. All $m$ and $n$ are integers $\ge 2$ except in $A_{3m^2-1}$, where $m$ can be 1 as well. For the last column, ``irreducible'' means that $L\subset D$ does not respect the decomposition of $D$ into two pieces, which implies that the resulting 6d theory is not a product of two non-interacting theories. Any $\frak{g}$ with an $E_8$ summand  are not included because having an $E_8$ will reduce to the $r=1$ case. For the case of $A_{m+1}\,\oplus$ $A_{n-1}$, the number of choices of $L$ is given by the function $F(m,n)$ in \eqref{FfunctionMN}. For a given $m$ and $n$, it is easy to enumerate all choices of $L$ following the algorithm introduced before, and whenever $F\neq 0$ and $M,N$ being non-square, there is always an irreducible $L$.}\label{tab:Tabler2}
\end{table}

Most of the entries, including $A\oplus A$, $A\oplus D_{4n-1}$, $A\oplus E$ and $D_{4n-1}\oplus D_{4n-1}$, directly follow from our analysis of the $D=\Z_M\times \Z_N$ case with $\kappa=\kappa'=1$. Notice that $E_7$ can be treated as if it were $A_1$,\footnote{However, the reader should not have the impression that they correspond to the same 7d TQFT: they are related by (\ref{eqn:7ddualityADE}).
Indeed, although $E_7$ and $A_1$ both have $D=\Z_2$, the quadratic functions on $D$ are different.} while $E_6$ behaves differently from $A_2$ even though they both have $D=\Z_3$ (but their pairings on $D$ are different). The cases that require a closer look are 
\begin{itemize}
    \item $D_{4n-3}$, for which the pairing on $D=\Z_4$ is given by $\kappa=-1$;
    \item $D_{4n}$, for which $D=\Z_2\times\Z_2$ has non-split pairing and has an additional outer-automorphism when $n=1$;
    \item and $D_{4n-2}$, for which $D=\Z_2\times\Z_2$ has split pairing. 
\end{itemize} 
Some of the cases (such as $D_{4m}\oplus D_{4n}$) were discussed earlier, while the rest can also be easily dealt with as $D$ is quite simple. For instance, one case where our previous analysis does not readily apply is the case of $A_{N-1}\oplus D_{4n-3}$. But factorizing $\Z_N$ into $p$-groups leads to the condition that $N=4m^2$. Then, besides $L=\Z_{2m}\times \Z_2$, there are two additional choices of $L=\Z_{4m}$, generated by $(m,1)$ or $(m,3)$. These two choices are related by an outer-automorphism of $D_{4n-3}$.

Some of the resulting theories are direct sums of absolute theories with $r=1$, but there are also (infinite families of) ``irreducible'' theories that cannot be decomposed. In the case of $A_{4m^2-1}\oplus D_{4n-3}$, $L=\Z_{2m}\times \Z_2$ is reducible, while $L=\Z_{4m}$ is irreducible.

\subsection{Boundary conditions}

In this subsection, we wish to study in concrete examples the relation between polarizations and topological boundary conditions. 

\subsubsection*{Example 1: $D=\Z_{n^2}$}

We start with the case of $\Z_n\subset\Z_{n^2}$. A 7d TQFT that has this particular defect group has action
\begin{equation}
    S_{\mathrm{CS}}=\frac{n^2}{4\pi}\int CdC.
\end{equation}
This is the bulk theory for the 6d $(2,0)$ theory labeled by $\frak{g}=\frak{su}(n^2)$.\footnote{In general, this statement is true only modulo an invertible theory in 7d. However, the effect is an overall phase for partition functions that we won't keep track of.} Then the isotropic subgroup $L=\Z_n$ corresponds to the boundary condition 
\begin{equation}
    C|_\partial = B_3
\end{equation}
where $B_3$ is a $\Z_n$-valued field on the boundary. This boundary condition can be imposed by the following boundary term
\begin{equation}\label{TopBoundaryZn}
    S_{\partial}=\frac{n}{4\pi}\int_\partial (C-B_3)dY
\end{equation}
where $Y$ is a U$(1)$-valued 3-form field. It is easy to check that this boundary condition is consistent and gauge invariant. An interesting fact is that this boundary condition has an 't Hooft anomaly, which can be cancelled by a bulk action
\begin{equation}
    S_{\mathrm{inv}}=\frac{n^2}{4\pi}\int B_3dB_3.
\end{equation}
 Therefore, it is more natural to view this as a domain wall between the 7d Chern--Simons theory and the invertible theory above. When a relative 6d theory is paired with this topological boundary condition, one finds a 6d ``projective'' theory, {\it i.e.}~an absolute theory except that it has an 't Hooft anomaly, see Figure~\ref{fig:collidewall}. 
In this paper,
the term ``absolute theory'' also includes projective theories, and in similar vein we also refer to such domain walls as boundary conditions of the 7d Chern--Simons theory for simplicity.\footnote{Requiring the invertible theory to be trivial leads to a stronger condition on $L$ than requiring that the quadratic function $\bf q$ on $L$ vanishes. This is an analogue of a similar condition for the existence of topological boundary conditions discussed in \cite{Kapustin:2010hk}.}

One can understand the fact that $D=\Z_N$ has such a subgroup $L$ only when $N$ is a square from the consistency of the boundary condition. For example, if one instead has
\begin{equation}
    S_{\mathrm{CS}}=\frac{N}{4\pi}\int CdC
\end{equation}
and
\begin{equation}
    S_{\partial}=\frac{m}{4\pi}\int_\partial (C-B_3)dY
\end{equation}
with $m|N$, then requiring the boundary variation to vanish when varying $C$ and $Y$ requires, respectively, that $C$ is valued in $\Z_{N/m}$ and in $\Z_{m}$ subgroup of $U(1)$. This can only be compatible when $N=m^2$.

\subsubsection*{Example 2: $D=\Z_N \times \Z_N$ with diagonal pairing}

Here, ``diagonal'' means that the pairing between two $\Z_N$ factors is zero. One 7d TQFT with this $D$ is given by 
\begin{equation}
    S_{\mathrm{CS}}=\frac{N}{4\pi}\int\left( C_1dC_1 +C_2dC_2\right).
\end{equation}
This is (again, modulo an invertible theory) the bulk theory for the relative 6d $(2,0)$ theory labeled by $\frak{g}=\frak{su}(N)\times \frak{su}(N)$. 

One can attempt to add a boundary term
\begin{equation}
    S_\partial = \frac{\ell N}{4\pi}\int_\partial C_1 C_2.
\end{equation}
This will impose the boundary conditions
\begin{equation}
    C_1|_\partial=-\ell C_2|_\partial
\end{equation}
and
\begin{equation}
    C_2|_\partial=\ell C_1|_\partial
\end{equation}
which will be consistent only if $\ell^2=-1\pmod N$. Therefore, $-1$ has to be a quadratic residue in $\Z_N$ for this boundary condition to make sense. When this is the case, a $\Z_N$-valued boundary field $B_3$ can be added by modifying the boundary action to 
\begin{equation}
    S_\partial = \frac{\ell N}{4\pi}\int_\partial C_1C_2 + B_3(C_1-\ell C_2).
\end{equation}
The boundary condition is now 
\begin{equation}
    C_1|_\partial=-\ell C_2|_\partial +B_3.
\end{equation}
This include examples of 6d $(2,0)$ theories labeled by $\frak{g}=\frak{su}(N)\times \frak{su}(N)$ with
$$ N \; =\; 2,~5,~10,~13,~17,~25,~26,\ldots $$ but not $$N \;= \; 3,~4,~7,~8,~9,~11,~12,~14,~15,~16,~18,~19,~20,~ 21,~22,~23,~27,\ldots$$
which can not be made absolute using topological boundary conditions of this type.\footnote{For some $N$, such as 4, 8, 9, 16 and 18 there are topological boundary conditions of different type, which may require introducing auxiliary fields to be explicitly written down.}

Another way to understand why the topological boundary conditions exist for such special $N$ is to realize that when $-1$ is a quadratic residue mod $N$, the 7d TQFT defined by the action $\frac{N}{4\pi}\int CdC $ is invariant under parity (modulo a possible shift of background field), $\overline{{\cal T}^\text{bulk}}\simeq{\cal T}^\text{bulk}$. Then, the folding trick can be used to give a topological boundary condition for two copies of the 7d theory. Namely, one can fold along an identity interface in the 7d theory $\CT^{\text{bulk}}$ to obtain a topological boundary condition for $\CT^{\text{bulk}}\otimes \bar{\CT^{\text{bulk}}}\simeq \CT^{\text{bulk}}\otimes \CT^{\text{bulk}}$. 

\subsubsection*{Example 3: $D=\Z_2\times \Z_2$ with non-split pairing}

When $D=\Z_2\times \Z_2$, there are two inequivalent choices of non-degenerate bilinear form. One of them is the diagonal form discussed earlier, which is relevant for 6d $(2,0)$ theories labeled by $\frak{g}=\frak{spin}(8n+4)$. The other is non-diagonal that corresponds to 6d $(2,0)$ theories with $\frak{g}=\frak{spin}(8n)$. For the case of Spin(8), the 7d theory is given by
\begin{equation}
    S_{\mathrm{CS}}=\frac{1}{4\pi}\sum_{I,J=1,2,3,4} K_{IJ} \int C_IdC_J
\end{equation}
with $K_{IJ}$ given by the Cartan matrix of Spin(8),
\begin{equation}
    K=\left(\begin{matrix} 2 & -1 & -1 & -1\\
    -1 & 2 & 0 & 0\\
    -1 & 0 & 2 & 0\\
    -1 & 0 & 0 & 2
    \end{matrix}\right).
\end{equation}
The equations of motion
$2C_1-C_2-C_3-C_4=0$, $-C_1+2C_2=0$, $-C_1+2C_3=0$, $-C_1+2C_4=0$, which imply
 $C_2+C_3+C_4=0$, $C_1=0$, $2C_2=2C_3=2C_4=0$.
Then there is a boundary condition given by
\begin{equation}
    C_2|_\partial=C_3|_\partial+C_4|_\partial=0,
\end{equation}
with the boundary action
\begin{equation}
    S_{\partial}=\frac{1}{2\pi}\int C_3C_4.
\end{equation}
This is well-defined using the quadratic refinement $Q$ for the $\Z_2$-valued intersection pairing on $H^3(M_6,\mathbb{Z}_2)$. This quadratic function has been used in the definition of the Arf--Kervaire invariant \cite{Kervaire:1969,Browder:1969,Brown:1972}. Since on the boundary $C_3|_\partial=-C_4|_\partial$, the above boundary term can be viewed as the quadratic function
\begin{equation}
    {\pi\over 2}Q(C^\text{dis})~,
\end{equation}
where $\oint C^\text{dis}=0,1$ is given by $C_3|_\partial=-C_4|_\partial=\pi C^\text{dis}$.

There are three boundary conditions of the similar form obtained by permuting $C_2,C_3$ and $C_4$. These corresponds to the three choices of $\Z_2$ subgroup inside $\Z_2\times \Z_2$. One can also add a background field $B_3$ on the boundary. To achieve this, it is more convenient to use a different but equivalent formulation of the 7d theory. For more details, please see Appendix~\ref{sec:7ddualities}.

\subsection{Quadratic refinement, discrete theta angles, and partition functions}\label{sec:6dPartition}

In Table~\ref{tab:Tabler2}, we classified absolute 6d $(2,0)$ theories with $r=2$ by classifying choices of certain subgroups of $D$.  As we have seen in the case of $\Pol(M_6)$, one often needs to specify a quadratic refinement in order to define the partition function of the theory. We shall explore such choices for $\Pol(\mathrm{pt})$ in this subsection and relate some of them to discrete theta angles of 6d theories. 

For $\L\in\Pol(M_6)$, the valid choices of quadratic function on $\L$ form a torsor over 2--Tors$(\L^\vee)$. All choices are allowed, and should be treated democratically. For a polarization $\CP\in\Pol(M_d)$ with $d<6$, a quadratic function becomes a non-trivial functorial cohomological operation that refines the intersection pairing. In the case of $d=0$, it is a quadratic function
\begin{equation}
   q:\quad H^3(M_6,L)\rightarrow \mathbb{Q}/\Z
\end{equation}
that can be defined for any $M_6$. Such functions are rare, and all the choices known to us are built from three basic examples with $L=\Z_2$, including
\begin{itemize}
    \item $q=0$ the trivial function,
    \item $q(x)=w_3x$ multiplication by $w_3$,\footnote{This is actually a quadratic function because $q(mx)=mw_3x=m^2w_3x =q(m^2x)$, since $m^2\equiv m\pmod2$ for any integer. On a non-orientable manifold, one can also consider multiplication by $w_1w_2$. }
    \item $q(x)=Q(x)$ the quadratic refinement for the $\Z_2$-valued intersection form used in defining the Arf--Kervaire invariant.
\end{itemize}
They are all valued in $H^6(M_6,\Z_2)=\Z_2$. The first two are possible when the symmetric pairing on $H^3(M_6,L)$ vanishes,\footnote{Recall, that one can define a symmetric (possibly, degenerate) pairing on $\L_{M_6}$ once a splitting of $H^3(M_6,D)$ is chosen. This also defines a pairing on $H^3(M_6,L)$ via pull-back along $H^3(M_6,L)\rightarrow \L_{M_6}$.} while the third is used when the pairing is non-trivial. In general, one can add a term $q(x)=w_3x$ to an existing quadratic function.

In fact, the second choice of the quadratic function in the above list can be generalized to $L=\Z_{2N}$ using the reduction of the integral Stiefel--Whitney class $W_3$ mod $2N$, and then generalized further to arbitrary $L$ via the decomposition into cyclic groups. In this way, one again arrives at the conclusion that 2-Tors($L^\vee$) acts on the space of quadratic functions. Conversely, the difference between two quadratic refinements on the same $\L_{M_6}$ can be viewed as a 2-torsion element in $H^3(M_6,L^\vee)$. It seems natural to expect that demanding it to be universal/functorial with respect to $M_6$ should imply that it has to be a reduction of $W_3$. Such reductions are classified by maps $\Z_2\rightarrow L^\vee$, or equivalently elements in 2-Tors($L^\vee$).

Physically, the difference between the two choices is characterized by an element $\gamma$ in 2-Tors($L^\vee$) that corresponds to turning on a background flux $B_\gamma\in H^3(M_6,L^\vee)$ which is a reduction of $W_3$ given by $\gamma: \Z_2\rightarrow L^\vee$. 

We will give some examples to illustrate the role played by a choice of quadratic function. First consider the case relevant to 6d $(2,0)$ theories with $\frak{g}=\frak{spin}(8)$ and $D=\Z_2\times\Z_2$.

\subsubsection*{The $SO(8n)$, $Ss(8n)$ and $Sc(8n)$ theories}

One can decompose $D=L_{Ss}\oplus \bar L_{Ss}$ with $L_{Ss}=\bar L_{Ss}=\Z_2$. This gives a basis to express the partition vector as a collection of 
\begin{equation}
    Z^{Ss}_{B_{Ss}}(M_6),\quad B_{Ss}\in H^3(M_6,\Z_2).
\end{equation}
The partition function of the $Ss(8n)$ theory is by definition $Z^{{Ss}}_{B=0}$. In addition, there is a ``$Ss(8n)_{w_3}$'' theory with partition function given by $Z^{{Ss}}_{B=w_3}$.

The partition function of the $Sc(8)$ theory with general background is 
\begin{equation}
    Z^{{Sc}}_{B_{Sc}}(M_6)=\sum_{B_{{Ss}}}(-)^{B_{Sc}B_{Ss}}Z^{{Ss}}_{B_{Ss}}(M_6),\quad B_{Sc}\in H^3(M_6,\Z_2).
\end{equation}
Again, shifting $B_{{Sc}}$ by $w_3$ gives the ``$Sc(8n)_{w_3}$'' theory with the partition function
\begin{equation}
    Z^{{Sc}_{w_2}}(M_6)=\sum_{B_{{Ss}}}(-)^{w_3B_{Ss}}Z^{{Ss}}_{B_{Ss}}(M_6).
\end{equation}
The $SO(8n)$ theory is more interesting. It corresponds to a choice of diagonal $\Z_2$ and, as a consequence, the symmetric bilinear form on $H^3(M_6,L_{{SO}})$ is non-zero, and the quadratic refinement can be chosen to be $Q$. The partition function is now
\begin{equation}
    Z^{{SO}}_{B_{SO}}(M_6)=\sum_{B_{{Ss}}}(-)^{Q(B_{{Ss}})+B_{SO}B_{Ss}}Z^{{Ss}}_{B_{Ss}}(M_6),\quad B_{SO}\in H^3(M_6,\Z_2).
\end{equation}
Again, there is a $SO(8n)_{w_3}$ theory obtained by shifting $B_{{SO}_{w_3}}=w_3+B_{{SO}}$ corresponding to choosing the different quadratic function $Q(x)+w_3 x$.

Of course, one can decompose $D$ in other ways; from the point of view of $D$ and the pairing on $D$, they are all on equal footing. This would indeed be the case for $n=1$, when $\mathrm{Out}(\frak{spin}(8n))=S_3$ can be used to relate different choices of $L$ and $\bar{L}$. However, when $n>1$, we only have a $\Z_2$ outer-automorphism and not all choices are equivalent. $L_{{Ss}}$ and $L_{{Sc}}$ are exchanged while $L_{{SO}}$ is left invariant. For the $Ss$ (or $Sc$) theory, it is natural to choose $\bar L_{{Ss}}=L_{{Sc}}$ (and $\bar L_{{Ss}}=L_{{Sc}}$), which is what we did in the above analysis. But, it is also possible to choose $\bar L_{{Ss}}=L_{{SO}}$, which amounts to normalizing the partition function by $(-)^{Q(B)}$. In the SO theory, there is no canonical choice between $\bar{L}_{{SO}}=L_{{Ss}}$ and $\bar{L}_{{SO}}=L_{{Sc}}$. The two choices differ by a 6d SPT given by the Arf--Kervaire invariant. Depending on the choice, gauging the $\Z_2$ symmetry of the $SO(8n)$ theory can either lead to $Ss(8n)$ or $Sc(8n)$ theory.

Put differently, the question whether gauging the $\Z_2$ 2-form symmetry of $SO(8n)$ gives $Ss(8n)$ or $Sc(8n)$ is not well-posed and depends on a choice of duality frame. After all, these two are equivalent as physical theories. At the level of charges of strings in the 6d theory, a new set of strings will emerge after gauging the $\Z_2$ symmetry of $SO(8n)$, and we can choose to label them either by the spinor or co-spinor representations of $\frak{spin}(8n)$.

In general, suppose a $\Z_2$ two-form symmetry is non-anomalous, one can gauge it and add an SPT phase for the two-form symmetry given by the Arf--Kervaire invariant. 
Some properties of the Arf--Kervaire invariant are reviewed in Appendix \ref{sec:ArfK}, where we also discuss $\mathbb{Z}_2$ higher-form gauge theory with action given by the corresponding quadratic function. The Arf--Kervaire invariant in $(1+1)$ dimensions is the effective action for the Kitaev chain \cite{Kitaev:2001kla} fermionic SPT phase \cite{Kapustin:2014dxa}.

We will come back to this example in Section~\ref{sec:SO(8)}, after discussion of 5d theories.

\subsubsection*{Absolute theories from $\frak{spin}(8m)\oplus\frak{spin}(8n)$}

In this case, $D=\Z_2^4$, and there are 15 different choices of $L=\Z_2\times\Z_2$ inside $D$. For each of them, there is 
$L^\vee=\Z_2\times \Z_2$ symmetry and the four quadratic refinements corresponds to shifting the background fields of $L^\vee$ by $(0,0)$, $(w_3,0)$, $(0,w_3)$ and $(w_3,w_3)$, respectively.

Nine out of the 15 theories are of the form $G\times G'$ with $G$ and $G'$ being SO, Ss or Sc. These corresponds to reducible $L$. The remaining six choices are more interesting and leads to irreducible theories.

Using the result from the previous example, one can write down the partition functions for all of these theories. We will work in the basis given by $D= L_{\text{ref}}\oplus \bar L_{\text{ref}}$ with the ``reference choice'' of $L_{\text{ref}}=L_{{Ss}}\oplus L_{\text{Ss'}}$ and $\bar L_{\text{ref}}=\bar L_{{Sc}}\oplus \bar L_{\text{Sc'}}$. Then for any other choice $L$ the partition function is always in the form
\begin{equation}
    Z^{L}=\sum_{B,B'}f(B,B')Z^{{Ss}}_B Z^{{Ss}'}_{B'}, \quad B,B'\in H^3(M_6,\Z_2\times \Z_2),
\end{equation}
for some function $f(B,B')$. For the reducible theories, $f(B,B')=g(B)g(B')$ can be factorized with $g$ and $g'$ given by the delta function $\delta_{B,0}$, the constant function $1$, or $(-)^{Q(B)}$. For the irreducible theories, this function is given below

\begin{center}
    
\begin{tabular}{c|c|c|c}
      label & construction & $L$ & $f(B,B')$  \\\hline
     ``vsc''& ($Ss\times Ss'$)/$\Z_2$ & $\{0,(vv),(ss),(cc)\}$& $\delta_{B,B'}$ \\\hline
     ``csv''&($Ss\times Ss'$)/$\Z_2$ &$\{0,(vc),(ss),(cv)\}$&$\delta_{B,B'}(-)^{Q(B)}$ \\\hline``vcs''&($Ss\times Ss'$)/$\Z_2\times \Z_2$ &$\{0,(vv),(sc),(cs)\}$& $(-)^{BB'}$  \\\hline``scv''&($Ss\times Ss'$)/$\Z_2\times \Z_2$ & $\{0,(vs),(sc),(cv)\}$
     & $(-)^{Q(B) +BB'}$  \\\hline``cvs''&($Ss\times Ss'$)/$\Z_2\times\Z_2$  &$\{0,(vc),(sv),(cs)\}$
          & $(-)^{Q(B') +BB'}$ \\\hline
     ``svc''&($Ss\times Ss'$)/$\Z_2\times \Z_2$ &$\{0,(vs),(sv),(cc)\}$&    $(-)^{Q(B) +BB'+Q(B')}$ 
     
\end{tabular}

\end{center}
Here we have used short-hand notation $v$, $s$ and $c$ for the three non-trivial elements in $\Z_2\times\Z_2$, and $BB'$ the $\Z_2$-valued function given by $\int_{M_6}B\cup B'$, indicating that the theory has a discrete theta angle, making an otherwise reducible theory irreducible. The second column describes how to physically construct these theories starting with the $Ss\times Ss'$ theory and gauging the whole for the diagonal of the $\Z_2\times\Z_2$ symmetry group. In this process one can choose a discrete theta angle which can be read off from the form of $f$ in the last column.

We will outline the computation for $f$ in the theory labeled by ``vcs'' and ``svc'' with the rest being very similar. 

This choice of $L_{\text{vcs}}$ leads to a subgroup $\L_{\text{vcs}}$ of $H^3(M_6,\Z_2^4)$ generated by $\alpha(cs)$ and $\beta(sc)$ with $\alpha,\beta\in H^3(M_6,\Z_2)$. The quadratic function can be chosen to be 
\begin{equation}
    q[\alpha(cs)+\beta(sc)]=\alpha\cdot\beta:=\int_{M_6}\alpha\cup\beta.
\end{equation}
It is easy to check that this refines the symmetric pairing on $\L$ given by\footnote{Recall that the symmetric pairing is defined by first decomposing $H^3(M_6,D)$ into $\L_{\text{ref}}\oplus \bar\L_{\text{ref}}$ using $D=L_{\text{ref}}\oplus\bar{L}_\text{ref}$, and then \[\langle x , y \rangle_{\text{sym}}:=\int_{M_6} p_{\bar\L_{\text{ref}}} (x)\cup p_{\L_{\text{ref}}} (y),\quad \text{for $x,y\in \L$} \] where $p$'s denote projections onto the $\L_{\text{ref}}$ and $\bar\L_{\text{ref}}$. As we commented before, there are three more quadratic refinements of this bilinear form given by shifting $q$ by $w_3\alpha$, $w_3\beta$ and $w_3(\alpha+\beta)$.} 
\begin{equation}
    \langle \alpha(cs),\alpha'(cs) \rangle_{\text{sym}}=\langle \beta(sc),\beta'(sc) \rangle_{\text{sym}}=0,
\end{equation}
and 
\begin{equation}
    \quad \langle \alpha(cs),\beta(sc) \rangle_{\text{sym}}=\alpha\cdot\beta.
\end{equation}
 Then applying the general formula for the partition function \eqref{6dPartition} leads to $f(B,B')=(-)^{BB'}$, as the projection $p_{\L_{\text{ref}}}(\alpha(cs)+\beta(sc))=(B=\alpha,B'=\beta)\in \bar{\L}_{\text{ref}}$.

Similarly, the choice of $L_{\text{svc}}$ leads to a subgroup $\L_{\text{svc}}$ generated by $\alpha(vs)$ and $\beta(sv)$. The quadratic function $q$ given by
\begin{equation}
    q[\alpha(vs)+\beta(sv)]=Q(\alpha)+Q(\beta)+\alpha\cdot\beta
\end{equation}
refines the symmetric pairing given by
\begin{equation}
    \langle \alpha(vs),\alpha'(vs) \rangle_{\text{sym}}=\alpha\cdot\alpha', \quad   \langle \beta(sv),\beta'(sv) \rangle_{\text{sym}}=\beta\cdot\beta',
\end{equation}
and
\begin{equation}
    \langle \alpha(vs),\beta(sv) \rangle_{\text{sym}}=\alpha\cdot \beta.
\end{equation}
And this $q$ leads to a discrete theta angle given by
\begin{equation}
    f(B,B')=(-)^{Q(B) +BB'+Q(B')}.
\end{equation}

For each choice of $L$, it is straightforward to turn on an $L^\vee$-valued background field. We will omit the answer here.

\section{Compactification to 5d}
\label{sec:compactification5d}

In this section, we study $\Pol(S^1)$ which leads to an absolute 5d theory. Unlike $\Pol(\mathrm{pt})$, where mixed polarizations do not exist, there are often many mixed polarizations in $\Pol(S^1)$ for sufficiently complicated choices of $D$.

We will start with a general discussion about classification of polarization on $S^1$ and then move to concrete examples.

\subsection{Classifying pure polarizations on $S^1$}

As $H^*(S^1,D)=D^{(0)}\oplus D^{(1)}$, with symmetric pairing between the two pieces in degree 0 and 1, it might appear at the first sight that the classification for pure polarization would be completely equivalent to the 6d story with replacement of $D$ by two copies of $D$. This is not completely correct, as $L\subset H^*(S^1,D)$ is assumed to be a sum of graded pieces $L^{(i)}\subset H^i(S^1,D)$. Therefore, the right way to classify $L$ is to find two subgroups $L^{(0)}$ and $L^{(1)}$ of $D$ that pair trivially with each other. Further, $L$ is also required to be maximal. This will turnout to be simpler than the classification problem we encountered in 6d.  

\subsubsection*{Two canonical choices of pure polarizations}

Such choices of $L$ always exist, and there are two canonical ones given by 
\begin{equation}
    L=D^{(0)}\quad \text{and }\quad L=D^{(1)}.
\end{equation}
They correspond to two polarizations $\CP_{0}$ and $\CP_{1}$, which in turn lead to two absolute theories. The theory $T[S^1,\CP_0]$ has 1-form $D$ symmetry with line operators carrying the charge, while the theory  $T[S^1,\CP_1]$ has 2-form $D$ symmetry with charged objects being strings.

Both polarizations are splittable, and splitting is given by $\bar L=D^{(1)}$ or $D^{(0)}$ respectively. As a consequence, there are always two sets of canonical basis for the partition vector of the $T[S^1]$ theory. To see this more clearly, consider the theory on  $M_6=S^1\times M_5$. The two choices of polarization correspond to decomposing
\begin{equation}
H^3(M_6,D)= H^2(M_5,D)\oplus H^3(M_5,D) =\Lambda\oplus \bar\Lambda~,
\end{equation}
with $\L=H^3(M_5,D)$ and $\bar{\L}=H^2(M_5,D)$ for $\CP_0$, and the opposite for $\CP_1$ with $\L=H^2(M_5,D)$ and $\bar{\L}=H^3(M_5,D)$.

Among them, $\CP_0$ is geometric, given by capping off $S^1$ with a disk $\Sigma$, as $D^{(0)}$ is the image of 
\begin{equation}
    H^*(\Sigma;D) \rightarrow H^*(S^1,D).
\end{equation}
The line operators in the theory $T[S^1,\CP_0]$ live on the boundary of 3d operators in $\CT^\text{bulk}$ wrapping $\Sigma$. To see this, it is more convenient to use the homology version of the map above,
\begin{equation}
    H_2(\Sigma,S^1;D) \rightarrow H_1(S^1,D).
\end{equation}

\begin{figure}
  \centering
    \includegraphics[width=0.7\textwidth]{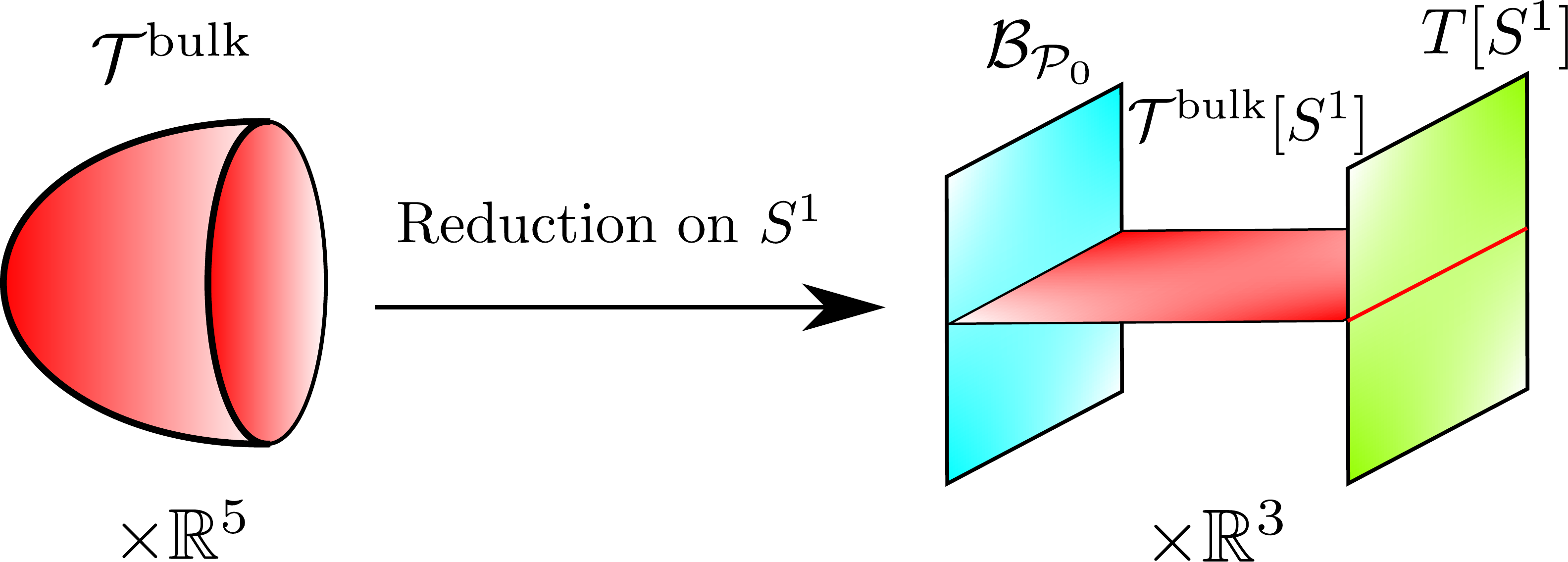}
    \caption{Three-dimensional operators in $\CT^{\text{bulk}}$ wrapping the disk $\Sigma$ become 2d surface operators stretched between the topological boundary $\mathfrak{B}_{\CP_0}$ and the relative theory $T[S^1]$. After colliding the two boundaries, they become line operators in $T[S^1,\CP_0]$.}\label{fig:disk}
\end{figure}

After reduction on $S^1$, one obtains $\CT^{\text{bulk}}[S^1]$ with 2d surface operators that can end on the topological boundary condition given by $\Sigma$. The line operators in $T[S^1,\CP_0]$ come from surface operators stretched between the two boundaries. See Figure~\ref{fig:disk} for an illustration.

\subsubsection*{Reduction of 6d $(2,0)$ theory}

These two polarizations can be understood rather concretely for the 6d $(2,0)$ theories. It is believed that their compactification leads to ${\cal N}=2$ supersymmetric Yang--Mills theories with some gauge group $G$ whose Lie algebra is ${\frak g}$ that labels the 6d theory \cite{Seiberg:1996bd,Seiberg:1997ax,Douglas:2010iu,Lambert:2010iw}. 

The precise form of the gauge group (and, as we will see soon, other discrete data such as possible discrete theta angles) depend on the choice of polarization. For the two canonical pure polarizations $\CP_0$ and $\CP_1$, they give rise to the $\tilde G$  gauge theory and the $G_{\mathrm{ad}}:=\tilde G/\CZ(\tilde G)$ gauge theory, respectively, where $\tilde G$ denotes the simply-connected form of a group with the Lie algebra ${\frak g}$. 

As an example, consider ${\frak g}=\mathfrak{su}(N)$. Then $\CP_0$ leads to an $\CN=2$ theory with gauge group $SU(N)$ while $\CP_1$ leads to the $PSU(N)$ gauge theory.

\subsubsection*{More general pure polarizations}

We do know that there are in general more forms of $G$ with the same Lie algebra between $G_{\mathrm{ad}}$  and $\tilde G$. For example, in 5d we can have theory with gauge group $SU(N)/\mathbb{Z}_k$ for any $k|N$. It is natural to wonder whether they also come from suitable choices of polarization. 

The answer to this question is affirmative, and the choice leading to $SU(N)/\mathbb{Z}_k$ is given by
\begin{equation}
    L=\mathbb{Z}^{(0)}_{N/k}\times \mathbb{Z}^{(1)}_k\subset \mathbb{Z}^{(0)}_N\times \mathbb{Z}^{(1)}_{N}.
\end{equation}

More generally, one can take $L^{(0)}$ to be any subgroup of $D^{(0)}$, and $L^{(1)}$ to be the subgroup of $D^{(1)}$ consist of elements that pair trivially with $L^{(0)}$. It is clear that $L$ is maximal, as $L^{(1)}$ is the Pontryagin dual of $D^{(0)}/L^{(0)}$ and anything in $D^{(0)}$ that is not in $L^{(0)}$ won't pair trivially with all of $L^{(1)}$.

In the context of 6d $(2,0)$ theories, such a polarization would leads to a 5d super-Yang--Mills theory with $\tilde G/L^{(0)}$ gauge theory. Therefore, one can obtain any compact global form with Lie algebra $\frak{g}$.

Such a polarization will leads to a theory with $D/L^{(1)}\simeq (L^{(0)})^\vee$ 1-form  symmetry and $D/L^{(0)}\simeq (L^{(1)})^\vee$ 2-form symmetry. Charges of line operators are classified by $L^{(0)}$ while charges of strings are classified by $ L^{(1)}$.

We can also turn on discrete theta angle in 5d theories. For example, if a theory is obtained by gauging a $L^{(0)}$ symmetry of $T[S^1,\CP_0]$, then there are a collection of $\Z_m$-valued discrete gauge field $ B_2\in H^2(M_5,\Z_m)$ for such cyclic subgroup $\Z_m$ of $L^{(0)}$ and one can introduce a topological term proportional to $
    \int_{M_5}B_2\cup\mathrm{Bock}(B_2)$
using the Bockstein of $B_2$, and the coefficient for this term leads to a discrete theta angle. As we will see later, this is only non-trivial when $m$ is even, and the discrete theta angle will be $\Z_2$-valued. Alternatively, one can mix different factors $\Z_m$ and $\Z_n$ via a term proportional to $
    \int_{M_5}B_2\cup \mathrm{Bock}(B'_2)$, with a $\mathrm{gcd}(m,n)$-valued discrete theta angle.

It turns out that these two types of theta angles have different origins, with the former related to quadratic refinement, and the latter related to mixed polarizations. 

\subsubsection*{Quadratic refinements for pure polarizations}

Similar to the 6d case that we discussed in the last section, the quadratic function is now a quadratic cohomological operation:
\begin{equation}
    q:\quad H^2(X_{5},L^{(1)})\oplus H^3(X_{5},L^{(0)}) \rightarrow \mathbb{Q}/\Z.
\end{equation}
As the two parts pair trivially, $q$ is determined by its value on $H^2$ and $H^3$ separately. After choosing $\L_{\text{ref}}=H^2(M_5,D)$ and $\bar{\L}_{\text{ref}}=H^3(M_5,D)$, the symmetric bilinear form on $\L$ for any $\L$ from pure polarization $\CP_L$ is trivial, so $q=0$ is a valid choice for the quadratic function. However, there are more choices such as\footnote{Again, they are quadratic $q(mB_{2,3})=m^2q(B_{2,3})$ because $m^2\equiv m\pmod2$.} 
\begin{equation}
    q(B_2)=\int_{M_5}w_3\cup B_2
\end{equation}
and 
\begin{equation}
q(B_3)=\int_{M_5} w_2\cup B_3
\end{equation}
for $B_{2,3}\in H^{2,3}(M_5,\Z_2)$.  Quadratic functions of this form are classified by homomorphisms from $\Z_2$ to $L$, which in turn are labeled by elements in 2-Tors$(L^\vee)$.

Physically, the role of $q$ in this form is to shift the background field $B_2$ or $B_3$ for the 1-form or 2-form symmetry by $w_2$ or $w_3$, respectively. As an example, consider the case of $D=\Z_4$ such as $(2,0)$ theory with $\frak{g}=\frak{su}(4)$. Then, there is a 5d theory given by $L=\Z_2^{(0)}\oplus \Z_2^{(1)}$ with $SU(4)/\Z_2\simeq SO(6)$ gauge group. This theory has $\Z_2$ 1-form electric symmetry and 2-form magnetic symmetry. There are four different choices of $q$ leading to four different theories obtained by shifting the background fields $(B_e,B_m)\in H^2(X_{5},\Z_2)\oplus H^3(X_{5},\Z_2)$ by $(0,0)$, $(w_2,0)$, $(0,w_3)$ and $(w_2,w_3)$.  
The first and the third are obtained from reduction of the 6d $SO(6)$ or $SO(6)_{w_3}$ theory.

\subsubsection*{Symmetries and anomalies}

We now analyze anomalies of the $T[S^1,\CP_L]$ theory. The theory has $(L^{(0)})^\vee = D/L^{(1)}$ 1-form symmetry and $(L^{(1)})^\vee = D/L^{(0)}$ 2-form symmetry. As discussed in previous sections, their anomaly can be captured by the ambiguity of lifting $B_2\in H^2(M_5,D/L^{(1)})$ and $B_3 \in H^3(M_5,D/L^{(0)})$ along the second map in
\begin{equation}
    H^2(M_5,L^{(1)})\oplus H^3(M_5,L^{(0)})\rightarrow H^2(M_5,D)\oplus H^3(M_5,D) \rightarrow H^2(M_5,D/L^{(1)})\oplus H^3(M_5,D/L^{(0)}).
\end{equation}
Any two lifts differ by the image of \begin{equation}
    (\delta  B_2,\delta B_3)\in H^2(M_5,L^{(1)})\oplus H^3(M_5,L^{(0)}),
\end{equation}
which gives a gauge transformation of the $L^\vee$ symmetry. Then a non-trivial pairing 
\begin{equation}
    \langle (B_2, B_3), (\delta B_2,\delta  B_3)\rangle =\int_{M_5}B_2\delta B_3 + B_3\delta B_2 :=\langle B_2, \delta  B_3\rangle + \langle \delta  B_2 , B_3\rangle 
\end{equation}
describes an 't Hooft anomaly of the $L^\vee$ symmetry. Here, the pairings on the right are the natural ones between $H^2(M_5,L^{(1)})$ and $H^3(M_5,(L^{(1)})^\vee)$, and between $H^3(M_5,L^{(0)})$ and $H^2(M_5,(L^{(0)})^\vee)$. This anomaly can be canceled by coupling to a 6d invertible TQFT, whose action can be expressed in the continuous notation as
\begin{equation}
    \sum\frac{K^{IJ}}{2\pi}\int_{6\text{d}} B_2^I d B_3^J ~
\end{equation}
with the $K^{IJ}$ from the coefficient matrix of the three-form Chern--Simons theory in 7d.

We will arrive at this action from a different point of view later after we relate polarizations on $S^1$ with boundary conditions of $\CT^{\text{bulk}}[S^1]$. Notice that in the special case of $L^{(0)}=L^{(1)}$, the 5d theory will be a reduction of a 6d absolute theory. From that perspective, the action above also arises as the reduction of the 7d invertible theory describing the 't Hooft anomaly for the 6d theory. 

\subsection{Mixed polarizations and discrete theta angles}

An interesting phenomenon for reduction of 6d theory on a circle is the possible existence of mixed polarizations. 

By definition, they are not obtained by choosing a subgroup $L$ of $H^*(S^1,D)$, but nonetheless a mixed polarization $\CP$ specifies a maximal isotropic subgroup $\L_{\CP}(M_5)$ of $H^3(M_5\times S^1,D)$ for each $M_5$.

All such polarizations will turn out to be generalizations of the following example with $D=\Z_m\times \Z_n$. We use $B_2$ and $B'_2$ to parametrize $H^2(M_5,\Z_m\times \Z_n)$. Then one can consider $\L_{\CP}(M_5)$  generated by 
\begin{equation}\label{Mixed5d}
    (B_2,B'_2,k\delta' B'_2,k\delta B_2)\in H^2(M_5,\Z_m)\oplus H^2(M_5, \Z_n)\oplus H^3(M_5,\Z_m)\oplus H^3(M_5, \Z_n)=H^3(M_5\times S^1,D)
\end{equation}
with $k\in\Z$, where 
\begin{equation}
    \delta:\quad H^2(M_5,\Z_m)\rightarrow H^3(M_5, \Z_n)
\end{equation}
 is the Bockstein associated with
 \begin{equation}
     \Z_n\rightarrow \Z_{mn} \rightarrow \Z_m
 \end{equation}
 or essentially\footnote{In other words, it only depends on $B_2$ modulo elements in $H^2(M_5,\Z_{m/\mathrm{gcd}(m,n)})$, and only hits elements in $H^3(M_5, \Z_n)$ via $H^3(M_5, \Z_{\mathrm{gcd}(m,n)})\rightarrow H^3(M_5, \Z_n)$.}
 \begin{equation}
     \Z_{\mathrm{gcd}(m,n)}\rightarrow \Z_{\mathrm{gcd}(m,n)^2} \rightarrow \Z_{\mathrm{gcd}(m,n)}.
 \end{equation}
 The homomorphism 
 \begin{equation}
    \delta':\quad H^2(M_5,\Z_n)\rightarrow H^3(M_5, \Z_m)
\end{equation}
is similarly defined. A short computation confirms that this $\L_{\CP}(M_5)$ is indeed maximal isotropic for any choice of $M_5$, and it only depends on $k$ modulo $\mathrm{gcd}(m,n)$.  Notice that in \eqref{Mixed5d}, it is important that the coefficient of $\delta B_2$ and $\delta'B'_2$ is the same. In other words, the subgroup generated by $(B_2,B'_2,k'\delta' B'_2,k\delta B_2)$ is only isotropic is $k\equiv k'\pmod{ \mathrm{gcd}(m,n)}$. 

This family of maximal isotropic subgroups will turn out to be functorial, and the mixed polarization will be denoted as $\CP(\Z_m\times\Z_n;k)$ with $k\in \Z_{\mathrm{gcd}(m,n)}$. For $k=0$, it will actually become the pure polarization $\CP_1$.

\subsubsection*{More general mixed polarizations and remaining symmetries}

To get all the other mixed polarizations, one only needs to generalize the above in two directions. The first is to allow taking a subgroup $L^{(1)}$ of $D$, while the second is to allow a $k\in \Z_{\mathrm{gcd}(m,n)}$ for any pairs of cyclic subfactors of $L^{(1)}$. Then it leads to the following maximal isotropic subgroup $\L_{\CP(L^{(1)};\mathbf{k})}(M_5)\subset H^3(M_5\times S^1,D)$ generated by 
\begin{equation}
    (B_2,\delta_{\mathbf{k}}B_2+B_3)\in H^2(M_5,L^{(1)})\oplus H^3(M_5,D).
\end{equation}
Here $B_3$ are elements in $H^3(M_5,D)$ that pair trivially with any $B_2$, which come from $H^3(M_5,L^{(0)})$ with $L^{(0)}:= (D/L^{(1)})^\vee$, while $\delta_{\mathbf{k}}$ is the Bockstein for the extension of $L^{(1)}$ by $(L^{(1)})^\vee$ given by $\mathbf{k}$.\footnote{Recall that $\mathrm{Ext}^1(\Z_m,\Z_n)=\Z_{\mathrm{gcd}(m,n)}$. Therefore, $\mathbf{k}$ determines an extension of $L^{(1)}$ by $(L^{(1)})^\vee$. } To interpret $\delta_{\mathbf{k}}B_2$ as an element in $H^3(M_5,D)$, one needs to choose a lift from $H^3(M_5,(L^{(1)})^\vee)$. There could be different choices, but they differ only by a $B_3\in H^3(M_5,L^{(0)})$ due to the exactness of
\begin{equation}
    H^3(M_5,L^{(0)})\rightarrow H^3(M_5,D) \rightarrow H^3\left(M_5,(L^{(1)})^\vee=D/L^{(0)}\right),
\end{equation}
and therefore lead to the same maximal isotropic subgroup $\L_{\CP(L^{(1)};k)}(M_5)$.

\begin{figure}
  \centering
    \includegraphics[width=0.5\textwidth]{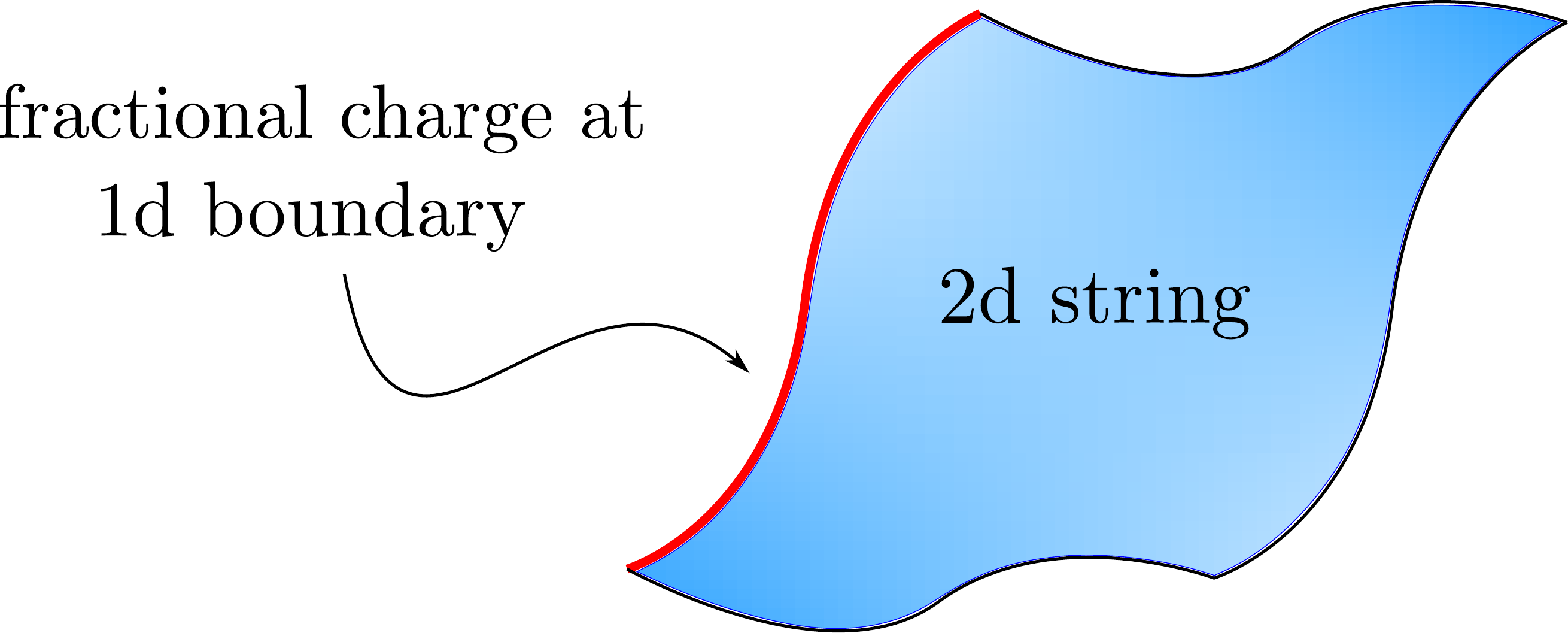}
    \caption{In the theory $T[S^1,\CP]$ given by a mixed polarization $\CP$, there are strings whose boundaries carry a ``fractional charge'' under the 1-form symmetry of the theory. This does not contradict the Dirac quantization condition, which applies only to independent objects.}\label{fig:string}
\end{figure}

When $\mathbf{k}$ is non-zero, we indeed obtain a genuine mixed polarization, as the projection of elements in  $\L_{\CP(L^{(1)};k)}(M_5)$ to $H^3(M_5,D)$ is no longer constrained in the image of $H^3(M_5,L^{(0)})$. In other words,
\begin{equation}
    L^{(0)}\oplus L^{(1)}\subset  \CS(\CP)
\end{equation}
is now expected to be a proper subset.  Furthermore, $\CS(\CP)$ is not be isotropic, and as we have explained, this is not a violation of the Dirac quantization condition, but instead signifies the existence of strings whose boundaries carry fractional charges under the 1-form symmetry. This is illustrated in Figure~\ref{fig:string}.

More precisely, $\CS(\CP)$ can be decomposed into two subgroups with fixed degree  $\CS(\CP)^{(0)}$ and  $\CS(\CP)^{(1)}$. The piece $\CS(\CP)^{(1)}$ labels the charge of strings, and the 2-form symmetry of the theory is $(\CS(\CP)^{(1)})^\vee$. On the other hand, charges for independent line operators are classified by $\CS(\CP)^{(0)}_{\text{ind}}$ which is a subgroup of $\CS(\CP)^{(0)}$ isomorphic to $(D/\CS(\CP)^{(1)})^\vee$. Elements of $\CS(\CP)^{(0)}$ that are not in $\CS(\CP)^{(0)}_{\text{ind}}$ label charges of line operators that can only live on the boundary of string worldsheets. And the 1-form symmetry is not $(\CS(\CP)^{(0)})^\vee$, but a smaller group $(\CS(\CP)^{(0)}_{\text{ind}})^\vee$ isomorphic to $D/\CS(\CP)^{(1)}$. 

As an example, consider the polarization $\CP(\Z_m\times \Z_n,k)$. When $k=0$, this is the pure polarization $\CP_1$ leading to a theory with $\Z_m\times\Z_n$ 2-form symmetry. For any $k$, the symmetry stays the same,  and only uncharged line can exist independently. In other words, $\CS(\CP)^{(1)}=\Z_m\times\Z_n$ and $\CS(\CP)^{(0)}_{\text{ind}}$ is trivial, while $\CS(\CP)^{(0)}=\Z^2_{\mathrm{gcd}(m,n,k)}$ labels charges of line operators that can appear on the boundary of strings. This phenomenon is discussed in further detail in Appendix~\ref{sec:5ddiscretetheta}.

More generally, for the polarization $\CP(L^{(1)};\mathbf{k})$, we have $\CS(\CP)^{(1)}=L^{(1)}$ classifying charges of strings, while $\CS(\CP)^{(0)}_{\text{ind}}=(D/L^{(1)})^\vee$ classifies independent line operators. In this case, we can still define 
\begin{equation}
    L^{(0)}:=(D/L^{(1)})^\vee \subset D,
\end{equation}
then $L:=L^{(0)}\oplus L^{(1)} \subset H^*(S^1,D)$ defines an associated pure polarization which is simply the limit of this family of mixed polarization for $\mathbf{k}=0$, \begin{equation}
    \CP_L=\CP(L^{(1)};\mathbf{k}=0).
\end{equation} 
We will see next that $\mathbf{k}$ can be interpreted as discrete theta angles that can be turned on in the theory $T[S^1,\CP_L]$.

\subsubsection*{Partition functions, discrete theta angles and quadratic refinements}

Since there is a canonical splitting of $H^3(M_5\times S^1,D)$ into $\L\oplus\bar\L$ associated with $\CP_0$, with $\L=H^3(M_5,D)$ and $\bar{\L}=H^2(M_5,D)$, one can use the explicit basis $Z_{\L,B}$ with $B\in H^2(M_5,D)$ to write down the partition function with a mixed polarization $\CP(L^{(1)};\mathbf{k})$. It takes the following form
\begin{equation}
    Z_{\CP(L^{(1)};\mathbf{k})}=\sum_{B_\alpha\in H^2(M_5,L^{(1)})}e^{\int_{M_5}B_\alpha\delta_{\mathbf k}B_\alpha} Z_{\L,B_\alpha},
\end{equation}
where the phase is given by the $U(1)$-valued pairing between $H^2(M_5,L^{(1)})$ and $H^3(M_5,L^{(1)})$ and can be identified with an SPT phase in $\Omega^{SO}_5(\mathbf{B}^2L^{(1)})$. This SPT phase can also be interpreted as a discrete theta angle term in the theory labeled by $\mathbf{k}$. One implication is that any theory  $T[S^1,\CP(L^{(1)};\mathbf{k})]$ can always be constructed from $T[S^1,\CP_0]$ by first gauging a $L^{(1)}$ subgroup of the $D$ 1-form symmetry and then turning on a discrete theta angle. This is a general feature of the family of absolute theories labeled by polarizations in all dimensions.

When $L^{(1)}$ and $L^{(0)}:=(D/L^{(1)})^\vee$ has 2-torsion, again there are different quadratic refinements for a mixed polarization. At the level of the partition function, they correspond to either turning on an element in $\Omega^{SO}_5(\mathbf{B}^2L^{(1)})$ that takes the form of $e^{\int_{M_5} B_\alpha \gamma_*(w_3)}$, or in addition, involving a shift $Z_{B_\alpha}\rightarrow Z_{B_\alpha+\gamma'_*(w_2)}$.\footnote{Recall that for $\Pol(M_6)$, we referred to the former as quadratic refinement of the first kind while the latter as quadratic refinement of the second kind.} Here, $\gamma: \Z_2\rightarrow (L^{(1)})^\vee$ and $\gamma_*$ is the induced map $H^3(M_5,\Z_2)\rightarrow H^3(M_5,(L^{(1)})^\vee)$, while $\gamma': \Z_2\rightarrow (L^{(0)})^\vee$ inducing $\gamma'_*: H^2(M_5,\Z_2)\rightarrow H^2(M_5,(L^{(0)})^\vee)$. The addition $B_\alpha+\gamma'_*(w_2)$ depends on a lift of $(L^{(0)})^\vee=D/L^{(1)}$ to $D$, but the entire sum doesn't depend on this choice.

To summarize, the quadratic refinement is labeled by elements in 2-Tors$(L^\vee)$, and doesn't depend on discrete theta angles $\mathbf{k}$. Another property of the theory that doesn't depend on $\mathbf{k}$ is the 't Hooft anomaly of the theory, as the analysis in the $\mathbf{k}=0$ case straightforwardly carries over.\footnote{Notice that this statement only concerns symmetries coming from the 2-form symmetry in 6d. There can be additional ``emergent symmtries'' in the 5d reduction, whose exact form and anomalies can depend on discrete theta angles. We will discuss this in more detail in later subsections and in Appendix~\ref{sec:5ddiscretetheta}.} At the level of partition function, the anomaly manifest itself as an ambiguity of the partition function in the presence of background fields $B_2$ and $B_3$ for the $(L^{(0)})^\vee$ 1-form and the $(L^{(1)})^\vee$ 2-form symmetry. Naively, one has
\begin{equation}
    Z_{\CP(L^{(1)};\mathbf{k}),B_2,B_3}=\sum_{B_\alpha\in H^2(M_5,L^{(1)})}e^{\int_{M_5}B_\alpha\delta_{\mathbf k}B_\alpha + B_3 (B_\alpha + B_2)} Z_{\L,B_\alpha+B_2},
\end{equation}
but the pairing between $B_2$ and $B_3$ depends on a choice of lift of $B_2$ and $B_3$ to $H^{*}(M_5,D)$, and a change of a lift, characterized by $\delta B_{2} \in H^2(M_5,L^{(1)})$ and $\delta B_3\in H^3(M_5,L^{(0)})$, will lead to an overall phase given by
\begin{equation}
    \langle B_2, \delta  B_3\rangle + \langle \delta  B_2 , B_3\rangle\in D\subset U(1).
\end{equation}
This is indeed independent of $\mathbf{k}$ and exactly the same as the case of $\mathbf{k}=0$ that we discussed before.

The existence of this family of partition functions for any $M_5$ labeled by $\CP(L^{(0)},\mathbf{k})$ provide strong evidence that they are functorial and well-defined as polarizations. One way to really show that these mixed polarizations exist is to construct the corresponding topological boundary conditions. We will see this explicitly next.

\subsection{Polarizations as topological boundary conditions}

Without loss of generality, we consider $D=\mathbb{Z}_N\times \mathbb{Z}_N$ with the 7d action
\begin{equation}
    \frac{N}{4\pi}\int_{\text{7d}}\left( C^1dC^1+C^2dC^2\right)~.
\end{equation}
Its circle reduction is obtained by writing $C=B_3+B_2\alpha$, with $H^1(S^1,\mathbb{Z}) = \langle \alpha \rangle \cong \Z$,
\begin{equation}
    \frac{N}{2\pi}\int_{6\text{d}}\left( B_3^1dB_2^1+B_3^2dB_2^2\right)~.
\end{equation}
What are consistent boundary conditions?
The boundary variation of the bulk action is
\begin{equation}
    -\frac{N}{2\pi}\int_{5\text{d}}\left( B_3^1\Delta B_2^1+B_3^2\Delta B_2^2\right)~.
\end{equation}
If there are no additional boundary terms, then the equations of motion for the boundary fields $B_2^1|_\partial$, $B_2^2|_\partial$ lead to the boundary condition $B_3^1|_\partial=0$, $B_3^2|_\partial=0$.
In the following we explore other possible boundary conditions and the corresponding 5d boundary terms. They are the corresponding 5d discrete theta angles.

For instance, the following boundary condition with $k\in\mathbb{Z}$
\begin{equation}
B_3^1|_\partial=\frac{k}{N} dB_2^2|_\partial,\quad
B_3^2|_\partial=-\frac{k}{N} dB_2^1|_\partial
\end{equation}
can be imposed by the equations of motion for the boundary fields with an additional boundary term
\begin{equation}
\frac{k}{2\pi}\int_{5\text{d}}B_2^1 d B_2^2~.
\end{equation}
This describes a mixed discrete theta angle in 5d $\mathbb{Z}_N\times \mathbb{Z}_N$ two-form gauge theories, and corresponds to the polarization $\CP(\Z_N\times\Z_N;k)$.

We have seen previously that when $N$ is even, $\CP(\Z_N\times\Z_N;k)$ will have four different refinements. This can also be seen at the level of boundary conditions. For simplicity, consider $k=0$. Then the boundary condition can be
\begin{equation}
    B_3^1|_\partial=\frac{2\pi \ell^1}{N} W_3,\quad B_3^2|_\partial=\frac{2\pi \ell^2}{N} W_3~
\end{equation}
with $\ell^I\in \mathbb{Z}_N$, $I=1,2$. It can be imposed by the equations of motion for the boundary fields by adding the following boundary term\footnote{
Note, the boundary term $\int_{5\text{d}}B_2^IdB_2^I$ does not contribute to the equations of motion since it is locally a total derivative.
}
\begin{equation}
\int_{5\text{d}}  W_3\left( \ell^1 B_2^1+\ell^2B_2^2\right)~.
\end{equation}
This describes the discrete theta angle for each $\mathbb{Z}_N$ two-form gauge field. In the discrete notation $B_2^I=\frac{2\pi}{N} B_2^{I,\text{dis}}$ for $I=1,2$, the boundary term is
\begin{equation}
\sum_{I=1,2}\frac{2\pi \ell^I}{N}\int_{5\text{d}} W_3B_2^{I,\text{dis}}
=\sum_{I=1,2} \pi\ell^I\int_{5\text{d}} w_2 \text{Bock}(B_2^{I,\text{dis}})
=\sum_{I=1,2}\frac{2\pi \ell^I}{N}\int_{5\text{d}} B_2^{I,\text{dis}}\text{Bock}\left(B_2^{I,\text{dis}}\right)~,
\end{equation}
where the first equality used $W_3=\text{Bock}(w_2)$ on orientable manifolds, and $\text{Bock}(B_2^{I,\text{dis}})=dB_2^{I,\text{dis}}/N$. As it only depends on the mod-2 reduction of $\ell^I$, we can effectively take $\ell^I\in\mathbb{Z}_2$. The second equality used (\ref{eqn:5dthetaidentity}). Note, when $N$ is odd, the boundary term is trivial since for $I=1,2$:
\begin{equation}
\frac{2\pi \ell^I}{N}\int_{5\text{d}} W_3B_2^{I,\text{dis}}=
\frac{2\pi \ell^I}{N}\int_{5\text{d}} (1+N)W_3B_2^{I,\text{dis}}=
(1+N)\pi\ell^I\int_{5\text{d}} w_2 \text{Bock}(B_2^{I,\text{dis}})=0\quad\text{mod }2\pi\mathbb{Z}~.
\end{equation}
This is consistent with having no discrete theta angle for each $\mathbb{Z}_N$ two-form gauge field when $N$ is odd. Equivalently, there is only a single quadratic function $q$ compatible with $\CP(\Z_N\times\Z_N,k)$ when $N$ is odd.

\subsubsection*{Boundary condition with 't Hooft anomaly}

One interesting class of boundary conditions consists of those that have 't Hooft anomaly, which we will study now.

Consider $D=\mathbb{Z}_N$ when $N$ is not a prime, and take the bulk action of the 6d TQFT $\CT^{\text{bulk}}[S^1]$ to be
\begin{equation}
\frac{N}{2\pi}\int_{6\text{d}}B_2dB_3~.
\end{equation}
For any divisor $k|N$, one can add the following boundary term
\begin{equation}
    \int_{5\text{d}}\frac{N}{2\pi}\left(y_2+B_2^\text{cl}\frac{1}{k}\right)B_3+\frac{k}{2\pi}y_2\left(dz_2+B_3^\text{cl}\right)~.
\end{equation}
Such an action will be relevant {\it e.g.}~for the reduction of 6d $(2,0)$ theory labeled by $\frak{g}=A_{N-1}$ to a 5d theory with gauge group $SU(N)/\Z_k$ via the pure polarization given by 
\begin{equation}
    L=\Z_{N/k}^{(0)}\times\Z_{k}^{(1)}\subset \Z_{N}^{(0)}\times\Z_{N}^{(1)}.
\end{equation}
The equation of motion from the boundary variation imposes $y_2$ to be a $\mathbb{Z}_k$-valued two-form gauge field, and
\begin{equation}
B_2|_\partial=y_2+\frac{1}{k}B_2^\text{cl},\quad
B_3|_\partial=\frac{k}{N}dz_2+\frac{k}{N}B_3^\text{cl}~.
\end{equation}
This means that the $\mathbb{Z}_k$ subgroup of $B_2$ is free on the boundary and the $\mathbb{Z}_{N/k}$ subgroup of $B_3$ is free on the boundary, while $B_2^\text{cl}$ and $B_2^\text{cl}$ are, respectively, background gauge fields for the $\mathbb{Z}_{N/k}$ one-form and $\mathbb{Z}_k$ two-form symmetries in the 5d $SU(N)/\mathbb{Z}_k$ gauge theory.
These symmetries have an 't Hooft anomaly; performing
\begin{equation}
 B_2^\text{cl}\rightarrow B_2^\text{cl}+d\lambda_1^\text{cl},\quad
 y_2\rightarrow y_2-\frac{1}{k}d\lambda_1^\text{cl},\quad
 B_3^\text{cl}\rightarrow B_3^\text{cl}+d\lambda_2^\text{cl},\quad
 z_2\rightarrow z_2-\lambda_2^\text{cl}
\end{equation}
leaves the bulk fields invariant but produce the phase on the boundary
\begin{equation}
-\frac{1}{2\pi}d\lambda_1^\text{cl}B_3^\text{cl}~,
\end{equation}
which is the signature of a mixed 't Hooft anomaly of the one-form and two-form symmetries. This mixed anomaly is controlled by the extension class of
\begin{equation}
    1\rightarrow\mathbb{Z}_k\rightarrow\mathbb{Z}_N\rightarrow \mathbb{Z}_{N/k}\rightarrow 1~,
\end{equation}
agreeing with the general description of the anomaly for $T[S^1]$ theories.

 We remark that one can think of the boundary $\mathbb{Z}_k$ two-form gauge field $y_2$ (the equation of motion of $z_2$ imposes it having $\mathbb{Z}_k$ holonomy) that couples to $B_3^\text{cl}$ as describing the $\mathbb{Z}_k$ discrete magnetic flux in the 5d $SU(N)/\mathbb{Z}_k$ gauge theory associated with $\pi_1(SU(N)/\mathbb{Z}_k)=\mathbb{Z}_k$.

\subsection{Symmetries in 5d gauge theory}

In general, a five-dimensional gauge theory with gauge group $G$ has the following symmetries
\begin{itemize}
    \item[(1)] $U(1)$ 0-form symmetry associated with the instanton number. 
    For 5d theories obtained from a 6d theory compactified on a circle, this is related to the isometry of the circle, {\it i.e.} the instanton charge can be identified with the Kaluza-Klein mode on the circle; both have mass of the order ${1\over g_5^2}={1\over R_6}$, where $g_5$ is the 5d gauge coupling and $R_6$ is the radius of the circle compactification \cite{Rozali:1997cb,Seiberg:1997ax,Douglas:2010iu,Lambert:2010iw,Kim:2011mv}.
    As we show in Section \ref{sec:5d3-groupfield} (see also Section \ref{sec:5d3-groupreduction}), this symmetry can mix with other symmetries to form a three-group.
    
    \item[(2)] For finite Abelian $\pi_1(G)$,    0-form symmetry $\Gamma(\pi_1(G))$, the universal quadratic group \cite{Whitehead:1950} of $\pi_1(G)$,
    generated by the codimension-one symmetry defect
\begin{equation}
\oint {\cal P}(w_2^G)~,
\end{equation}
where ${\cal P}$ is the generalized Pontryagin square operation, $w_2^G$ is the obstruction to lifting the $G$ gauge bundle to a $\tilde G$ bundle for the universal covering group $\tilde G$. 
    \item[(3)] $\CZ(G)$ electric one-form symmetry transforms the Wilson lines, with the one-form symmetry charge given by evaluating the representation of the Wilson line on $\CZ(G)$ \cite{Gaiotto:2014kfa}.
    
    \item[(4)] $\pi_1(G)$ magnetic two-form symmetry transforms the magnetic strings, with the two-form symmetry charge given by $\oint w_2^G$ for $S^2$ surrounding the magnetic string. 
    
\end{itemize}

Depending on $G$, the action of the 0-form symmetries (1) and (2) in the above list factors through the quotient
\begin{equation}
{U(1)\times \Gamma(\pi_1(G))\over {\cal A}}~,
\end{equation}
for some finite Abelian group ${\cal A}\subset \Gamma(\pi_1(G))$. For ${\frak g}=A_n,D_n,E_n$ the 0-form symmetry is summarized in Table \ref{tab:5dinstantonsymmetry}.
This comes from the relation between the instanton number and ${\cal P}(w_2^G)$, and it is related to the periodicity of theta angle in the 4d $G$ gauge theory (see {\it e.g.} \cite{Witten:2000nv,Aharony:2013hda,Cordova:2019uob}). It is also related to the anomaly in $\tilde G$ Chern--Simons theory for gauging $\pi_1(G)$ one-form symmetry which is isomorphic to a subgroup of $Z(\tilde G)$.

\begin{table}[t]
    \centering
    \begin{tabular}{c|c|c|c}
    $G$  &  $\Gamma(\pi_1(G))$ & ${\cal A}$ & relation \\ \hline
    $SU(N)/\mathbb{Z}_k$ odd $k$ & $\mathbb{Z}_{k}$
    &$\mathbb{Z}_{\ell}$        &  $Q_1=((k-1)/2)(N/k)Q_2$ mod $k$.\\
    $SU(N)/\mathbb{Z}_k$ even $k$  & $\mathbb{Z}_{2k}$ 
    &$\mathbb{Z}_{\ell'}$        &  $Q_1=(k-1)(N/k)Q_2$ mod $2k$.\\
    $SO(2n)$ & $\mathbb{Z}_4$ 
    & $\mathbb{Z}_2$        & $Q_1=2Q_2$ mod $4$\\
    $PSO(4n+2)$ & $\mathbb{Z}_8$ 
    & $\mathbb{Z}_8$        & $Q_1=(2n+1)Q_2$ mod $8$.\\
    $PSO(4n)$ & $\mathbb{Z}_4\times \mathbb{Z}_4\times \mathbb{Z}_2$ &
    $\mathbb{Z}_4$ or $\mathbb{Z}_2$ for odd/even $n$ & $Q_1=2Q_2^{11}+nQ_2^{22}+2Q_2^{12}$ mod $4$.\\
    $E_7/\mathbb{Z}_2$ & $\mathbb{Z}_4$ & $\mathbb{Z}_4$ & $Q_1=-Q_2$ mod $4$\\
    $E_6/\mathbb{Z}_3$ & $\mathbb{Z}_3$ & $\mathbb{Z}_3$ & $Q_1=-Q_2$ mod $3$.
    
    \end{tabular}
    \caption{The instanton (continuous and discrete) 0-form symmetry in 5d gauge theory with gauge group $G$ and without Chern--Simon term, given by $\left(U(1)\times \Gamma(\pi_1(G))\right)/{\cal A}$. In the fist row $\ell=\gcd(((k-1)/2)(N/k),k)$, $\ell'=\gcd((k-1)(N/k),2k)$.}
    \label{tab:5dinstantonsymmetry}
\end{table}

In addition, one can define $i$-dimensional symmetry defects using the SPT phase in $i$ space-time dimensions with $G$ symmetry as classified by $G^{(4-i)}=\Omega^i(BG)$, where the classification gives the symmetry group with group action given by stacking the SPT phases. The symmetry defect hosts $G$ gauge field on its worldvolume, with topological action specified by gauging the $G$ symmetry in the SPT phase. The symmetry (2) is a special case of this, and we will not focus on other defects in this family of defects in the discussion.

We remark that the symmetry in 5d compatification is also discussed in \cite{Morrison:2020ool,Albertini:2020mdx}. Here we mainly focus on the compactification of 6d ${\cal N}=(2,0)$ theory. However, since the discussion does not rely on supersymmetry in any essential way, it can be straightforwardly generalized to more general 6d theories discussed in \cite{Morrison:2020ool,Albertini:2020mdx}. Moreover, here we clarify the precise global structure such as the higher-group symmetry.

In Appendix \ref{sec:5ddiscretetheta}, we show that the discrete theta angle does not affect the global symmetry discussed above (including its 't Hooft anomaly).
On the other hand, it affects the correlation function of the magnetic strings charged under the magnetic two-form symmetry.\footnote{
This can be seen as follows. Take the space-time to be $S^5$ and consider magnetic strings inserted at $\gamma_2,\gamma_2'$, with $\gamma_2=\partial V_3$, $\gamma_2'=\partial V_3'$.
Then $w_2^k=\delta(V_3^\perp)+\delta(V_3'^\perp)$, and the discrete theta angle gives the phase
\begin{equation}
\exp\left(\frac{2\pi i p}{k^2}\int \delta(V_3^\perp)\delta(\gamma_2')+\delta(V_3'^\perp)\delta(\gamma_2)\right)
=\exp\left(\frac{2\pi i p}{k^2}\int \#(V_3,\gamma_2')\right)
=\exp\left(\frac{2\pi i p}{k^2}\text{Link}(\gamma_2,\gamma_2')\right).
\end{equation}
}

In general, there can also be other discrete theta angles (such as those associated with $\pi_4(G)$), but they may not come from any choice of polarization.
In what follows we will not discuss such discrete theta angles.
Here, this can be achieved by choosing suitable ${\frak g}$ such that $\pi_4({\tilde G})=0$ for the corresponding simply connected group ${\tilde G}$ (for instance, any classical Lie algebras that are not $\mathfrak{sp}(n)$).

Since $Sp(n)$ has a $\mathbb{Z}_2$ center, one might wonder whether in $Sp(n)$ gauge theory there is a mixed anomaly or symmetry extension that involves the electric one-form symmetry and the discrete theta angle classified by $\pi_4(Sp(n))=\mathbb{Z}_2$. Since $\pi_4(Sp(n)/\mathbb{Z}_2)=\mathbb{Z}_2$, the discrete theta angle is well-defined when the bundle is modified to be an $Sp(n)/\mathbb{Z}_2$ bundle by the background gauge field of the electric one-form symmetry. Thus the discrete theta angle $\pi_4(Sp(n))$ does not modify other global symmetries or their 't Hooft anomalies.

\subsubsection*{${\cal P}(w_2^G)$ 0-form symmetry and magnetic strings}

Let us make a few more comments on the symmetry generated by ${\cal P}(w_2^G)$. In some cases, the charged operator can be identified with the instanton operator (modulo an integer).
Generally, the charged objects are not identified with the instanton operator. They can be understood as follows. Take two linked magnetic strings that carry magnetic flux $\oint w_2^G$ and un-link them. In the process there appears a singular point that belongs to both magnetic strings. This point supports an operator that is charged under the 0-form symmetry generated by $\oint {\cal P}(w_2^G)$.

What happen if we gauge the 0-form symmetry ${\cal P}(w_2^G)$? This modifies the gauge bundle to have trivial ${\cal P}(w_2^G)$.

\subsection{'t Hooft anomaly and 3-group symmetry in 5d gauge theory}\label{sec:5d3-groupfield}

Let us begin with the case of a simply connected $G$, {\it i.e.}~$\pi_1(G)=0$.
In the presence of a background field $B_2$ for the electric one-form symmetry, the quantization of the gauge field is modified. 
It is known that for simply connected Lie groups the instanton number becomes fractional in $G/\CZ(G)$ bundles with $B_2$ the obstruction to lifting them to $G$-bundles. The fractional part is related to the Pontryagin square ${\cal P}(B_2)$, where ${\cal P}:H^2(M,\CZ(G))\rightarrow \Gamma(\CZ(G))$ with $\Gamma(\CZ(G))$ the quadratic group for $\CZ(G)$.
Thus, the 0-form instanton symmetry current needs to be modified to
\begin{equation}
j_1=\star \left(p_1-\ell{\cal P}(B_2)\right)~,
\end{equation}
where $\ell$ is a map from the quadratic group $\Gamma(\CZ(G))$ to $\mathbb{R}$ that determines the fractional part of the instanton number, we pick a lift of ${\cal P}(B_2)$ to an integral cochain. It is not an integral cocycle, which means that the current is no longer conserved, but violated by an operator proportional to the identity and depends on the background gauge fields:
\begin{equation}
d\star j_1=-\ell d{\cal P}(B_2)~,
\end{equation}
where the right hand side can be expressed using the image of Bockstein homomorphism for ${\cal P}(B_2)$ in $\mathbb{Z}^r\rightarrow \mathbb{Z}^r\rightarrow \CZ(G)$ where $B_2$ takes value in the finite Abelian group $\CZ(G)=\prod_{i=1}^r\mathbb{Z}_{n_i}$.
The anomaly can be described by the 6d SPT phase
\begin{equation}
    \ell\int B_1 d{\cal P}(B_2)~,
\end{equation}
where $B_1$ is the background gauge field for the $U(1)$ instanton number symmetry.

Now let us consider the case $\pi_1(G)\neq 0$. Then, the above discussion applies with $B_2$ replaced by
\begin{equation}
\tilde B_2= \iota(w_2) + B_2~,
\end{equation}
where $\iota$ is the inclusion of $C = \pi_1(G)$ in $\CZ(\tilde G)$ and $\tilde G$ is the simply connected form of $G=\tilde G/C$. $w_2$ is the obstruction to lifting the $G=\tilde G/C$ bundle to a $\tilde G$ bundle.
Then, depending on the quadratic function in the Pontryagin square the conservation of the 0-form symmetry current is violated by an operator, and there is a 3-group symmetry. In the following we will discuss several examples: $SU(N)/\mathbb{Z}_k$ gauge theory (with $k$ a divisor of $N$), $SO(2n)$ gauge theory and $U(N)$ gauge theory.

The 6d origin of the three-group symmetry and its anomaly in 5d was explained in section \ref{sec:5d3-groupreduction} from the point of view of the dimensional reduction of the 2-form symmetry in 6d.

During the completion of this work we notice another paper \cite{BenettiGenolini:2020doj} appeared that also discussed a mixed anomaly between the instanton number symmetry and the center one-form symmetry in $SU(N)$ gauge theory. In such case, there is no three-group symmetry.

\subsubsection{Three-group symmetry in 5d $SU(N)/\mathbb{Z}_k$ gauge theory}

Consider a compactification of the 6d theory with ${\frak g}=\mathfrak{su}(N)$. We choose the pure polarization given by $L=\mathbb{Z}^{(0)}_{N/k}\times \mathbb{Z}_k^{(1)}\subset H^*(S^1,\Z_N)$, which leads to an $SU(N)/\mathbb{Z}_k$ gauge theory in 5d.

The theory has $\mathbb{Z}_{N/k}$ electric one-form symmetry and $\mathbb{Z}_k$ magnetic two-form symmetry, and we denote their backgrounds by $B_2,B_3$.
The gauge bundle can be characterized by $w_2^{k}$ which is the obstruction to lifting the $SU(N)/\mathbb{Z}_k$ bundle to an $SU(N)$ bundle.
In the presence of the background $B_2$, the bundle is modified to a $PSU(N)$ bundle, where $w_2^k$ is no longer closed but satisfies
\begin{equation}
\delta w_2^k = \text{Bock}(B_2)~,
\end{equation}
and $\delta$ is the differential (the coboundary operator for $C^*(M_5,\mathbb{Z}_k)$ on space-time $M_5$), and Bock is the Bockstein homomorphism for $1\rightarrow\mathbb{Z}_{k}\rightarrow\mathbb{Z}_N\rightarrow\mathbb{Z}_{N/k}\rightarrow 1$.
It describes the $PSU(N)$ bundle with the $\mathbb{Z}_N$ magnetic flux
\begin{equation}
w_2^N= \frac{N}{k}w_2^k-B_2
\end{equation}
which is the $\mathbb{Z}_N$ cocycle that describes the obstruction to lifting the $PSU(N)$ bundle to an $SU(N)$ bundle.

The $PSU(N)$ bundle has fractional instanton number compared to the normalizaion of $SU(N)/\mathbb{Z}_k$ bundle.
In comparison to the $SU(N)$ bundle, the fractional part is given by
\begin{equation}
 \frac{-1}{2N}\int {\cal P}(w_2^N)=
{N(-1)/k\over 2k}\int {\cal P}(w_2^k)
-\frac{-1}{k}\int w_2^k\cup B_2+\frac{-1}{2k}\int {\cal P}(B_2)~.
\end{equation}
Thus, the instanton number for $SU(N)/\mathbb{Z}_k$ corresponds to the above multiplied by $\frac{k}{\gcd(k,N/k)}$, so that it is an integer when $B_2=0$.

In the presence of the background $B_1$ for the 0-form instanton symmetry, the coupling to $B_1$ is not well-defined. We can extend the fields to a bulk
\begin{equation}
\frac{k}{8\pi^2 \gcd(k,N/k)}\int \text{Tr }F\wedge F dB_1
=\int\left(
-\frac{-1}{\gcd(k,N/k)} w_2^k\cup B_2+\frac{-1}{2\gcd(k,N/k)} {\cal P}(B_2)
\right)dB_1~.
\end{equation}
The terms involving $w_2^k$ represents a gauge-global anomaly, which can be cancelled by
\begin{equation}\label{eqn:sunkmag}
\frac{2\pi}{k}\int w_2^k B_3
\end{equation}
with the condition
\begin{equation}
\delta B_3={k\over \gcd(k,N/k)}B_2\frac{dB_1}{2\pi}\bmod k~.
\end{equation}
The above equation describes a three-group symmetry.
Note the three-group symmetry is non-trivial only when $\gcd(k,N/k)\neq 1$, namely when the extension $\mathbb{Z}_k\rightarrow\mathbb{Z}_N\rightarrow \mathbb{Z}_{N/k}$ does not split.

The three-group symmetry has an 't Hooft anomaly described by the SPT phase
\begin{equation}
\int \frac{2\pi}{k}\text{Bock}(B_2)B_3
+\frac{-1}{2\gcd(k,N/k)}\int {\cal P}(B_2)dB_1~.
\end{equation}
where the first term comes from $\int \delta w_2^k B_3$ in (\ref{eqn:sunkmag}).

One can also turn on $\mathbb{Z}_{2k}$ background $B_1^{\cal P}$ for the 0-form symmetry generated by $\oint {\cal P}(w_2^k)$.
Then, from \cite{Hsin:2020nts} we find the 3-group symmetry has an additional term
\begin{equation}
\delta B_3={k\over \gcd(k,N/k)}B_2\frac{dB_1}{2\pi}
+B_1^{\cal P}\text{Bock}(B_2)
\bmod k~.
\end{equation}
The 't Hooft anomaly also has an additional term
\begin{equation}
\int \frac{2\pi}{k}\text{Bock}(B_2)B_3
+\frac{-1}{2\gcd(k,N/k)}\int {\cal P}(B_2)dB_1
+\frac{2\pi}{2k}\int B_1^{\cal P}\text{Bock}(B_2)\cup_1 \text{Bock}(B_2)
~.
\end{equation}

\subsubsection{Three-group symmetry in 5d $SO(2n)$ gauge theory}
\label{sec:5dSO3group}

Consider the compactification of the 6d theory with ${\frak g}=\mathfrak{so}(2n)$. As we have seen in Section~\ref{sec:compactification6d}, in 6d, one can choose a polarization given by $\Z_2\subset D=\Z_2\times \Z_2$ which leads to the 6d $SO(2n)$ theory. Its dimensional reduction gives an $SO(2n)$ gauge theory in 5d, labeled by the pure polarization $\CP_L$ with $L=\Z^{(0)}_2\times\Z^{(1)}_2\subset H^*(S^1,\Z_2\times \Z_2)$. 

The theory has $\mathbb{Z}_2$ electric one-form symmetry and $\mathbb{Z}_2$ magnetic two-form symmetry, with background fields $B_2$ and $B_3$ respectively.
In the presence of a background $B_2$ the instanton number over a general closed four-manifold in space-time can become fractional,
\begin{equation}
\frac{1}{8\pi^2}\oint\text{Tr }F\wedge F \equiv \frac{1}{2}\oint w_2^{SO}\cup B_2+{2n\over 16}\oint {\cal P}(B_2) \pmod 1~.
\end{equation}
where $w_2^{SO}$ is the obstruction to lifting the $SO(N)$ bundle to an Spin$(N)$ bundle.

Denote the background for the 0-form instanton number symmetry by $B_1$. Due to the fractional instanton number, the coupling to $B_1$ is not well-defined. We can extend the fields to the bulk
\begin{equation}
\frac{1}{8\pi^2}\int \left(\text{Tr }F\wedge F\right) dB_1
=\int\left(
{1\over 2} w_2^{SO}\cup B_2+\frac{2n}{16}\int {\cal P}(B_2)
\right)dB_1~.
\end{equation}
The term involving $w_2^{SO}$ represents a gauge-global anomaly, and it can be cancelled by
\begin{equation}\label{eqn:so2nmag}
\pi\int w_2^{SO}B_3
\end{equation}
with the condition
\begin{equation}
\delta B_3=B_2\frac{dB_1}{2\pi}~.
\end{equation}
Thus, the symmetries combine into a three-group symmetry.

The three-group symmetry has an 't Hooft anomaly described by the SPT phase
\begin{equation}
    n\pi\int \text{Bock}(B_2)B_3
+    \frac{2n}{16}\int {\cal P}(B_2)dB_1~,
\end{equation}
where the first term comes from $\int \delta w_2^{SO} B_3$ in (\ref{eqn:so2nmag}).

One can also turn on $\mathbb{Z}_{4}$ background $B_1^{\cal P}$ for the 0-form symmetry generated by $\oint {\cal P}(w_2^{SO})$.
Then from \cite{Hsin:2020nts} we find the 3-group symmetry has an additional term
\begin{equation}
\delta B_3=B_2\frac{dB_1}{2\pi}+B_1^{\cal P}\text{Bock}(B_2)~.
\end{equation}
The 't Hooft anomaly has an additional term
\begin{equation}
    n\pi\int \text{Bock}(B_2)B_3
+    \frac{2n}{16}\int {\cal P}(B_2)dB_1
+\frac{\pi}{2}\int B_1^{\cal P}\text{Bock}(B_2)\cup_1 \text{Bock}(B_2)
~.
\end{equation}

\subsubsection{Three-group symmetry in 5d $U(N)$ gauge theory}
\label{sec:3group}

Consider the compactification of the 6d theory with ${\frak g}=\mathfrak{u}(N)$. The theory is absolute in 6d and reduces to a $U(N)$ gauge theory in 5d.

The $U(N)$ gauge theory has instanton 0-form symmetry associated with rotation of the circle used in the reduction.
The corresponding current is
\begin{equation}
j=\star \frac{1}{8\pi^2}\text{Tr}\, F\wedge F~.
\end{equation}
In addition, the theory has $U(1)$ magnetic two-form symmetry with the current
\begin{equation}
j_3=\star \text{Tr }\frac{F}{2\pi}~.
\end{equation}
The theory also has $U(1)$ center one-form symmetry. If we turn on background $B_2$ for this symmetry, the quantization of the gauge field is modified,
\begin{equation}
\text{Tr }F\equiv NB_2\pmod {2\pi\mathbb{Z}}~.
\end{equation}
This modifies the currents $j_1$ and $j_3$ as follows.
Since the magnetic charges become fractional, we need to modify the current $j_3$ as
\begin{equation}
j_3=\star \text{Tr }\frac{F-B_2\mathbf{1}_N}{2\pi}~.
\end{equation}
However, the current is no longer conserved
\begin{equation}
d\star j_3=-N\frac{dB_2}{2\pi}~.
\end{equation}
This represents an 't Hooft anomaly.

In the presence of the background $B_2$, the instanton number becomes fractional,
\begin{equation}
\frac{1}{8\pi^2}\int\text{Tr } (F-B_2\mathbf{1}_N)\wedge (F-B_2\mathbf{1}_N)
=\frac{1}{8\pi^2}\int\text{Tr } F\wedge F
-\frac{1}{4\pi^2}\text{Tr }FB_2+\frac{N}{4\pi^2}B_2B_2
\equiv 0\pmod {2\pi\mathbb{Z}}~.
\end{equation}
Thus, the current for the 0-form symmetry needs to be modified as
\begin{equation}
j_1=\star\left( \frac{1}{8\pi^2}\text{Tr} (F-B_2\mathbf{1}_N)\wedge (F-B_2\mathbf{1}_N)
\right)~.
\end{equation}
However, the current is no longer conserved: it is violated by a non-trivial operator $j_3$ 
\begin{equation}
d\star j_1=\frac{1}{4\pi^2}\text{Tr }(F-B_2\mathbf{1}_N)dB_2-\frac{N}{4\pi^2}B_2dB_2
={1\over 2\pi}\star j_3 dB_2-\frac{N}{4\pi^2}B_2dB_2~.
\end{equation}
The classical term in the above represents an 't Hooft anomaly, while the term with non-trivial operator represents a modification of global symmetry: the symmetries combine into a 3-group.
If we turn on the backgrounds for the 0-form and 2-form symmetries $B_1,B_3$ that couple to the currents as $\int B_1\star j_1+B_3\star j_3$, then the current conservation implies the backgronds obey the relation
\begin{equation}\label{eqn:3-group5dU(1)}
dB_3=-\frac{dB_2}{2\pi}B_1
\end{equation}
with the 0-form symmetry gauge transformation
\begin{align}
B_1\rightarrow B_1+d\lambda,\quad
B_3\rightarrow B_3 - {dB_2\over 2\pi}\lambda~.
\end{align}
The 't Hooft anomaly for the 3-group symmetry is described by the bulk SPT phase
\begin{equation}
N\int \frac{dB_2}{2\pi}B_3+\frac{N}{(2\pi)^2}dB_2B_2B_1~.
\end{equation}

\subsection{Effect of discrete theta angle on symmetry}

Depending on the polarization chosen, the 5d $G$ gauge theory can have discrete theta-angles for the torsion part of $\pi_1(G)$, written as $\prod_I\mathbb{Z}_{N_I}$:
\begin{equation}
\sum_{I<J}\frac{2\pi q_{IJ}}{N_I}\int w_2^I\cup \frac{\delta w_2^J}{N_J}~,
\end{equation}
where $w_2^I$ is the $\mathbb{Z}_{N_I}$ class that measures the discrete magnetic flux.
As discussed in Appendix \ref{sec:5ddiscretetheta}, the discrete theta angle is non-trivial only for $I\neq J$. Moreover, it does not affect the symmetry or anomaly for the backgrounds (that modify the cocycle condition $\delta w_2^I=Y_3^I$)
\begin{equation}
    Y_3^I=\text{Bock}(B_2^I)~,
\end{equation}
where $B_2^I$ is the background for electric one-form symmetry, and Bock is the Bockstein homomorphism for the exact sequence $1\rightarrow \CZ(G)\rightarrow \pi_1(G)\rightarrow \pi_1(G)/\CZ(G)\rightarrow 1$. We note that such $Y_3^I$ can be lifted to an integral cocycle, and $\int_{6\text{d}} Y_3^IY_3^I$ is trivial on orientable manifolds, and thus it does not contribute to the symmetry extension (\ref{eqn:5ddiscretethetasym}) and to the 't Hooft anomaly (\ref{eqn:5ddiscretethetaanomaly}).

If the background $Y_3^I$ is not in the above form, then the discrete theta angle can modify the symmetry and anomaly.
This is the case in $G'\times SO(2n)$ gauge theory for the torsion part of $\pi_1(G')$ containing a $\mathbb{Z}_2$ subgroup, and we include charge conjugation 0-form symmetry that acts on the $SO(2n)$ gauge theory. As discussed in \cite{Hsin:2020nts}, the magnetic flux in $SO(2n)$ gauge theory obeys
\begin{equation}
\delta w_2^{SO}=\frac{n}{2}\text{Bock}(B^e)+B^eB^{\cal C}~,
\end{equation}
where $B^{\cal C}$ is the background for the charge conjugation symmetry, and $B^e$ is the background for the electric one-form symmetry.
Comparing with (\ref{eqn:5ddiscretetheta2formgaugeY3}) we can identify the background
\begin{equation}
    Y^{SO}_3=\frac{n}{2}\text{Bock}(B^e)+B^eB^{\cal C}~.
\end{equation}
Then, the mixed discrete theta angle $q_{I,SO}$ between the discrete magnetic fluxes in the $SO$ gauge theory and the $G'$ gauge theory ($I$ labels the cyclic factors in the torsion part of $\pi_1(G')$) modifies the three-group symmetry in $G'$ gauge theory by an additional term in (\ref{eqn:5ddiscretethetasym}),
\begin{equation}
\delta B_3^I \; = \; \big( \text{as in } q_{IJ}=0\text{ theory} \big) + q_{I,SO}\text{Bock}(B^eB^{\cal C})~.
\end{equation}
The discrete theta angle also contributes to an additional 't Hooft anomaly (\ref{eqn:5ddiscretethetaanomaly}), with $Y^{SO}_3=\frac{n}{2}\text{Bock}(B^e)+B^eB^{\cal C}$:
\begin{equation}
-\sum_I\frac{2\pi q_{I,SO}}{2N_I}\int_{6\d} \left(Y_3^I Y_3^{SO}-\delta Y_3^I\cup_1 Y_3^{SO}\right)~.
\end{equation}

For instance, consider a $SO(2n)\times SO(2n)$ gauge theory in 5d. 
There is a discrete theta angle
\begin{equation}
    p{\pi\over 2}\int w_2^{(1)}\delta w_2^{(2)}~,
\end{equation}
where $w_2^{(1)},w_2^{(2)}$ are the second Stiefel--Whitney classes for the $SO(2n)\times SO(2n)$ gauge bundles.
The theory has a three-group symmetry, for which the background fields satisfy
\begin{align}
    &\delta B_3^{(1)}=B_2^{(1)}\frac{dB_1^{(1)}}{2\pi}+\text{Bock}\left(B_2^{(2)}B^{{\cal C},(2)}\right)\cr
    &\delta B_3^{(2)}=B_2^{(2)}\frac{dB_1^{(2)}}{2\pi}+\text{Bock}\left(B_2^{(1)}B^{{\cal C},(1)}\right)~,
\end{align}
where $B_2^{(I)}$  are the backgrounds for the electric one-form symmetry, $B^{{\cal C},(I)}$ are the backgrounds for the charge conjugaion 0-form symmetry, $B_1^{(I)}$ is the background for the instanton symmetry, and $I=1,2$ labels the two $SO(2n)$ gauge bundles.

\subsection{Implications for RG flows}
\label{sec:3-groupdynamics}

Let us explore some consequences of the higher-group symmetry involving 0, 1, and 2-form symmetries for an RG flow to a UV fixed point.

If a symmetry is broken explicitly, we must fix its background gauge field.
Then, since the three-group symmetry implies that a background gauge transformation for 0-form and 1-form symmetries induce an additional background for the 2-form symmetry, this places constraints on the breaking of 2-form symmetry and the breaking of 0-form and 1-form symmetries.

\begin{figure}
  \centering
    \includegraphics[width=0.5\textwidth]{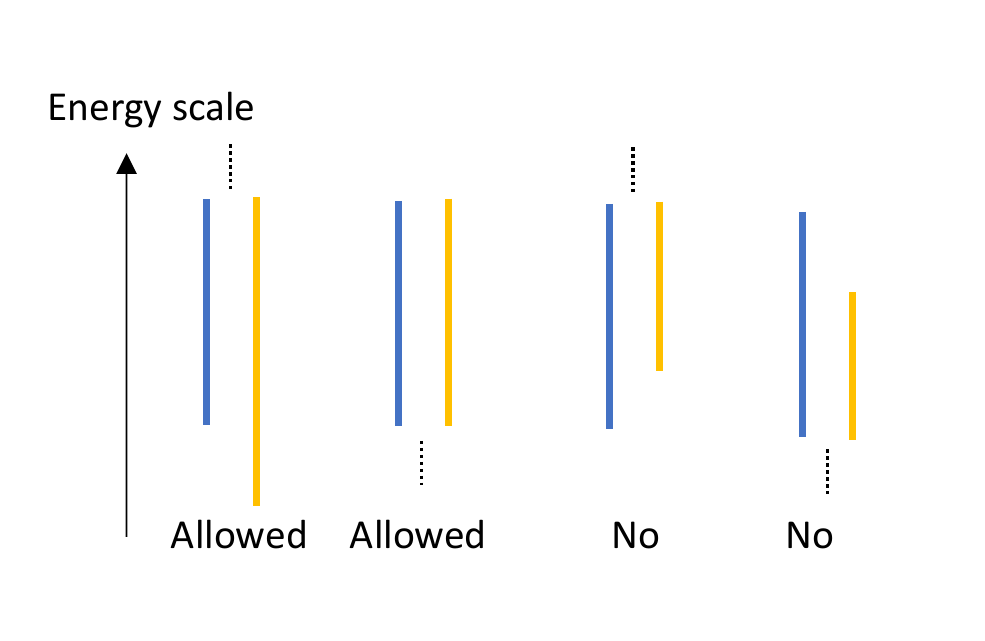}
    \caption{Examples of possible and impossible symmetry breaking patterns constrained by the 3-group symmetry. The blue (left) segment is the energy scale where the 2-form symmetry is broken, while the orange (right) segment is the energy scale where the 0-form or 1-form symmetry is broken. Namely, when the 2-form symmetry is broken the 0-form or 1-form symmetry must also be broken, but not vice versa.}\label{fig:3groupRG}
\end{figure}

For instance, the 3-group symmetry excludes the following scenarios: (see also Figure~\ref{fig:3groupRG})
\begin{itemize}
    \item The 2-form symmetry is broken at a scale below $\Lambda$,
    while the 0-form and 1-form symmetries are broken only at lower scale $\Lambda'<\Lambda$.
    Then, in the window between $\Lambda$ and $\Lambda'$ we have a trivial background for the 2-form symmetry but non-trivial backgrounds for 0-form and 1-form symmetries, which is inconsistent with the 3-group symmetry.
    
    \item The 2-form symmetry is broken at energy scales above $\Lambda$, 
    while the 0-form and 1-form symmetries are broken at higher scales $\Lambda'>\Lambda$. Then, in the window between $\Lambda$ and $\Lambda'$ we again have a broken 2-form symmetry but unbroken 0-form and 1-form symmetries, inconsistent with the 3-group structure.

    \item The 0-form and 1-form symmetries are unbroken at all energy scales while the 2-form symmetry is broken at some energy interval (explicitly or spontaneously).
    
    \item The 2-form symmetry is broken only above energy scale $\Lambda$, while the 0-form and 1-form symmetries are broken only below energy scale $\Lambda'<\Lambda$.
    
    \item The 2-form symmetry is broken only below energy scale $\Lambda$, while the 0-form and 1-form symmetries are broken only above energy scale $\Lambda'>\Lambda$.
    
\end{itemize}

\subsection{Matter that breaks one-form symmetry by screening}
\label{sec:screeningmatter}

When there are matter fields in some representation of the gauge group, it can screen the line operator in the same representation. In particular, for representations transforming under the center of the gauge group the matter fields break the center one-form symmetry explicitly. However, when the matter field itself also transforms under a global flavor symmetry, one might be able to assign a quantum number to the line operator that includes also the quantum number of the flavor symmetry. For instance, if the flavor symmetry is $H=\tilde H/C$ with $C\subset \CZ(\tilde H)$ identified with a gauge rotation in the center of gauge group, then activating a background $H$ gauge field that is not an $\tilde H$ gauge field, as distinguished by discrete magnetic flux $w_2^H\in H^2(M,C)$, amounts to identifying the one-form symmetry background
\begin{equation}
    B_2=\iota (w_2^H)~,
\end{equation}
where $\iota$ is the inclusion $C\rightarrow \CZ(G)$. Then, all of the previous discussion applies with this identification.

For instance, consider $SO(2n)$ gauge theory with even $N_f$ matter fields in the vector representation.
The theory has $SO(N_f)$ flavor symmetry, and $\mathbb{Z}_2\subset SO(N_f)$ is identified with a $\mathbb{Z}_2$ gauge rotation.
Thus, the faithful flavor symmetry is $SO(N_f)/\mathbb{Z}_2=PSO(N_f)$. We can turn on a background gauge field $B_1^f$ for the faithful flavor symmetry that is not a background for $SO(N_f)$, controlled by the obstruction $w_2^f\in H^2(BPSO(N_f),\mathbb{Z}_2)$.

From the discussion in Section \ref{sec:5dSO3group}, the theory has a 3-group symmetry, with backgrounds that satisfy
\begin{equation}\label{eqn:3gropfund}
\delta B_3=
(B_1^f,B_1)^*\Theta_4
=(B_1^f)^*w_2^f\frac{dB_1}{2\pi}+B_1^{\cal P}\text{Bock}((B_1^f)^*w_2^f)~.
\end{equation}
where $\Theta_4\in H^4(BG^{(0)},G^{(2)})$,
with $G^{(0)}$ the entire 0-form symmetry and $G^{(2)}$ the 2-form symmetry.
The 't Hooft anomaly is given by
\begin{equation}
    \int n\pi\text{Bock}((B_1^f)^*w_2^f)B_3
+    \frac{2n}{16}\int {\cal P}((B_1^f)^*w_2^f)dB_1
+\frac{\pi}{2}\int B_1^{\cal P}\text{Bock}((B_1^f)^*w_2^f)\cup_1 \text{Bock}((B_1^f)^*w_2^f)
~.
\end{equation}

An important difference is that, in this class of examples, we have more local counterterms to cancel the potential 't Hooft anomaly.
For instance, we can add
\begin{equation}
    \frac{k_f}{8\pi^2}\int \text{Tr }\left( F^{PSO(N_f)}\wedge F^{PSO(N_f)}\right)B_1~,
\end{equation}
which cancels 't Hooft anomaly
\begin{equation}
{k_f\over 2} \int (w_2')^2 dB_1 +
 {k_f\over 2}\int w_2'w_2^{PSO(N_f)}dB_1+
 k_f\frac{N_f}{16}\int{\cal P}(w_2^{PSO(N_f)})dB_1~,
\end{equation}
where $w_2^{PSO}=(B_1^f)^*w_2^f$, and $w_2'$ is the obstruction to lifting $SO(N_f)$ bundle to $Spin(N_f)$ bundle. Unlike the 't Hooft anomaly, though, the 3-group symmetry is not modified by local counterterms of the background fields. 

Here (and further below, in Section \ref{sec:exmpSOQCD}) the discussion also applies to the ${\cal N}=1$ supersymmetric version of the gauge theory, where the flavor symmetry is $PSp(N_f/2)=Sp(N_f/2)/\mathbb{Z}_2$ instead of $PSO(N_f)$. The same argument applies and the conclusion above remains the same.

Another example is $U(N)$ gauge theory with $N_f$ matter fields in the fundamental representation.
The theory has $SU(N_f)/\mathbb{Z}_{N_f}=PSU(N_f)$ flavor symmetry in addition to the instanton number symmetry, where the $\mathbb{Z}_{N_f}$ in the center of $SU(N_f)$ is identified with a gauge rotation.
The background gauge field $B_1^f$ for $PSU(N_f)$ has $\mathbb{Z}_{N_f}$ discrete magnetic flux $w_2^f$ which is the obstruction to lifting the bundle to an $SU(N_f)$ bundle.
From the discussion in Section \ref{sec:3group}, the theory has three-group symmetry with background $B_3$ for the $U(1)$ 2-form magnetic symmetry, $B_1$ for the $U(1)$ instanton number symmetry, and $B_1^f$ for $PSU(N_f)$ flavor symmetry, they satisfy
\begin{equation}\label{eqn:UNNf}
    dB_3=-(B_1^f)^*\text{Bock}(w_2^f)B_1~.
\end{equation}

The 3-group symmetry has implications to the dynamics and renormalization group flows, as discussed in Section \ref{sec:3-groupdynamics}, with the 1-form symmetry replaced by flavor 0-form symmetry $PSp(N_f)$ or $PSU(N_f)$.

\subsection{Constraint on symmetry enhancement}

In a 5d gauge theory with a UV fixed point, it can happen that the symmetry is enhanced at the fixed point, see for instance \cite{Seiberg:1996bd,Tachikawa:2015mha} and references therein.

For instance, suppose the flavor symmetry and the instanton symmetry enhance to a larger group $G^{{0},\text{UV}}$.
In this case, there is an inclusion map
\begin{equation}
    \iota^{(0)}:\quad G^{{0},\text{IR}}\rightarrow G^{{0},\text{UV}}~.
\end{equation}
Furthermore, suppose that the 2-form symmetries in the IR and UV are related by
\begin{equation}
    \iota^{(2)}:\quad G^{(2),\text{IR}}\rightarrow G^{(2),\text{UV}}~.
\end{equation}
Then, the 3-group symmetry in the IR satisfies
\begin{equation}\label{eqn:UVIRenhancement3group}
    \iota^{(2)}_\star\Theta_4^{\text{IR}}=(\iota^{(0)})^* \Theta_4^{\text{UV}}~.
\end{equation}
In particular, for a 5d gauge theory that has a non-trivial 3-group symmetry with non-trivial $\Theta_4^\text{IR}$, the above equation implies that the UV symmetry must also be a non-trivial 3-group with a non-trivial $\Theta_4^{\text{UV}}$. In particular, this rules out the symmetry enhancement to $G^{{0},\text{UV}}$ if there is no $\Theta_4^\text{UV}$ that can satisfy (\ref{eqn:UVIRenhancement3group}).

\subsubsection{Example: $SO(2n)$ gauge theory with $N_f$ vector flavors}
\label{sec:exmpSOQCD}

To illustrate that the condition (\ref{eqn:UVIRenhancement3group}) is not vacuous, consider the IR theory to be $SO(2n)$ gauge theory with $N_f$ matter fields in the vector representation with $N_f$ even. As discussed in Section \ref{sec:screeningmatter}, the theory has 3-group symmetry that combines the 0-form symmetry
$G^{{0},\text{IR}}=PSO(N_f)\times U(1)$ and the 2-form symmetry  $G^{(2),\text{IR}}=\mathbb{Z}_2$ with the Postnikov class $\Theta_4^{IR}=w_2^{PSO(N_f)}\frac{dx}{2\pi}$, where $x$ is a generator of $H^1(U(1),U(1))$, and $w_2^{PSO(N_f)}$ is the obstruction to lifting the $PSO(N_f)$ bundle to an $SO(N_f)$ bundle.The $\mathbb{Z}_2\subset SO(N_f)$ transformation that flips the sign of the matter field is identified with an element in the center of the gauge group $SO(2n)$, and the faithful flavor symmetry is $(SO(N_f)/\mathbb{Z}_2)\times U(1)=PSO(N_f)\times U(1)$.

Then in the UV suppose the 2-form symmetry is enhanced from $\mathbb{Z}_2$ to $U(1)$, there cannot be enhanced 0-form symmetry $G^{(0),\text{UV}}$ that is a Lie group which is both connected and simply connected, since for such group the 3-group symmetry in the UV has vanishing Postnikov class that lives in $H^4(G^{(0),\text{UV}},U(1))=0$, which is inconsistent with (\ref{eqn:UVIRenhancement3group}).
Thus, the three-group symmetry in the IR forbids the above symmetry enhancement.


\subsubsection{Example: $U(N)$ gauge theory with $N_f$ fundamental flavors}

We can repeat the above discussion for $U(N)$ gauge theory with $N_f$ fundamental flavors. From the discussion in Section \ref{sec:screeningmatter}, the theory has three-group symmetry that combines the 0-form symmetry $G^{(0),\text{IR}}=PSU(N_f)\times U(1)$ and two-form magnetic symmetry $G^{(2),\text{IR}}=U(1)$. By a similar reasoning we find that the 3-group symmetry in the IR implies that if the 2-form symmetry remains $U(1)$, the 0-form symmetry cannot enhance in the UV to be any connected and simply connected Lie group.

\subsection{Adding 5d Chern--Simons term: symmetry and anomaly}

In principle, the 5d gauge theory can have Chern--Simons terms.  String theory realizations of 5d Chern--Simons terms are discussed {\it e.g.}~in~\cite{Intriligator:1997pq}. Such terms are absent in straightforward compactifications of 6d theories we are most interested in, but they can be generated along the RG flow by integrating out massive fermions coupled to ordinary gauge field (see {\it e.g.}~\cite{Bonetti:2013ela}), which we will discuss from the anomaly inflow viewpoint in Section~\ref{sec:5dCSinduced}. Therefore it is useful to understand them in order to get a better picture of the IR physics.  Here, we would like to focus on how Chern--Simons terms affect the global symmetry and its anomaly.

\subsubsection{1-form symmetry and 't Hooft anomaly in $SU(N)$ Chern--Simons theory}

The 5d Chern--Simons term is proportional to the symmetric 3-index tensor of the gauge algebra, $\text{Tr }t_a\{t_b,t_c\}$, where $t_a$ denotes a generator of the algebra.
Let us consider $SU(N)$ gauge theory with $N>2$. 
The 5d Chern--Simons term is
\begin{equation}
    k\text{CS}_5=2\pi\cdot\frac{k}{6(2\pi)^3}\int \text{Tr }F'^3
\end{equation}
where the integral is over a 6-manifold that bounds the 5-manifold.\footnote{
There are, in fact, non-spin$^c$ 5-manifolds that are not null-cobordant. In what follows the bulk can be viewed as merely a convenient way to package the anomalous transformations in 5d by inflow.}
This is a 5d theory since the right-hand side is trivial mod $2\pi\mathbb{Z}$ on closed 6-manifolds.
We will distinguish the two cases of even and odd $k$.
As we will show in the following, for even values of $k$ the theory is well-defined on general orientable manifold, while for odd $k$ additional structure such as spin$^c$ structure is required to define the Chern--Simons term.
In both cases, the one-form center symmetry is broken explicitly by the Chern--Simons term to $\mathbb{Z}_{\gcd(k,N)}$.
This agrees with \cite{Morrison:2020ool,Albertini:2020mdx,Bhardwaj:2020phs,BenettiGenolini:2020doj}.
We also compute the 't Hooft anomaly of the one-form symmetry, which is new.

For suitable matter content (such as the adjoint representation of $SU(N)$), the theory has electric $\mathbb{Z}_N$ one-form center symmetry, and we denote its background by $B_2$.
We embed $SU(N)$ gauge field into a $U(N)$ gauge field as
\begin{equation}
    a=a'^{SU}+{1\over N}\mathbf{1}_N B~,
\end{equation}
where $B$ is a background $U(1)$ gauge field that satisfies $B_2=\frac{1}{N}dB$. The one-form symmetry is $a\rightarrow a+\mathbf{1}_N \lambda,B_2\rightarrow B_2+d\lambda$.
In the presence of a general background $B_2$, the 5d Chern--Simons term is not well-defined, but instead depends on 6d bulk as
\begin{equation}
\frac{k}{6(2\pi)^3}\text{Tr}(F-B_2\mathbf{1})^3
=
\frac{k}{6(2\pi)^3}\left(\text{Tr }F^3-3\text{Tr }F^2B_2+2NB_2^3\right)~.
\end{equation}
We note that the second and third Chern classes for $U(N)$ bundle are
\begin{align}
&c_2=\frac{1}{2(2\pi)^2}\left(
-\text{Tr }F^2+(\text{Tr }F)^2
\right)
\cr
&c_3=\frac{1}{6(2\pi)^3}\left(
2\text{Tr }F^3-3\text{Tr }F^2(\text{Tr }F)+(\text{Tr }F)^3
\right)~.
\end{align}
Thus the bulk action $\frac{k}{6(2\pi)^3}\text{Tr}(F-B_2\mathbf{1})^3$ can be expressed as
\begin{equation}
2\pi\int \left( {k\over 2} c_3 + \Big( \frac{k}{2}N-k \Big) \frac{B_2}{2\pi} \Big( -c_2+{N^2B_2^2\over 2(2\pi)^2} \Big) +\frac{1}{6(2\pi)^3}\left(-\frac{k}{2}N^3+2kN\right)B_2^3\right)~.
\end{equation}
The gauge anomaly is
\begin{equation}\label{eqn:5dCSgaugeanom}
 2\pi\int\left( {k\over 2} c_3 -\Big( \frac{k}{2}N-k \Big) \frac{B_2}{2\pi}c_2\right).
\end{equation}

\subsubsection*{Even level}

For even $k$, the first term in (\ref{eqn:5dCSgaugeanom}) is trivial, and the second term is trivial for\footnote{
One can also introduce a one-form background $B_1$ coupled to the instanton number $c_2$ to cancel the gauge anomaly, but the cancellation requires $k B_2=dB_1$, so it still satisfies (\ref{eqn:5dCS1-form}).
}
\begin{equation}\label{eqn:5dCS1-form}
    \oint B_2\in \frac{2\pi}{\gcd(N,k)}\mathbb{Z}\subset \frac{2\pi}{N}\mathbb{Z}~.
\end{equation}
Thus, for even $k$, the center one-form symmetry is broken explicitly by the Chern--Simons term to $\mathbb{Z}_{\gcd(N,k)}\subset\mathbb{Z}_N$.

The 't Hooft anomaly is given by
\begin{equation}
2\pi\cdot\frac{kN(N-1)(N-2)/6}{(2\pi)^3}\int B_2^3~.
\end{equation}
It is generally non-vanishing for $\mathbb{Z}_{\ell}$ one-form symmetry with $\ell=\gcd(N,k)$. We remark that the anomaly is consistent with vanishing Chern--Simons term for $N\leq 2$.

\subsubsection*{Odd level}

For odd $k$, in order for the gauge anomaly (\ref{eqn:5dCSgaugeanom}) to vanish, the manifold must be equipped with additional structures.

Instead of defining the 5d Chern--Simons term using the bulk, we can study how it transforms under large gauge transformations, which produces a Wess--Zumino term in 4d of the form $\text{Tr}(g^{-1}dg)^5$ for transformation by $SU(N)$ valued field $g$.
As discussed in \cite{Lee:2020ojw}, for $N\geq 3$ the unit Wess--Zumino--Witten term is not an integer on general 5-manifold but requires a spin$^c$ structure. The background spin$^c$ connection $A$ satisfies $\oint dA/2\pi=\frac{1}{2}w_2$  (on a spin manifold we can set $A=0$); then, the following combination of Wess--Zumino--Witten terms is an integer on spin$^c$ 5-manifold
\begin{equation}
    \int_{5\d}\left(\Gamma_5+\Gamma_3\frac{dA}{2\pi}\right),\quad
    \Gamma_{2n-1}=\frac{1}{(2\pi)^n}\frac{(n-1)!}{(2n-1)!}\text{Tr}\left(g^{-1}dg\right)^{2n-1}~.
\end{equation}
This means that the following combination of Chern--Simons terms is gauge invariant --- but not separately gauge invariant! --- for odd level $k$:
\begin{equation}
k\int \left(\text{CS}_5+\frac{dA}{2\pi}\text{CS}_3\right)~.
\end{equation}
Let us repeat the analysis of coupling to background $B_2$, now with the second term $\text{CS}_3\frac{dA}{2\pi}$ included.
There is additional bulk dependence
\begin{equation}
2\pi k\int \left(
-c_2+\frac{N(N-1)}{8\pi^2}B_2^2
\right)\frac{dA}{2\pi}=\pi k \int c_2 w_2(TM)+\frac{Nk(N-1)}{4\pi}\int B_2B_2\frac{dA}{2\pi}~.
\end{equation}
Note
\begin{equation}
\pi\int (c_3+c_2w_2(TM))=\pi\int (c_3+Sq^2c_2)=\pi\int (w_6+Sq^2(w_4))=\pi\int w_2w_4=\pi\int c_2 {NB_2\over 2\pi}~.  
\end{equation}
Thus, the total gauge anomaly is
\begin{equation}
    kN\pi\int c_2 \frac{B_2}{2\pi}
    -2\pi (\frac{k}{2}N-k)\int \frac{B_2}{2\pi}c_2=k\int B_2 c_2 ~.
\end{equation}
The vanishing of gauge anomaly again requires (\ref{eqn:5dCS1-form}), namely the one-form symmetry is broken explicitly to $\mathbb{Z}_\text{gcd}(k,N)$.
The 't Hooft anomaly is
\begin{equation}
    2\pi\cdot\frac{kN(N-1)(N-2)/6}{(2\pi)^3}\int B_2^3+\frac{kN(N-1)}{8\pi^2}\int B_2^2 dA~.
\end{equation}

\subsubsection{1-form symmetry and 't Hooft anomaly in $U(1)$ Chern--Simons theory}

We can repeat the analysis for $U(1)$ Chern--Simons theory.
The Chern--Simons term can be written as
\begin{equation}
    k\text{CS}_5[U(1)]=2\pi\cdot \frac{k}{6(2\pi)^3}\int F^3~,
\end{equation}
where the integral is over a 6-manifold that bounds the 5-manifold.
On a general manifold, $k$ has to be a multiple of 6 for the Chern--Simons term to be well-defined.

If we turn on background gauge field $B_2$ for the $\mathbb{Z}_\ell\subset U(1)$ one-form symmetry, the bulk dependence is modified to be
\begin{equation}
    2\pi\cdot\frac{k}{6(2\pi)^3}\int (F-B_2)^3=
    2\pi\cdot\frac{k}{6(2\pi)^3}\int \left(F^3-3F^2B_2+3FB_2^2-B_2^3\right)~.
\end{equation}
The gauge anomaly is
\begin{equation}
    -\frac{k}{2(2\pi)^2}\int F^2B_2+\frac{k}{2(2\pi)^2}\int FB_2^2~.
\end{equation}
They can be cancelled by introducing a background field $B_1$ for the instanton number symmetry and $B_3$ for the magnetic two-form symmetry:
\begin{align}
&\frac{1}{2(2\pi)^2}\int_{5\d} (F-B_2)^2 B_1+\frac{1}{2\pi}\int_{5\d}(F-B_2)B_3\cr
&=\int_{6\d}\frac{1}{2(2\pi)^2}F^2dB_1-\frac{1}{(2\pi)^2}Fd(B_2B_1)+\frac{1}{2(2\pi)^2}d(B_2^2B_1)\cr
&+\int_{6\d}\frac{1}{2\pi}FdB_3-\frac{1}{2\pi}d(B_2B_3)~.
\end{align}
Thus, we find $\oint B_2\in \frac{2\pi}{k}\mathbb{Z}$, namely the one-form symmetry is broken by the Chern--Simons term to be $\mathbb{Z}_k$, and it participates in a three-group symmetry
\begin{equation}
dB_3=-\frac{k}{4\pi}B_2^2+\frac{1}{(2\pi)^2}d(B_2B_1)~.
\end{equation}

The 't Hooft anomaly of the symmetries is given by the bulk term
\begin{equation}
    -2\pi\cdot \frac{k}{6(2\pi)^3}\int_{6\d} B_2^3~.
\end{equation}

\subsubsection{Relation with chiral anomaly in 4d}
\label{sec:5dCSinduced}

The perturbative chiral anomaly in 4d is described by a 5d Chern--Simons term by anomaly inflow (or, from the modern point of view, the 4d 't Hooft anomaly is described by the 5d SPT phase given by the Chern--Simons term).
Thus, a well-defined $SU(N)/\mathbb{Z}_L$ Chern--Simons term describes the chiral anomaly for fermions in representation of $SU(N)/\mathbb{Z}_L$. This means that the 4d chiral fermions are in a representation of $SU(N)$ with $r$ boxes in the Young tableaux that satisfy $\gcd(r,N)=L$.
Well-defined 5d Chern--Simons term for $SU(N)/\mathbb{Z}_L$ requires 5d $SU(N)$ Chern--Simons term to have $\mathbb{Z}_L\subset \mathbb{Z}_N$ subgroup one-form symmetry and it is anomaly-free.
This imposes the following conditions on the 4d chiral anomaly coefficient $k$ (= the 5d Chern--Simons level):
\begin{equation}\label{eqn:quantizationsunmodL}
    \gcd(k,N)/L\in \mathbb{Z},\quad
    {kN(N-1)(N-2)\over 6L^3}\in \mathbb{Z}~,
\end{equation}
where we consider 5d Chern--Simons terms on spin manifolds since they describe the anomaly for 4d fermions defined on spin manifolds.

For instance, the 5d Chern--Simons term for 
$PSU(3)$ describes the perturbative anomaly for $PSU(3)$ symmetry in 4d, and one can check that the 4d fermions in $PSU(3)$ representations produce perturbative anomaly with coefficient that is always a multiple of $27$, with the quantization of the $PSU(3)$ Chern--Simons level in (\ref{eqn:quantizationsunmodL}) for $N=3,L=3$.\footnote{
For a representation with $SU(3)$ Dynkin labels $\lambda_1,\lambda_2$, the anomaly coefficient is 
\begin{equation}
    k^{(\lambda_1,\lambda_2)}=\frac{1}{120}(\lambda_1 + 1) (\lambda_1 + \lambda_2 + 2) (\lambda_2 + 1) (2 \lambda_1 + \lambda_2 + 3) (\lambda_1 + 2 \lambda_2 + 3) (\lambda_1 -\lambda_2)~.
\end{equation}
The $PSU(3)$ representations satisfy $2\lambda_1+\lambda_2\in 3\mathbb{Z}$. Let $\lambda_2=3n-2\lambda_1$, then 
\begin{equation}
    40 k^{(\lambda_1,\lambda_2)}=9(\lambda_1 + 1) (3n-\lambda_1+ 2) (3n-2\lambda_1+1) (n+1) (2n-\lambda_1+1) (\lambda_1-n)
\end{equation}
is divisible by 27 since at least one of $n+1$, $n-\lambda_1$, $2n-\lambda_1+1$ and $3n-\lambda_1+2$ is divisible by 3. (For example, if $n+1$ is not divisible by 3, then the rest three will have different residues mod 3, and one of them has to be divisible by 3).}
Similarly, the 5d Chern--Simons term for 
$PSU(4)$ describes the perturbative anomaly for $PSU(4)$ symmetry in 4d, and one can check that the 4d fermions in $PSU(4)$ representations such as  $\mathbf{20},\mathbf{35},\mathbf{15}$\footnote{
The corresponding Young tableaux are  $$\yng(2,2) \;, \qquad \yng(4) \;, \qquad \yng(2,1,1)  \;.$$
}
produce perturbative anomaly with a coefficient that is a multiple of 16, consistent with the quantization of the $PSU(4)$ Chern--Simons level in (\ref{eqn:quantizationsunmodL}) for $N=4,L=4$.

\subsubsection*{Induced Chern--Simons term from massive 5d fermion}

\begin{figure}[t]
  \centering
    \includegraphics[width=0.7\textwidth]{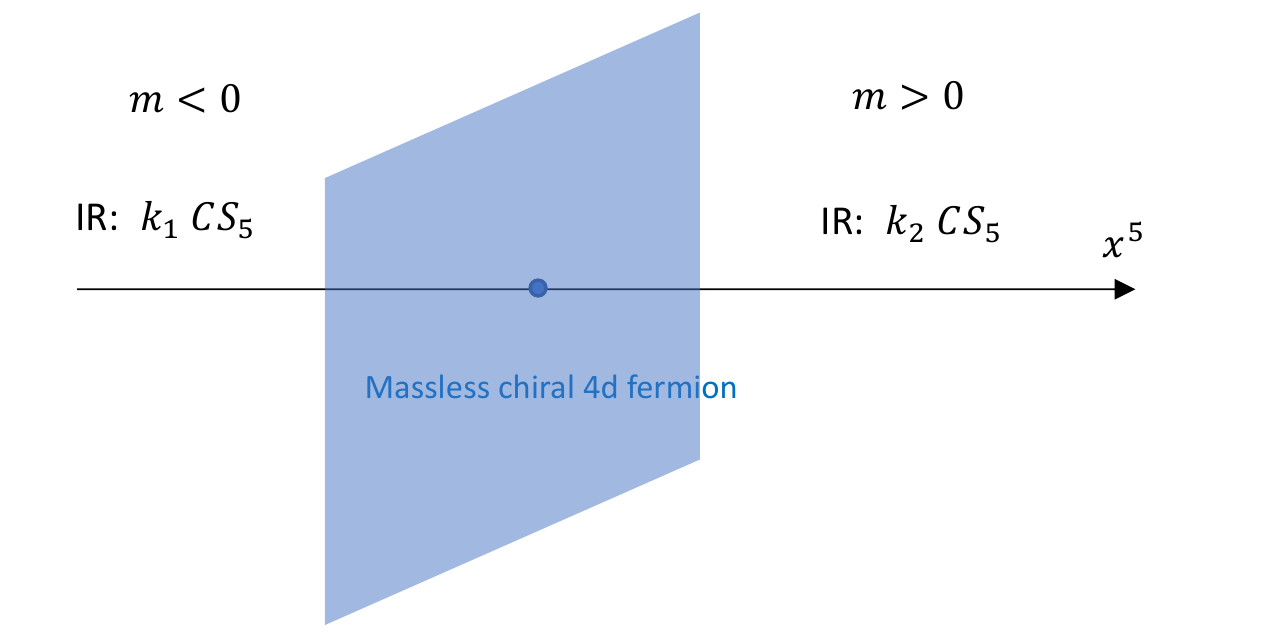}
      \caption{5d Chern--Simons term from a massive fermion in representation $R$ is given by the anomaly of a 4d chiral fermion in the same representation by anomaly inflow: $k_2-k_1=$ 4d chiral anomaly.}\label{fig:5dCSwall}
\end{figure}

The 4d chiral anomaly for fermion in representation $R$  also gives the 5d Chern--Simons term generated by integrating out massive 5d fermions in the same representation.
This can be understood as follows. Consider 5d fermion with a real mass profile interpolating between positive and negative value along some direction $x^5$:
\begin{equation}
    m(x^5)=\left\{\begin{array}{cl}
    -m_0     &\text{for } x^5<0 \\
    m_0     &\text{for } x^5>0
    \end{array}
    \right. ~.
\end{equation}
Then, the mass will vanish at the 4d interface $x^5=0$ (the profile can be smoothed out, so that there is some point where the fermion becomes massless but may not be exactly at $x^5=0$), which then hosts massless 4d chiral fermions.
The chiral anomaly on the interface should be compensated by anomaly inflow from the bulk on the two sides of the interface.
Thus after integrating out the massive 5d fermions with masses of opposite signs on the two sides of the interface, they differ by the induced Chern--Simons term at level $k$ that corresponds to the 4d chiral anomaly (see Figure~\ref{fig:5dCSwall}):
\begin{equation}
    \left\{\begin{array}{cl}
     m<0: & k_1\text{ CS}_5      \\
    m>0: & k_2\text{ CS}_5     
    \end{array}
    \right. ,\quad k_2-k_1=k=\text{chiral anomaly coefficient for }R~.
\end{equation}
Since the fermion partition function is given by the $\eta$-invariant, the result also follows from the Atiyah--Patodi--Singer index theorem \cite{atiyah_patodi_singer_1975} (see also \cite{AlvarezGaume:1984nf,Witten:2015aba}).

\subsubsection{$SU(N)/\mathbb{Z}_L$ Chern--Simons term}

Consider, for instance, an $SU(4)$ Chern--Simons term at even level with the non-anomalous one-form symmetry $\mathbb{Z}_2$. Then, one can gauge the one-form symmetry to obtain $SU(4)/\mathbb{Z}_2=SO(6)$ gauge theory with a Chern--Simons term.

When the level is a multiple of $4$, $k=4r$, the one-form symmetry is $\mathbb{Z}_4$. Its 't Hooft anomaly is
\begin{equation}
    \frac{r\pi}{2}\int (B_2^\text{dis})^3~,
\end{equation}
where $B_2^\text{dis}$ has holonomy $0,1,2,3$ mod $4$.
The $\mathbb{Z}_2$ subgroup with even $B_2^\text{dis}$ is non-anomalous as discussed above. 
When $r\not\in 4\mathbb{Z}$ the full $\mathbb{Z}_4$ one-form symmetry is anomalous.
We can gauge the $\mathbb{Z}_2$ subgroup one-form symmetry by writing $B_2^\text{dis}=2b_2+\tilde B_2$, where $\delta b_2=\text{Bock}(B_2)$ with $B_2=\tilde B_2$ mod 2.
For odd $r$ the original 't Hooft anomaly gives rise to a gauge anomaly
\begin{equation}
    \pi r \int b_2 B_2 B_2~.
\end{equation}
The gauge anomaly can be cancelled by new 2-form symmetry with background $B_3'$ that couples to $b_2$ as $\pi\int b_2 B_3'$. 
Then the 't Hooft anomaly converts into additional 3-group symmetry
\begin{equation}
    \delta B_3'=r B_2\cup B_2~.
\end{equation}

When $r\in 4\mathbb{Z}$ {\it i.e.}~level $4r\in 16\mathbb{Z}$, one can gauge the $\mathbb{Z}_4$ one-form symmetry to obtain $SU(4)/\mathbb{Z}_4=PSO(6)$ gauge theory with Chern--Simons term. Namely, consistent $PSU(4)$ Chern--Simons term has level quantized to be a multiple of $16$.

The discussion can be generalized to other values of $N$. For instance, if the level is $k=N^2r$ with $N=1,2,4,5$ mod 6 (so $(N-1)(N-2)\equiv 0$ mod 6), we can gauge the $\mathbb{Z}_N$ 1-form symmetry for any integer $r$ and obtain a properly quantized Chern--Simons term at level $N^2r$ for $SU(N)/\mathbb{Z}_N$.

\subsubsection{'t Hooft anomaly in 5d Chern--Simons matter theory}

When there are matter fields that breaks the one-form symmetry, we can apply the discussion of Section \ref{sec:screeningmatter}.
For instance, Consider $SU(4)$ gauge theory with Chern--Simons level $4r\in 4\mathbb{Z}$ and $N_f=8$ scalars in the fundamental representation $\mathbf{4}$.
The theory has $SU(N_f)/\mathbb{Z}_4$ flavor symmetry, whose background has $\mathbb{Z}_{4}$-valued discrete magnetic flux $w_2^{f}$ that obstructs lifting the bundle to an $SU(N_f)$ bundle.
We can apply the previous discussion with the identification $B_2^\text{dis}=w_2^{f}$.
The 't Hooft anomaly for the flavor symmetry is
\begin{equation}\label{eqn:5dcstHooft}
    {r\pi\over 2} \int (w_2^{f})^3~.
\end{equation}
We can partially cancel the 't Hooft anomaly by adding a Chern--Simons term for $SU(N_f)$ at level $4r'\in 4\mathbb{Z}$ (so $\mathbb{Z}_4$ is a subgroup of $\mathbb{Z}_{\gcd(4r',N_f)}$ with $N_f=8$), which contributes a bulk term
\begin{equation}
    r'\pi \int (w_2^{f})^3~.
\end{equation}
Thus, when $r$ is odd, the 't Hooft anomaly cannot be cancelled by a local counterterm and represents a genuine anomaly. In contrast, for even $r$ the 't Hooft anomaly can be cancelled by a local counterterm with $r'=-r/2$.

Another example is $SU(6)_{3k}$ with $N_f=9$ flavors in the fundamental representation. The flavor symmetry is $SU(9)/\mathbb{Z}_3$.
The Chern--Simons term contributes the bulk term
\begin{equation}
    2\pi\cdot \frac{2k}{9}\int (w_2^f)^3~.
\end{equation}
We can reduce the anomaly by turning on Chern--Simons level $3k_f$ for the background of the flavor symmetry, which contributes the bulk term
\begin{equation}
2\pi\cdot \frac{k_f}{3}\int  (w_2^f)^3~.
\end{equation}
Thus we find that when $k\in 3\mathbb{Z}$ the 't Hooft anomaly can be cancelled by a local counterterm of the flavor background gauge field, while for $k\not\in 3\mathbb{Z}$ the theory has an 't Hooft anomaly for the flavor symmetry.

When there are fermions, we can compute the anomaly by giving the fermions a mass that preserves the flavor symmetry, which leads to additional Chern--Simons term in the IR as discussed in section \ref{sec:5dCSinduced}.
We can also include a $U(1)$ baryon number symmetry as in \cite{Benini:2017dus} in the discussion of anomaly. We leave this to future work.

An 't Hooft anomaly constrains RG flows. For instance, if two 5d theories have different anomalies, they cannot arise from the same UV fixed point by RG flows that preserve the symmetries. This applies to the above examples, for instance $SU(4)$ gauge theories with Chern--Simons level $4r$: the two theories with even $r$ and odd $r$ cannot arise from the same UV fixed point by flows that preserve the $SU(N_f)/\mathbb{Z}_4$ flavor symmetry.

\section{A case study: $\mathfrak{so}(8)$ theories in 6d and 5d}
\label{sec:SO(8)}

In this section, we will combine what we learned in the previous sections to study in further details theories associated with the Lie algebra $\frak{so}(8)$ in 6d and 5d. 

We first recall a few facts from Section~\ref{sec:6dPartition}. The 6d $SO(8),Sc(8),Ss(8)$ theories can be obtained from the relative $\mathfrak{so}(8)$ theory by choosing one of the three polarizations given by the three $L=\Z_2$ subgroups of $D=\Z_2\times \Z_2$. For each choice there are two compatible $\Z_2$-valued quadratic functions on $H^3(M_6,L=\Z_2)$ given by $q(x)=0$ and $q(x)=w_3x$. For simplicity, we will choose the former while the discussion in this section can be directly generalized to the latter. 

For any choice of $L$, there is a $\bar L$ such that $D=L\oplus \bar L$. Therefore, there is always a non-anomalous $L^\vee=\mathbb{Z}_2$ 2-form symmetry.

One can also see this by studying the corresponding boundary conditions of the 7d Chern--Simons theory, which has action
\begin{equation}\label{eqn:Cartan}
\frac{2}{4\pi}C^1dC^1+\frac{2}{4\pi}C^2dC^2+\frac{2}{4\pi}C^3dC^3+\frac{2}{4\pi}C^4dC^4-\frac{1}{2\pi}C^1d\left(C^2+C^3+C^4\right)~.
\end{equation}
The on-shell fields satisfy $C^1=0,2C^2=2C^3=2C^4=0$, $C^4=C^2+C^3$, where the equations are in $\mathbb{R}/2\pi\mathbb{Z}$.
The on-shell action is
\begin{equation}
\frac{4}{4\pi}C^2dC^2+\frac{4}{4\pi}C^3dC^3+\frac{2}{2\pi}C^2dC^3~,
\end{equation}
which takes value in $2\pi\mathbb{Z}$, and thus the boundary condition $C^1|_\partial=0,C^2|_\partial=B,C^3|_\partial=B',C^4|_\partial=B+B'$ for $\mathbb{Z}_2$ gauge fields $B,B'$ corresponds to symmetry with trivial anomaly. 

However, there are two different lifts of $L^\vee$ to $D$, or in other words, two choices of $\bar L$. The difference between them is an SPT for the $\Z_2$ background field given by $Q$, the quadratic refinement for the $\Z_2$-valued intersection form on $M_6$. Such a difference is unimportant for many purposes, but to uniquely answer the question ``what happens after gauging the $\Z_2$ symmetry," one needs to make a choice (or in physics terminology, a choice of a ``local counter term'').

Denote the three $\Z_2$ subgroups of $D$ as $L_{ SO}$, $L_{ Ss}$ and $L_{ Sc}$. For the $SO(8)$ theory, $L=L_{ SO}$ and if one choose $\bar{L}=L_{ Sc}$, then gauging the 2-form symmetry with or without local counterterm gives respectively $Sc(8)$ or $Ss(8)$ theories. 

\begin{figure}[t]
  \centering
    \includegraphics[width=0.5\textwidth]{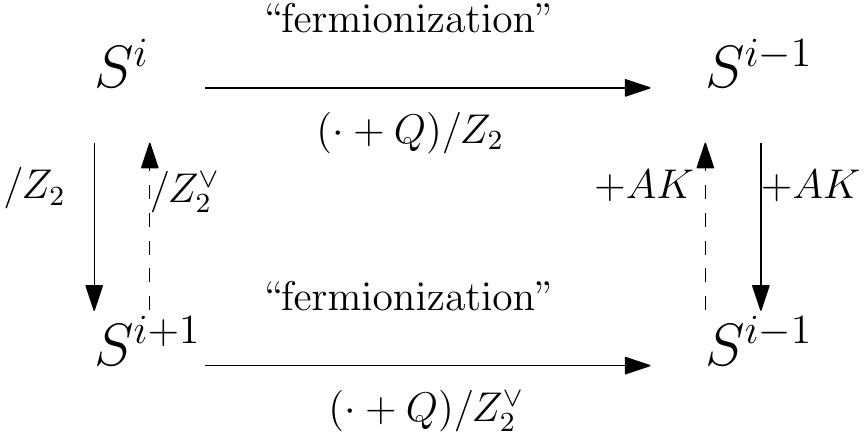}
    \caption{6d SCFTs with $\mathfrak{g}=\mathfrak{so}(8)$ from three polarizations $S^i=SO(8),Sc(8),Ss(8)$ for $i=0,\pm1$ mod 3 respectively.
    These three theories are all dual to one another by triality with the same partition function.
    The theories on the right are treated as a $v_4$-theory by tensoring with trivial $v_4$-SPT phase (for a detailed discussion of such SPT phases and their generalizations to $2n$ space-time dimension, see \cite{Hsin:2021qiy}), similar to Appendix C of \cite{Hsin:2019gvb}.
    The vertical arrows correspond to changing the polarizations of the 7d three-form Chern--Simons theory. This is a special case of the general relation between gauging $\mathbb{Z}_2$ $(n-1)$-form symmetry and stacking higher fermionic SPT phases in $2n$-dimensional space-time discussed in \cite{Hsin:2021qiy}.
    }\label{fig:SO(8)gauging}
\end{figure}

We will make a choice that is compatible with the ``cyclic'' symmetry. Namely, $L=L_{ Sc}$ and $\bar{L}=L_{Ss}$ for the $Sc(8)$ theory, and $L=L_{ Ss}$ and $\bar{L}=L_{SO}$ for the $Ss(8)$ theory. Then the behavior under gauging is summarized in Figure~\ref{fig:SO(8)gauging}, where $S^i$ with $i=0,1,-1$ stands for $SO(8),Sc(8),Ss(8)$.

The right-hand side of Figure~\ref{fig:SO(8)gauging} can be explained by looking at the partition function.
Denote the partition function of $S^i$ coupled to background $B_3$ for the two-form symmetry by $Z_i[B_3]$.
Then the upper horizontal arrow leads to the partition function for the theory at the upper-right:
\begin{equation}
Z_{\rm ur}=\sum_{b_3} Z_i[b_3](-1)^{Q(b_3)}~.
\end{equation}
This is a special case of the ``fermionization" procedure in $2n$ space-time dimensions discussed in \cite{Hsin:2021qiy} by gauging $\mathbb{Z}_2$ $(n-1)$-form symmetry with additional higher fermionic local counterterm.
The downward arrow on the right leads to the partition function $\sum_{b_3}Z_i[b_3]$, and the partition function coupled to background $B_3'$ for the dual two-form symmetry $\mathbb{Z}_2^\vee$ is $\sum_{b_3}Z_i[b_3](-1)^{\int b_3 B_3'}$.
The lower horizontal arrow gives the partition function for the theory at the lower-right
\begin{equation}
Z_{\rm lr}=\sum_{b_3,b_3'}Z_i[b_3](-1)^{\int b_3b_3'+Q(b_3')}=
\sum_{b_3,b_3'}Z_i[b_3](-1)^{Q(b_3)+Q(b_3'+b_3)}=Z_{\rm ur}\sum_{b_3''}(-1)^{Q(b_3'')}~,
\end{equation}
where $b_3''=b_3'+b_3$ and the factor $\sum_{b_3''}(-1)^{Q(b_3'')}$ is the Gaussian sum of the quadratic function, which is the Arf--Kervaire invariant of the manifold and thus is a $v_4$-fermionic SPT phase \cite{Hsin:2021qiy}.
The discussion is a special case of the computation discussed in \cite{Hsin:2021qiy} for gauging $\mathbb{Z}_2$ $(n-1)$-form symmetry in $2n$ space-time dimension.

In the 2d-3d analogue of Figure~\ref{fig:SO(8)gauging}, the theories $S^i,S^{i+1}$ are Ising and Ising$/\mathbb{Z}_2$ which are dual, while $S^{i-1}$ is the Majorana fermion obtained by orbifolding the Ising model with additional local countertem given by  $\mathbb{Z}_2$ fermion SPT phase.
The vertical arrows on the right are given by staking a $(1+1)$-dimensional invertible spin TQFT with effective action given by the 2d Arf invariant \cite{Ji:2019ugf}. In the 3d bulk $\mathbb{Z}_2$ gauge theory
 the vertical arrows correspond to electromagnetic duality  \cite{Freed:2018cec} that is equivalent to changing the polarization.

How do we verify this prediction about gauging directly at the level of quantum field theory? One should notice that $SO,Sc$, or $Ss$ is not the ``gauge group'' of the 6d theory and these theories are expected to be ``non-Lagrangian.'' Instead, the names describe the spectrum of strings in each theory. However, once compactified to 5d via $S^1$, they become familiar gauge theories, which we analyze next.

\subsubsection*{Analysis via 5d gauge theories}

Let us start with 6d theory with polarization given by $L_{SO}\subset D$. After reduction to 5d, the theory is characterized by a pure polarization in $\Pol(S^1)$ given by
\begin{equation}
    L_{SO}^{(0)}\oplus L_{SO}^{(1)}\subset D^{(0)}\oplus D^{(1)}=H^*(S^1,D).
\end{equation}
This is neither the geometric polarization $\CP_0$ nor its opposite $\CP_1$, but instead in between. And it has both one-form  $D^{(1)}/L_{SO}^{(1)}=\Z_2^{(1)}$ symmetry and 2-form $D^{(0)}/L_{SO}^{(0)}=\Z_2^{(2)}$ symmetry.

From the point of view of reduction from 6d non-gauge theory to 5d gauge theory, the 2-form symmetry in 6d becomes 1-form $\CZ(SO(8))=\Z_2^{(1)}$ electric center symmetry and 2-form $\pi_1(SO(8))=\Z_2^{(2)}$ magnetic symmetry.
The 1-form symmetry $\mathbb{Z}_2^{(1)}$ acts on Wilson lines as $(-1)^r$ where $r$ denotes the number of boxes in the Young tableaux of the representation.

Gauging the 6d 2-form symmetry sums over the 3-form gauge fields, and that translates into summing over the 3-form and 2-form gauge fields in the 5d theory, namely in 5d gauging {\it both} the one-form and two-form symmetries in 5d. Let us carry out this gauging procedure in 5d in two steps: first gauge the magnetic 2-form symmetry and then the one-form symmetry.
Gauging the 2-form symmetry changes the gauge bundle from $SO(8)$ to Spin$(8)$, and it gives a dual $\mathbb{Z}_2$ one-form symmetry: it acts as $-1$ on all Wilson lines in the spinor representations and $+1$ on all tensor representations, denoted by $(-1)^s$. This $\mathbb{Z}_2'^{(1)}=(\mathbb{Z}_2^{(2)})^\vee$ one-form symmetry is different from the uplift of the original $\mathbb{Z}_2^{(1)}$ one-form symmetry since the latter acts as $-1$ on tensor representations with odd number of boxes in the Young tableaux instead of $+1$. To determine how $\mathbb{Z}_2'^{(1)}$ acts on Spin$(8)$ representations, one in fact needs to choose a lift of $D^{(1)}/L_{SO}^{(1)}$ into $\bar L^{(1)}\subset D^{(1)}$. In our convention, this subgroup is chosen to be $L_{Sc}$. Then, 
gauging the original $\mathbb{Z}_2^{(1)}$ one-form symmetry $(-1)^r$ in the Spin$(8)$ gauge theory gives $Sc(8)$ gauge theory. Notice that the lift of $D^{(0)}/L_{SO}^{(0)}$ to $\bar L^{(0)}\subset D^{(0)}$ is now fixed by isotropy condition on $\bar L=\bar L^{(0)}\oplus \bar L^{(1)}$, so the second step involves no further choices.

The $SO(8)$ and $Sc(8)$ theories are dual to each other by triality with the same partition function, similar to the Kramers--Wannier duality in 2d. Some of the 't Hooft operators in $SO(8)$ gauge theories that are non-genuine line operators in the spinor projective representations are mapped to Wilson lines in the $Sc(8)$ gauge theory similar to the mapping between order-disorder operators in 2d.

How do we get $Ss$ theory? In 5d, when gauging both the one-form and two-form symmetry, one has the choice of adding a local counterterm
\begin{equation}
\pi\int b_2b_3~,    
\end{equation}
where $b_2,b_3$ are the gauge fields for the one-form and two-form symmetries, respectively. This is equivalent to switching to the other choice of $\bar L$ given by $L_{Ss}^{(0)}\oplus L_{Ss}^{(1)}\in H^*(S^1,D)$. After gauging the two-form symmetry in the first step, the dual one-form symmetry $(-1)^s$ is generated by $\int b_3$. Thus, adding this counterterm is equivalent to gauging the diagonal $\mathbb{Z}_2$ one-form symmetry $(-1)^{r+s}$ instead of $(-1)^r$. In other words, adding the local counterterm changes how the original one-form symmetry (that acts on tensor representations) extends to act on spinor representations of Spin$(8)$. Gauging the one-form symmetry then gives $Ss(8)$ gauge theory. 

To get back to the 6d story, one just needs to realize that the 5d local counterterm above is the dimensional reduction of the 6d local counterterm $\pi Q(B^{6\d}_3)$, since with $B_3^{6\d}=B_3+B_2\alpha$ and $\alpha$ the generator of $H^1(S^1,\mathbb{Z})$, we have
\begin{equation}
    \pi Q(B_3^{6\d})=\pi\left( Q(B_3)+Q(B_2\alpha)+\int B_3B_2\alpha\right)=\pi\int_{5\d} B_3B_2~,
\end{equation}
where we used the property that $Q$ is the quadratic refinement of the intersection pairing in 6d.
Thus, we find that, starting with 6d theories with polarization $L_{SO}$ that gives $SO(8)$,
\begin{itemize}
    \item gauging the $\mathbb{Z}_2$ 2-form symmetry gives $Sc(8)$,
    \item gauging $\mathbb{Z}_2$ 2-form symmetry with local counterterm $\pi Q(B_3^{6\d})$ for the 3-form gauge field $B_3^{6\d}$ gives $Ss(8)$.
\end{itemize}
This is in perfect agreement with the prediction by the general theoretic framework for polarization that we developed in this paper.

\subsubsection*{Generalization to higher rank $\mathfrak{so}(8k)$}

As we have emphasized, one benefit of this approach is that it applies \emph{universally}  to any boundary 6d theory as long as then can be coupled to the same $\CT^{\text{bulk}}$. In the present case, it means that the discussion above applies to any 6d theory labeled by $\mathfrak{so}(8k)$ for $k\geq 1$. Namely, the two-form $\Z_2$ symmetry in $SO(8k),Sc(8k)$, or $Ss(8k)$ is again non-anomalous by a similar computation as equation (\ref{eqn:Cartan}), and when it is gauged, these theories behave exactly the same as in the $k=1$ case. 
Notice that for $k>1$, there is no triality and $SO(8k)$ is no longer dual to either $Sc(8k)$ or $Ss(8k)$. Nonetheless, universality implies that the diagram~\ref{fig:SO(8)gauging}, which enjoys triality, still applies with $S^i=\{SO(8k),Sc(8k),Ss(8k)\}$ for $i=0,1,-1$ mod 3.

\section*{Acknowledgement}

We would like to thank Dan Freed, Anton Kapustin, Pavel Putrov, Nathan Seiberg,  Cumrun Vafa, Juven Wang, Edward Witten, and Shing-Tung Yau for illuminating discussions and
comments. We would like to especially thank Nikita Sopenko for participation at the early stage of this project.  
The work of S.G.~is supported by the U.S.~Department of Energy, Office of Science, Office of High Energy Physics, under Award No.~DE-SC0011632, and by the National Science Foundation under Grant No.~NSF DMS 1664227.
The work of P.-S.\ H.\ is supported by the U.S. Department of Energy, Office of Science, Office of High Energy Physics, under Award Number DE-SC0011632, and by the Simons Foundation through the Simons Investigator Award. The work of D.P.~is supported by the Center for Mathematical Sciences and Applications at Harvard
University, and by an NSF grant DMS-0932078, administered by the Mathematical Sciences Research Institute while the author was in residence at MSRI for the program ``
Holomorphic differentials in mathematics and physics'' during the Fall of 2019.

\appendix

\section{Quantization of 7d three-form Chern--Simons level}
\label{sec:7dCSlevel}

If the coefficients $K_{IJ}$ in (\ref{eqn:7dTQFT}) have odd diagonal entries (which can arise for boundary ${\cal N}=(1,0)$ theories), then the 7d theory requires a ``Wu$^c$ structure'' analogous to the more familiar spin$^c$ structure. Let us briefly review it here. Consider
\begin{equation}
\frac{1}{4\pi}\int_{\partial M_8} C_3dC_3=\pi \int_{M_8} \frac{dC_3}{2\pi}\frac{dC_3}{2\pi}~.
\end{equation}
For $C_3$ properly quantized, $\oint dC_3\in 2\pi\mathbb{Z}$, the theory depends on the choice of a $M_8$ that bounds the seven-manifold. The dependence is given by
\begin{equation}
\pi \int_{M_8} \frac{dC_3}{2\pi}\frac{dC_3}{2\pi}=\pi \int_{M_8} \frac{dC_3}{2\pi}v_4\quad\text{mod }2\pi\mathbb{Z}~,
\end{equation}
where $v_4=w_1^4+w_2^2+w_1w_3+w_4$ is the fourth Wu class. For spin manifolds $w_1=w_2=0$ and $w_3=Sq^1(w_2)+w_1w_2=0$, so that we have $v_4=w_4$.
To have a well-defined theory we introduce a background ``Wu$^c$ structure" $X_3$ that satisfies
\begin{equation}
\frac{dX_3}{2\pi}=\frac{1}{2}v_4=\frac{1}{2}w_4\text{ mod }\mathbb{Z}~,
\end{equation}
and couples to the theory as
\begin{equation}
\frac{1}{4\pi}C_3dC_3+\frac{1}{2\pi}C_3dX_3~.
\end{equation}
Then, the 7d theory is independent of $M_8$.\footnote{
This is also consistent with the fact that, when the boundary is an M5-brane, the boundary can be defined with a Wu$^c$ structure as discussed in {\it e.g.} \cite{Sati:2011rw}.
}

We can also define $C_3'=C_3+X_3$. Then, the action can be written as
\begin{equation}
\frac{1}{4\pi}C_3'dC_3'+\cdots
\end{equation}
where $\cdots$ are gravitational corrections independent of $C_3'$. The three-form $C_3'$ obeys a modified quantization condition \cite{Witten:1996md}
\begin{equation}
\oint\frac{dC_3'}{2\pi}\equiv\frac{1}{2}w_4\pmod{\mathbb{Z}}~.
\end{equation}

If the manifold has a spin structure (in particular, is orientable), then there is a natural Wu$^c$ structure given by the gravitational Chern--Simons form.
To see this, note \cite{Wu:1954_3,Thomas:1960}
\begin{equation}
p_1\equiv{\cal P}(w_2)-\text{Bock}(w_1w_2)-2\left(w_1\text{Bock}(w_2)+w_4\right) \pmod 4.
\end{equation}
For spin manifolds we have $w_4=p_1/2$ mod 2, and thus 
\begin{equation}\label{eqn:7dCSquantspin}
\oint\frac{dC_3'}{2\pi}=\frac{1}{4}p_1\equiv\frac{1}{2}v_4\pmod{\mathbb{Z}}~.
\end{equation}
Such shifted quantization condition was already discussed in \cite{Witten:1996md}.

In fact, we can define theories on more general spin$^c$ manifolds, which also have a natural Wu$^c$ structure.
To see this, note that for orientable manifolds \cite{Wu:1954_3,Thomas:1960}:
\begin{equation}
{\cal P}(w_2)=p_1+2w_4\text{ mod }4~,
\end{equation}
where we used the inclusion $\mathbb{Z}_2\rightarrow \mathbb{Z}_4$ by multiplication by 2.
For a spin$^c$ manifold there is a line bundle with $c_1\text{ mod }2=w_2$, and thus ${\cal P}(w_2)=c_1^2$ mod 4. Then, the fourth Wu class satisfies
\begin{equation}
v_4=w_4+w_2^2=\left(p_1+c_1^2\right)/2\text{ mod 2}~,
\end{equation}
The right hand side is an integral class and gives the corresponding Wu$^c$ structure, {\it i.e.} 
\begin{equation}
\oint\frac{dX_3}{2\pi}=\frac{1}{4}\left(p_1+c_1^2\right)\equiv\frac{1}{2}v_4 \pmod {\mathbb{Z}}~,
\end{equation}
with $X_3$ given by the combination of the corresponding gravitational and $U(1)$ Chern--Simons terms. Then, the shifted $C_3'$ obeys 
\begin{equation}\label{eqn:7dCSquantspinc}
\oint\frac{dC_3'}{2\pi}=\frac{1}{4}\left(p_1+c_1^2\right)\equiv\frac{1}{2}v_4 \pmod {\mathbb{Z}}~.
\end{equation}
For spin manifolds, since $c_1$ is even, $c_1=2c_1'$ for another line bundle with first Chern class $c_1'$ and the formula reduces to $v_4=p_1/2$ mod 2. Then, the shifted quantization condition (\ref{eqn:7dCSquantspinc}) reduces to (\ref{eqn:7dCSquantspin}) obtained in \cite{Witten:1996md,Witten:1996hc}.

\section{Discrete theta angle from quadratic refinement}
\label{sec:ArfK}

Consider an oriented manifold $M$ of dimension $4n+2$. There is an intersection form on $H^{4n+2}(M,\mathbb{Z}_2)$ given by
\begin{equation}
H^{2n+1}(M,\mathbb{Z}_2)\times H^{2n+1}(M,\mathbb{Z}_2)\rightarrow\mathbb{Z}_2:\quad
(\alpha,\alpha') \;\mapsto\; \int_M \alpha\cup \alpha'\bmod 2~.
\end{equation}
It has a quadratic refinement\footnote{
The quadratic function on a general manifold is a $\mathbb{Z}_4$-valued function. For orientable manifolds, since $\alpha\cup\alpha =\alpha \cup v_{2n+1}$ mod 2 and the $(2n+1)$-th Wu class $v_{2n+1}$ is trivial, the quadratic function is $\mathbb{Z}_2$-valued.}
\begin{equation}
Q:\quad H^{2n+1}(M,\mathbb{Z}_2)\rightarrow\mathbb{Z}_2~
\end{equation}
that obeys
\begin{equation}
Q(B_{2n+1}+B_{2n+1}')=
Q(B_{2n+1})+Q(B_{2n+1}')+\int_M B_{2n+1}\cup B_{2n+1}'\bmod 2~.
\end{equation}
From the quadratic refinement one can define the Arf--Kervaire invariant \text{AK} as follows \cite{Arf:1941,Kervaire:1969,Browder:1969,Brown:1972}: first take a symplectic basis $(\alpha^I,\beta^I)$ in $H^{2n+1}(M,\mathbb{Z}_2)$, such that $\int_M \alpha^I\alpha^J=\int_M \beta^I\beta^J=0$, $\int_M \alpha^I\beta^J=\delta^{IJ}$. Then, the Arf--Kervaire invariant \text{AK} is
\begin{equation}
\text{AK}=\sum_{I} Q(\alpha^I)Q(\beta^I)~.
\end{equation}
The AK invariant is independent of the choice of the symplectic basis \cite{Arf:1941}. It is also given by a Gaussian sum of the quadratic function
\begin{equation}\label{eqn:AKSPT}
    \pi \text{AK}=\text{Arg}\sum_{b_{2n+1}\in H^{2n+1}(M,\mathbb{Z}_2)}(-1)^{Q(b_{2n+1})}~.
\end{equation}

From the quadratic function $Q$ we can define a 
 $(2n+1)$-form $\mathbb{Z}_2$ gauge theory with the following topological action: denote the gauge field by $B_{2n+1}$, the action is
\begin{equation}
\pi pQ(B_{2n+1}),\quad p=0,1~.
\end{equation}
The physical properties of the $\mathbb{Z}_2$ gauge theory will be discussed in detail in \cite{Hsin:2021qiy}.

\subsubsection*{Symmetry extension}

The above $\mathbb{Z}_2$ $(2n+1)$-form gauge theory has the following property. Suppose $b_{2n+1}$ obeys the twisted cocycle condition
\begin{equation}
\delta b_{2n+1}=\mu_{2n+2}~,
\end{equation}
which happens if $B_{2n+1}$ participates in a non-trivial group extension.
Denote the background of the dual $2n$-form $\mathbb{Z}_2$ symmetry by $\hat B_{2n+1}$, then its coupling to $b_{2n+1}$ together with the discrete theta angle combines into $\pi Q(b_{2n+1}+\hat B_{2n+1})$, where
\begin{equation}
\delta \hat B_{2n+1}=\mu_{2n+2}
\end{equation}
and thus $\delta (b_{2n+1}+\hat B_{2n+1})=0$ and the quadratic function $Q(b_{2n+1}+\hat B_{2n+1})$ is well-defined. This implies that the dual $2n$-form symmetry participates in a group extension,
The discussion generalizes the 2d situation found in \cite{Hsin:2020nts}.

\section{Discrete theta angles in 5d gauge theories}
\label{sec:5ddiscretetheta}

\subsection*{Continuous notation}

Consider the $\mathbb{Z}_N$ two-form gauge theory
\begin{equation}
\frac{k}{4\pi}b_2db_2+\frac{N}{2\pi}b_2da_2~.
\end{equation}
The first term does not contribute to the equations of motion since locally it can be written as a total derivative. The equations of motion imply
\begin{equation}
    e^{iN\oint a_2}=1,\quad
    e^{iN\oint b_2}=1~.
\end{equation}

If we consider $(\mathbb{Z}_N)^r$ two-form gauge theory with mixed discrete theta term
\begin{equation}
    \sum_{I}\frac{N}{2\pi}b_2^Ida_2^I+\sum_{I<J}\frac{q_{IJ}}{2\pi}b_2^Idb_2^J~,
\end{equation}
where $q_{IJ}=-q_{JI}$, and $I,J=1,\cdots r$.
The equations of motion imply
\begin{equation}\label{eqn:5d2form}
e^{iq_{IJ}\oint b_2^J+iN\oint a_2^J}=1,\quad
e^{Ni\oint b_2^I}=1~.
\end{equation}

We can couple the two-form gauge theory to 5d gauge theory with gauge group $G$ and $\pi_1(G)=\mathbb{Z}_N$ by identifying $e^{i\oint b_2^I}$ with the generator of the magnetic two-form symmetry $\pi_1(G)$ \cite{Hsin:2020nts}.
Then the magnetic string is attached to $e^{i\oint a_2^I}$, and thus from (\ref{eqn:5d2form}) it carries electric charge $q_{IJ}/N$ for the electric one-form gauge symmetry corresponding to $b_2^J$ analogous to the fractional quantum Hall effect.

We remark that since the magnetic string is not topological for continuous $G$, the two-form symmetry generated by $e^{i\oint a_2^I}$ is explicitly broken by coupling to the $G$ gauge theory, and it is left with the $\pi_1(G)$ two-form symmetry generated by $e^{i\oint b_2^I}$.

We can also reduce the 5d discrete theta angle to 4d by decomposing $b_2=b_2'+b_1'{d\varphi\over 2\pi}$, where $\varphi$ is the coordinate on the compactified $S^1$, $b_2',b_1'$ are fields in 4d. 
Then the 5d discrete theta angle $(q_{IJ}/2\pi)\int b_2^Idb_2^J$ becomes 
\begin{equation}
\frac{q_{IJ}}{2\pi}\int \left(b_2'^I+b_1'^I{d\varphi\over 2\pi}\right)
 \left(db_2'^J+db_1'^J{d\varphi\over 2\pi}\right)
=\frac{q_{IJ}}{2\pi}\int_{4d} \left(b_2'^Idb_1'^J-(I\leftrightarrow J)\right)
~.
\end{equation}
Thus, the 4d mixed discrete theta angle, to be discussed in 
\cite{Gukov:2020inprepare}, can be obtained from reduction of the 5d mixed discrete theta angle.

\subsection*{Discrete notation}

\paragraph{Odd N}
The discrete theta angle for $\mathbb{Z}_N$ with odd $N$ is trivial:
\begin{equation}
    \frac{2\pi k}{N}\int B_2{\delta B_2\over N}=
    \frac{2\pi k(1+N)/2}{N}\int 2B_2{\delta B_2\over N}
    ={2\pi k(1+N)/2\over N}\int {\delta {\cal P}(B_2)\over N}\equiv 0\pmod {2\pi\mathbb{Z}}~.
\end{equation}
Thus we will focus on even $N$.

\paragraph{Even N}
For even $N$, ${\cal P}(B_2)=B_2B_2-\delta B_2\cup_1 B_2$ and we have the formula between cochains:
\begin{equation}
\delta {\cal P}(B_2)=2\delta B_2B_2+\delta B_2\cup_1\delta B_2~.
\end{equation}
Thus
\begin{equation}
B_2\frac{\delta B_2}{N}=\frac{1}{2N}\delta {\cal P}(B_2)
-\frac{N}{2}
    \left(\frac{\delta B_2}{N}\right)\cup_1 \left(\frac{\delta B_2}{N}\right)~,
\end{equation}
where we note each term is expressed using Bockstein which can be lifted to an integral class (note ${\cal P}(B_2)$ is a $\mathbb{Z}_{2N}$-valued cocycle), and it is well-defined mod $N$.\footnote{However, due to the $\cup_1$ product, the last term cannot be lifted to an integral cocycle: 
\begin{equation}
    \frac{N}{2}\delta \left(\left(\frac{\delta B_2}{N}\right)\cup_1 \left(\frac{\delta B_2}{N}\right)\right)
    =-N\left(\frac{\delta B_2}{N}\right)\left(\frac{\delta B_2}{N}\right)~.
\end{equation}
Note that one cannot write the action $\frac{\pi}{N^2}\int B_2\delta B_2$ since it depends on the lift of $B_2$: changing $B_2\rightarrow B_2+N y_2$ for some integral 2-cochain $y$ changes the action by
\begin{equation}
\pi\int y_2\delta y_2+
{\pi\over N}\int\left( y\delta B_2-\delta B_2 y\right)=
\pi\left(\int y_2\delta y_2-\left({\delta B_2\over N}\right)\cup_1 \delta y_2\right)~.
\end{equation}

}
Thus
\begin{equation}\label{eqn:5dthetaidentity}
\frac{2\pi}{N}\int B_2\frac{\delta B_2}{N}
=\pi\int \left(\frac{\delta B_2}{N}\right)\cup_1 \left(\frac{\delta B_2}{N}\right)
=
\pi\int w_2\left(\frac{\delta B_2}{N}\right)~,
\end{equation}
where we used $\pi \int x_3\cup_1 x_3=\pi\int Sq^2(x_3)=\pi\int w_2 x_3$ on an orientable manifold for $\mathbb{Z}_2$ cocycle $x_3=\delta B_2/N$ mod $2$.
We can rewrite it as
\begin{equation}
{2\pi\over N}\int {\delta w_2\over 2} B_2=
{2\pi\over N}\int W_3 B_2~,
\end{equation}
where $W_3=\delta w_2/2=\text{Bock}(w_2)$ is the third integral Stiefel--Whitney class on an orientable manifold. The case of $N=2$ is also obtained in \cite{Kapustin:2017jrc}.
This means that the discrete theta angle does not affect the spectrum of operators, and what it does is attaching to the magnetic string that carries unit flux of $\int B_2$ with
\begin{equation}
\frac{2\pi}{N}\int W_3~.
\end{equation}
When $N=2$, this is interpreted as the magnetic string being ``fermionic'' in the sense that it depends on the framing specified by a trivialization of $w_3=W_3$ mod 2 \cite{Thorngren:2014pza}.

\paragraph{General case}

Let us consider the discrete theta angle for ${\cal A}=\prod_I\mathbb{Z}_{N_I}$
\begin{equation}\label{eqn:5ddiscretetheta1}
\sum_{I<J}\frac{2\pi q_{IJ}}{N_I}\int b_2^I \frac{\delta b_2^J}{N_J}~,
\end{equation}
where $q_{IJ}=-q_{JI}=0,1,\cdots \gcd(N_I,N_J)-1$.
Note $q_{IJ}=\gcd(N_I,N_J)$ is equivalent to $q_{IJ}=0$ since there exits integers $m,n$ such that $\gcd(N_I,N_J)=nN_I+mN_J$.

In the following we will compute how symmetry and its 't Hooft anomaly depend on the discrete theta angle (\ref{eqn:5ddiscretetheta1}).
Consider turning on a $\mathbb{Z}_{N_I}$ background $Y_3^I$ such that
\begin{equation}\label{eqn:5ddiscretetheta2formgaugeY3}
    \delta b_2^I = Y_3^I~.
\end{equation}
The discrete theta angle action needs modification to ensure that it is independent of the lift of $B_2^I$. Changing the lift $b_2\rightarrow b_2^I+N_I u_2^I$ for integer cochain $u_2^I$ changes the action by
\begin{equation}
2\pi\sum_{I<J} q_{IJ}\int\left( {u_2^IY_3^J\over N_J}-{Y_3^Ju_2^I\over N_I}\right)~.
\end{equation}
Thus we add the following term 
\begin{equation}\label{eqn:5ddiscretetheta2}
-2\pi\sum_{I<J}\frac{q_{IJ}}{N_IN_J}\int
    \left(b_2^IY_3^J-Y_3^Ib_2^J\right)~.
\end{equation}
The discrete theta angle in the presence of background $Y_3^I$ is the sum of (\ref{eqn:5ddiscretetheta1}),(\ref{eqn:5ddiscretetheta2}).
In addition, we can add the following coupling to the $\mathbb{Z}_{N_I}$ background $B_3^I$ for the two-form symmetry generated by $\oint B_2^I$,
\begin{equation}
\sum_I\frac{2\pi}{N_I}\int b_2^I B_3^I~.
\end{equation}

Next, we examine the gauge invariance for the background gauge field. We extend the fields to a 6d bulk and demand that the dynamical field $b_2^I$ to be independent of the bulk
\begin{align}
&\sum_{I<J}\frac{2\pi q_{IJ}}{N_IN_J}\int_{6\d} \left(\delta b_2^I-Y_3^I\right)
\left(\delta b_2^J-Y_3^J\right)
+\sum_{I<J}\frac{2\pi q_{IJ}}{N_IN_J}\int_{6\d} \left(b_2^J\delta Y_3^I-b_2^I\delta Y_3^J\right)\cr
&-\sum_{I<J}\frac{2\pi q_{IJ}}{N_IN_J}\int_{6\d} \left(Y_3^IY_3^J-\delta Y_3^I\cup_1 Y_3^J\right)
+\sum_I\frac{2\pi}{N_I}\int_{6\d} 
    \left( b_2^I\delta B_3^I+Y_3^IB_3^I\right)~.
\end{align}
The first is trivial since $\delta b_2^I=Y_3^I$ mod $N_I$, and we will drop it from the bulk action.
The terms involving $b_2^I$ need to cancel, and thus leading to extension of symmetries with modified background fields
\begin{equation}\label{eqn:5ddiscretethetasym}
\delta B_3^I=\sum_J q_{IJ}\text{Bock}(Y_3^J)~,
\end{equation}
where $q_{IJ}=-q_{JI}$ and Bock is the Bockstein homomorphism for $1\rightarrow\mathbb{Z}_{N_J}\rightarrow\mathbb{Z}_{N_J^2}\rightarrow\mathbb{Z}_{N_J}\rightarrow 1$.
The remaining terms represent an 't Hooft anomaly for the two-form symmetry
\begin{equation}\label{eqn:5ddiscretethetaanomaly}
    -\sum_{I<J}\frac{2\pi q_{IJ}}{N_IN_J}\int_{6\d} \left(Y_3^IY_3^J-\delta Y_3^I\cup_1 Y_3^J\right)
+\sum_I\frac{2\pi}{N_I}\int_{6\d} Y_3^IB_3^I~,
\end{equation}
where the first term comes from the discrete theta  angle $q_{IJ}$, while the second term comes from the coupling $\int b_2^I B_3^I$ and it is universal for all discrete theta angles.
The equations (\ref{eqn:5ddiscretethetasym}) and (\ref{eqn:5ddiscretethetaanomaly}) are the main results of this appendix.

5d gauge theory with discrete theta angle can be obtained from the theory without discrete theta angle by coupling to the two-form gauge theory  (\ref{eqn:5ddiscretetheta1}) as in \cite{Hsin:2020nts}. If the zero discrete theta angle theory has symmetry extension with modified $\delta B_3^I\neq 0$, which means there is gauge-global anomaly for $b_2^I$, then a nonzero discrete theta angle gives extra gauge-global anomaly that  modifies the symmetry extension by additional term in $\delta B_3^I$ given by the right-hand side of (\ref{eqn:5ddiscretethetasym}). The discrete theta angle contributes to an additional 't Hooft anomaly given by (\ref{eqn:5ddiscretethetaanomaly}).
 
For instance, consider $SO(6)\times SO(6)'$ gauge theory (with a prime to distinguish the two $SO(6)$) in 5d with the mixed discrete theta angle $\pi\int w_2(SO(6))\cup \text{Bock}( w_2(SO(6)'))$. Denote the background gauge fields for the residual electric one-form symmetry by $B,B'$ and the background gauge fields for the charge conjugation 0-form symmetry by $B_1,B_1'$.
Then this corresponds to $Y^1=\text{Bock}(B)+BB_1$ and $Y^2=\text{Bock}(B')+B'B'_1$. When $B_1=B_1'=0$ the mixed discrete theta angle $q_{12}=1$ does not modify the symmetry or 't Hooft anomaly (on an orientable manifold {\it i.e.} without time-reversal symmetry) since $\text{Bock}(\text{Bock}(B))=0$ and $\int Y^1Y^2=\int \text{Bock}(B\text{Bock}(B'))=0$.
For $B_1,B_1'$ non-trivial there is modification in the symmetry and anomaly as in (\ref{eqn:5ddiscretethetasym}) and (\ref{eqn:5ddiscretethetaanomaly}).

\section{A mixed discrete theta angle}
\label{sec:mixeddiscretetheta}

Consider the discrete theta angle for a $\mathbb{Z}_N$ $n$-form gauge field $x_n$ and $(d-n)$-form gauge field $y_{d-n}$
\begin{equation}\label{eqn:xytheta}
\frac{2\pi p}{N}\int x_n y_{d-n},
\quad p=0,1,\cdots N-1~.
\end{equation}
Suppose we are given theory 1 with non-anomalous $(n-1)$-form $\mathbb{Z}_N$ symmetry, and theory 2 with non-anomalous $(d-n-1)$-form $\mathbb{Z}_N$ symmetry, then we can gauge these symmetries and there are different ways of gauging the symmetries correspond to adding (\ref{eqn:xytheta}).
\begin{equation}
\sum_{x,y}Z_1[x]Z_2[y]e^{2\pi i p/N\int xy}=\sum_x Z_1[x]Z_2'[px],\quad
Z_2'[px]=\sum_y Z_2[y]e^{2\pi i p/N\int x y}.
\end{equation}
This can be expressed as
\begin{equation}
\sum_{x,u,v}Z_1[x]Z_2[u]e^{{2\pi i\over N}\int v(u-px)}~.
\end{equation}
With the partition function of $
T_1/\mathbb{Z}_N$ and $T_2$ being respectively $ \sum_x Z_1[x]$ and $Z_2[0]$,
we will denote the resulting theory as
\begin{equation}
{(T_1/\mathbb{Z}_N)\times T_2\over \mathbb{Z}_N}~
\end{equation}
where the quotient denotes gauging the diagonal $\mathbb{Z}_N$ symmetry that acts on $T_2$ as the $\mathbb{Z}_N$ $(d-n-1)$-form symmetry and on $T_2/\mathbb{Z}_N$ as the $\mathbb{Z}_L\subset \mathbb{Z}_N$ dual $(d-n-1)$-form symmetry with $L=\gcd(p,N)$. 

Here it worth emphasizing that in the context of discrete theta angles that arise from polarizations, the above mixed discrete theta angle (\ref{eqn:xytheta}) actually often comes from a \emph{pure} polarization. 

To give an example of such theta angles, one can take two copies of $(1+1)$-dimensional Ising model (denoted by Ising$_2$) having $\mathbb{Z}_2\times \mathbb{Z}_2$ symmetry. Gauging the symmetries with mixed discrete theta angle given by the non-trivial element in $H^2(\mathbb{Z}_2\times \mathbb{Z}_2,U(1))=\mathbb{Z}_2$ produces
\begin{equation}
    {\text{Ising}_2/\mathbb{Z}_2\times \text{Ising}_2\over \mathbb{Z}_2}\leftrightarrow    { \text{Ising}_2\times \text{Ising}_2\over \mathbb{Z}_2}\leftrightarrow U(1)_4~,
\end{equation}
where the first duality used the property $    \text{Ising}_2/\mathbb{Z}_2\leftrightarrow \text{Ising}_2$,  and the second duality used the property that the gauging the $\mathbb{Z}_2$ symmetry in the compact boson theory $U(1)_4$ produces Ising$_2$ $\times $ Ising$_2$ \cite{Dijkgraaf:1989hb}, and thus gauging the dual $\mathbb{Z}_2$ symmetry in Ising$_2$ $\times $ Ising$_2$ (which is the diagonal $\mathbb{Z}_2$ symmetry, since $\text{Ising}_2/\mathbb{Z}_2\leftrightarrow\text{Ising}_2$) gives back $U(1)_4$.
This reproduces the result in Appendix D of \cite{Hsin:2020nts}.

\section{Linking forms}
\label{sec:linkingform}

As a warm-up, let us consider  a pair of closed, oriented, disjoint manifolds $M_n$, $M_m \subset \mathbb{R}^{m+n+1}$. The linking number Lk$(M_n,M_m)$ is defined as the degree of the map
\begin{equation}
\begin{matrix}
f: & M_n \times M_m & \to & S^{m+n} \\
& (x,y) & \mapsto & \frac{x-y}{|x-y|}.
\end{matrix}
\end{equation}
It is easy to see that, for any $m$ and $n$, there are examples of submanifolds with non-zero linking number, {\it e.g.}~a pair of spheres $S^n \subset \mathbb{R}^{n+1}$ and $S^m \subset \mathbb{R}^{n+1}$ in $S^{m+n+1}$ (= sphere in $\mathbb{R}^{n+1} \times \mathbb{R}^{m+1} = \mathbb{R}^{m+n+2}$).

This can be generalized to linking number on arbitrary manifold as follows.
In $d$ dimensions we can define linking form for torsion cycles of degrees $\ell$ and $d-\ell-1$ as follows.
Denote two such cycles by $\alpha_\ell,\alpha'_{d-\ell-1}$.
Then there exists an integer $K$, (for $\mathbb{R}^{m+n+1}$ it is the identity)
\begin{equation}
    K\alpha_\ell=\partial \gamma_{\ell+1}~.
\end{equation}
Then the linking form is defined using the intersection number between $\gamma_{\ell+1}$ and $\alpha'_{d-\ell-1}$,
\begin{equation}
    \text{Link}(\alpha_\ell,\alpha'_{d-\ell-1}):=K^{-1}\#(\gamma_{\ell+1},\alpha'_{d-\ell-1}).
\end{equation}
can be expressed as follows.
Denote 
\begin{equation}
{K\over 2\pi}B_{d-\ell-1}=\text{PD}(\gamma_{\ell+1}),\quad
{dB'_{\ell}\over 2\pi}=\text{PD}(\alpha'_{d-\ell-1})~,
\end{equation}
then the linking form is
\begin{equation}
\frac{1}{2\pi}\int B_{d-\ell-1}dB_{\ell}
=
(-1)^{(d-\ell)(\ell+1)}\frac{1}{2\pi}\int B_{\ell}dB_{d-\ell-1}~,
\end{equation}
where the second expression used integration by parts.
Thus the linking form on torsion cycles has the symmetry property
\begin{equation}
    \text{Link}(\alpha_\ell,\alpha_{d-\ell-1})=(-1)^{d(\ell+1)}
    \text{Link}(\alpha_{d-\ell-1},\alpha_\ell)~.
\end{equation}
If in the definition of linking form we take the intersection with $\gamma'_{d-\ell}$ with $\partial\gamma'_{d-\ell}=K'\alpha_{d-\ell-1}$, then the sign is replaced with $(-1)^{d\ell-1}$ from the additional contribution $(-1)^{(d-\ell)\ell-(d-\ell-1)(\ell+1)}$.

For instance, the linking between torsion one-cycles in three space-time dimension $d=3$ is symmetric, while the linking between torsion 2-cycles in $d=5$ is antisymmetric.

\section{Dualities for 7d three-form Chern--Simons theory} \label{sec:7ddualities}

Just as in the usual Chern--Simons theory, the three-form Chern--Simons theory for $A_{N-1}$ enjoys a similar version of level-rank duality \cite{Hsin:2016blu}: it is dual to the theory similar to $U(1)_{-N}$ {\it i.e.}~the level-$(-N)$ Chern--Simons theory with a single three-form gauge field, up to an invertible TQFT.
To see this, we can write the theory as
\begin{equation}\label{eqn:An-1theory}
\frac{2}{4\pi}C^1dC^1+\sum_{I=2}^{N-1}\left(-\frac{1}{2\pi}C^{I-1}dC^I+\frac{2}{4\pi}C^IdC^I\right)~.
\end{equation}
The first term can be rewritten using
\begin{equation}
\frac{2}{4\pi}C^1dC^1-\frac{1}{4\pi}C^0dC^0\quad\longleftrightarrow\quad \frac{1}{4\pi}C^1dC^1-\frac{1}{2\pi}C^1dx-\frac{1}{4\pi}xdx~,
\end{equation}
where $x,C^0$ are three-forms related by $x=C^0-C^1$ and $-\frac{1}{4\pi}C^0dC^0$ is an invertible TQFT.
Using this duality we can rewrite (\ref{eqn:An-1theory}) as a level-one Chern--Simons theory and
\begin{equation}
-\frac{1}{4\pi}xdx+\frac{1}{4\pi}C^1dC^1-\frac{1}{2\pi}C^1d(C^2+x)
+\sum_{I=3}^{N-1}\left(-\frac{1}{2\pi}C^{I-1}dC^I+\frac{2}{4\pi}C^IdC^I\right)~.
\end{equation}
Redefining $C^1\rightarrow C^1+C^2+x$ gives
\begin{equation}
-\frac{2}{4\pi}xdx+\frac{1}{4\pi}C^1dC^1
+\frac{1}{4\pi}C^2dC^2-\frac{1}{2\pi}C^2d(C^3+x)
+\sum_{I=4}^{N-1}\left(-\frac{1}{2\pi}C^{I-1}dC^I+\frac{2}{4\pi}C^IdC^I\right)~.
\end{equation}
Repeating the steps for $C^2,\cdots C^{N-1}$ we find
\begin{equation}
-\frac{N}{4\pi}xdx+\sum_{I=0}^{N}\frac{1}{4\pi} C^IdC^I~,
\end{equation}
where the later $\sum_{I=0}^{N}\frac{1}{4\pi} C^IdC^I$ is an invertible TQFT.
The duality is discussed in \cite{Monnier:2017klz}. This is consistent with the effective action from supergravity \cite{Witten:1998wy}.

Similarly, the $D_N$ theory is dual to
\begin{equation}
-\frac{N}{4\pi}xdx+\frac{2}{2\pi}xdy
\end{equation}
for three-forms $x,y$, up to an invertible TQFT. This is an analogue of the ordinary Chern--Simons duality discussed in \cite{Aharony:2016jvv,Cordova:2017vab}.
Another duality is between the $E_N$ theory and the time-reversal image of $A_{8-N}$ tensored with the $E_8$ theory, which is the counterpart of the duality in the usual Chern--Simons theory discovered in \cite{Cordova:2018qvg}.

Another class of dualites comes from the analogue of fractionalization map discussed in \cite{Hsin:2019gvb}. There the operators are lines, and their spin can be changed by $\frac{1}{2}$ if one turns on a background $B_2=w_2$ (which is $v_2$ on orientable manifolds) for the $\mathbb{Z}_2$ one-form symmetry that acts on the lines; here the operators are three-dimensional, and their spins can be changed by $\frac{1}{2}$ by turning on a background $B_4=v_4$ for the $\mathbb{Z}_2$ three-form symmetry that acts on the three-dimensional operators. This amounts to inserting the generator of the $\mathbb{Z}_2$ subgroup three-form symmetry (which is a combination of $\oint C$) at the Poincar\'e dual of $v_4$.

For instance, the theory associated with ${\frak g}=\mathfrak{so}(8)$ can be expressed as a $\mathbb{Z}_2$ three-form gauge theory coupled to $B_4=v_4$ for the one-form symmetry generated by $em$.
Concretely, if we write the $\mathbb{Z}_2$ gauge theory as $\frac{2}{2\pi}CdC'$ with three-form $U(1)$ gauge fields $C,C'$, then the theory for ${\frak g}=\mathfrak{so}(8)$ is obtained by turning on the background $B_4=v_4$ for the one-form symmetry generated by the fermion $\oint (C+C')$. This changes the spin of the $e$ operator $\oint C$ and $m$ operator $\oint C'$ from boson to fermion. It is the analogue of the ``efmf'' theory in 2+1 dimensions.

\section{Factorization of fermionic Abelian Chern--Simons theory}\label{sec:factorize}

In this appendix we show that a fermionic Abelian Chern--Simons theory in 7d or 3d (with three-form or one-form gauge field, respectively) always factorizes into the product of a bosonic Abelian TQFT and an invertible fermionic TQFT that consists of two objects: the identity and the transparent fermionic object $f$.

The proof is by contradiction.
Suppose the contrary, {\it i.e.}~the $\mathbb{Z}_2$ generated by $f$ participates in a non-trivial extension of the operator algebra. Then, there exists another object $x$ that obeys $\mathbb{Z}_4$ fusion rule such that
\begin{equation}
    x^2=f,\quad x^3=x^{-1}=fx~.
\end{equation}
Let us compare the spin of $fx$ with the spin of $x^{-1}$; they should agree according to the above equation.
In an Abelian TQFT, the spin of $fx$ can be obtained from the spin of $f$ and the spin of $x$ and their braiding, {\it i.e.}~the spin is a quadratic refinement of the braiding. The spin of the fermionic object $f$ is $1/2$. The object $f$, being transparent, does not have braiding with $x$, so the spin of $fx$ is the sum of the spins of $f$ and $x$, {\it i.e.}~$1/2 +$(spin of $x$).
On the other hand, the spin of $x^{-1} =f x$ is the same as the spin of $x$, being its inverse. Then we have a contradiction
\begin{equation}
 \text{spin of }x = 1/2+\text{spin of }x~.   
\end{equation}
Namely, no such $x$ exists. In other words, an arbitrary Abelian fermionic Chern--Simons theory in 3d or $7\d$ factorizes into the product of a bosonic Abelian TQFT and an invertible fermionic TQFT.

\bibliographystyle{utphys}
\bibliography{Draft1}{}

\end{document}